\documentclass[onecolumn,notitlepage,nofootinbib,amsmath,amssymb,showkeys,aps]{revtex4-1} % one-column style
\usepackage[utf8]{inputenc}
\usepackage{textgreek}
\usepackage{siunitx}
\usepackage{amssymb}
\usepackage{url}
\usepackage{hyperref}
\usepackage{array,multirow}
\usepackage{enumerate}
\usepackage{bm}
\usepackage{mathtools}
\usepackage{makecell}

\usepackage{amsmath}
\usepackage{graphicx}
\usepackage{xcolor}

\usepackage{ulem}
\usepackage{cancel}

\def\B{\mathrm{B}}
\def\F{\mathrm{F}}
\definecolor{grey}{rgb}{0.5,0.5,0.5}

\usepackage[flushleft]{threeparttable}

\usepackage{comment}
\usepackage{hyperref}

\begin{document}

\begingroup
\renewcommand\thefootnote{a}
\footnotetext{Z.Z. and M.S. contributed equally to this work.}
\endgroup

\title{Testing the cosmological Euler equation: viscosity, equivalence principle, and gravity beyond general relativity} 

\author{Ziyang Zheng$^{1,a}$}
\email{Zheng@thphys.uni-heidelberg.de}
\author{Malte Schneider$^{1,a}$}
\email{m.schneider@thphys.uni-heidelberg.de}
\author{Luca Amendola$^{1.2}$}
\email{l.amendola@thphys.uni-heidelberg.de}
\affiliation{$^1$ Institut f\"ur Theoretische Physik, Ruprecht-Karls-Universit\"at Heidelberg, Philosophenweg 16, D-69120 Heidelberg, Germany}
\affiliation{$^2$ New York University Abu Dhabi, PO Box 129188, Abu Dhabi, United Arab Emirates and Center for Astrophysics and Space Science (CASS), New York University Abu Dhabi}

\begin{abstract}
We investigate how the cosmological Euler equation can be tested in the presence of viscous dark matter, violations of the equivalence principle (EP), and modifications of gravity, while relying on minimal theoretical assumptions. Extending the analysis of \cite{Castello:2024lhl}, we generalize the observable $E_P$, which quantifies EP violation, to $\tilde{E}_P$, discuss the degeneracy between bulk and shear viscosities and EP-violating effects, and explicitly show that the EP can still be tested in the small-viscosity limit. In addition, we identify a model-independent observable, $C_{\rm vis,0}$, which characterizes the present-day dark matter viscosity and can be measured from relativistic galaxy number counts by cross-correlating two galaxy populations. We perform forecasts for three forthcoming Stage-IV surveys: DESI, Euclid, and SKA Phase 2 (SKA2), and find that $C_{\rm vis,0}$ can be tightly constrained, at the level of $\mathcal{O}(10^{-6})$ or better in all cases. Among these surveys, SKA2 provides the tightest constraint, with a $1\sigma$ uncertainty of $1.08 \times 10^{-7}$ on $C_{\rm vis,0}$. 
\end{abstract}

\keywords{Cosmology: theory-- Cosmology: observations -- dark matter -- modified gravity}

\maketitle

%%%%%%%%%%%%%%%%%%%%%%%%%%%%%%%%%%%%%%%%%%%%%
%%%%%%%%%%%%%%%%%%%%%%%%%%%%%%%%%%%%%%%%%%%%%
%%%%%%%%%%%%%%%%%%%%%%%%%%%%%%%%%%%%%%%%%%%%%

\section{Introduction}\label{sec:intro}

The cosmological Euler equation is a cornerstone of large-scale structure (LSS) formation, as it governs the evolution of velocity perturbations and determines how matter responds to gravitational potentials. Moreover, it encodes the weak equivalence principle, which states that all test particles fall identically in a gravitational field. When testing this relation, it is commonly assumed that dark matter (DM) behaves as a ``cold'', pressureless, perfect fluid. This assumption is made both in forecasts  (e.g.\ \cite{Euclid:2019clj,Castello:2024lhl})
and in the analysis of real data (e.g.\ \cite{Grimm:2025fwc})
%\cite{Grimm:2025fwc}. 
However, the microscopic nature of dark matter remains unknown, leaving room for more general scenarios when testing gravity on cosmological scales. In this work, we lift the perfect-fluid assumption and allow dark matter to exhibit dissipative properties, focusing on dark matter viscosity. 

Deviations from the standard Euler and Poisson equations can lead to scenarios involving EP violation and modified gravity. These modifications can generally be parameterized by several phenomenological free functions that may depend on both time and scale. For instance, deviations in the Euler equation can be captured by an additional friction term $\Theta$ and a fifth-force term $\Gamma$ acting on dark matter (see e.g.\ \cite{Amendola:2003wa,Bonvin:2018ckp}). On the other hand, modifications to the Poisson equations are commonly described by an effective gravitational coupling $\mu_{\rm G}$ and a gravitational slip $\eta$ ( see e.g.\ \cite{Pogosian:2010tj, Amendola:2007rr, Amendola:2013qna}). The main goal of this paper is to identify the effective combinations of parameters of the Euler equation that can be extracted from LSS observations.
As a consequence, we will also find to what extent the parameters that encode the modified gravity and EP violation effects are degenerate with dark matter viscosity. In order to obtain robust conclusions, we remain agnostic about the specific form of the free functions, aiming to cover as broad a class of models as possible. 

Viscosity in dark matter may arise from various microscopic processes and can affect cosmic evolution in different ways. At the background level, only the isotropic bulk viscosity contributes, while shear viscosity becomes important at the perturbation level. Previous studies (see e.g.\ \cite{Barbosa:2017ojt, Velten:2012uv, Velten:2013pra, Velten:2014xca}) have shown that viscous dark matter can suppress the growth of structure on small scales. We revisit this result and derive the corresponding perturbation equations in a more general setup. Specifically, we focus on how viscosity modifies the perturbed Euler and Einstein equations in the broader context of both EP violation and modified gravity. 

Compared to the ``standard" way of testing gravity with redshift-space distortions (see e.g.\ \cite{Zheng:2023yco, Sakr:2025kgq}), relativistic effects in galaxy number counts  \cite{Bonvin_2011, Challinor:2011bk} provide a powerful probe for testing gravity on large cosmological scales, especially as they offer a way to test gravity without assuming the standard Euler equation \cite{Sobral-Blanco:2021cks, Tutusaus:2022cab, Castello:2024jmq}. These include corrections such as gravitational redshift, Doppler corrections, (Integrated) Sachs-Wolfe effects, and lensing contributions in the observed galaxy power spectra. Including these corrections becomes increasingly important for upcoming surveys that will be able to detect much larger cosmic volumes, where relativistic effects can no longer be neglected. Forthcoming Stage-IV surveys, such as DESI \cite{DESI:2016fyo}, Euclid \cite{Euclid:2024yrr}, and SKA2 \cite{Bull:2015lja}, will enhance the detectability of relativistic signals and potential deviations from standard gravity. 
In \cite{Castello:2024lhl}, a new observable $E_P$ was defined as a linear combination of $\Theta$ and $\Gamma$, where $E_P \neq 1$ denotes EP violation. 
It provides a null and model-independent test of EP and is sensitive to scenarios such as coupled quintessence \cite{Amendola:1999er}. 
In this work, we investigate whether the observable $E_P$, or its viscous generalization $\tilde{E}_P$, remains a smoking gun for EP violation in the presence of dark matter viscosity.

This work goes beyond previous studies by developing, for the first time, a pipeline to measure the dimensionless viscosity parameter $C_{\rm vis,0}$ at cosmological scales in a significantly more model-independent way. While previous studies have made efforts to constrain dark matter viscosity on both cosmological and smaller astrophysical scales \cite{Velten:2012uv, Velten:2013pra, Velten:2014xca, Barbosa:2017ojt, Ashoorioon:2023jwf, AparicioResco:2020shd, Anand:2017wsj, Goswami:2016tsu}, they typically rely on strong assumptions, such as the validity of  specific parametrizations of structure growth. In contrast, our approach avoids assumptions about the background expansion, galaxy bias functions, or initial conditions of the Universe that determine the shape of the power spectrum. As we will show, despite these minimal assumptions, upcoming Stage-IV surveys will enable precise constraints on viscosity, mainly through the scale-dependent imprint of viscosity on structure growth. This provides a robust test for detecting viscosity, even in the presence of modified gravity and EP violations, without requiring knowledge of their specific physical mechanisms.

%%%%%%%%%%%%%%%%%%%%%%%%%%%%%%%%%%%%%%%%%%%%%
%%%%%%%%%%%%%%%%%%%%%%%%%%%%%%%%%%%%%%%%%%%%%
%%%%%%%%%%%%%%%%%%%%%%%%%%%%%%%%%%%%%%%%%%%%%

%%%%%%%%%%%%%%%%%%%%%%%%%%%%%%%%%%%%%%%%%%%%%
%%%%%%%%%%%%%%%%%%%%%%%%%%%%%%%%%%%%%%%%%%%%%   

\section{Cosmological Perturbations with Viscous Dark Matter} \label{sec:eqs_vis}  

We consider a perturbed flat Friedmann-Lema\^itre-Robertson-Walker (FLRW) spacetime and treat the dark matter component as a pressureless fluid with constant bulk and shear viscosities, denoted by $\zeta$ and $\eta_s$, respectively. 
Within the framework of Eckart's theory of relativistic viscous fluids (cf. \cite{Eckart:1940te,Weinberg:1971mx}), the energy-momentum tensor of the dark matter fluid is given by:
\begin{equation}\label{eq:EMT}
    T^{\mu}_{~\nu} = \,\rho_m\,u^{\mu}u_{\nu} + p_m(\delta^{\mu}_{\nu}+u^{\mu}u_{\nu})- (\zeta-\tfrac{2}{3}\,\eta_s) \nabla\!\!\cdot\!u(\delta^{\mu}_{\nu}+u^{\mu}u_{\nu}) -\eta_s\,\bigl(\nabla^{\mu}u_{\nu}+\nabla_{\!\!\nu}u^{\mu}+u^{\alpha}\nabla_{\!\!\alpha}(u^{\mu}u_{\nu})\bigr) \,  .
\end{equation}
Here, $\rho_m$ is the matter density, $p_m$ is the pressure (which we set to zero from now on), and $u^{\mu}$ is the four-velocity of the fluid, assumed to be comoving with the Hubble flow at the background level. For convenience, we define two positive, dimensionless parameters associated with the viscous terms:
\begin{equation}\label{eq:def_ZE}
    \mathcal{Z}\equiv\frac{3\zeta\mathcal{H}}{a\rho_m}=\frac{1}{\Omega_m}\frac{\zeta}{H}; \qquad \qquad \mathcal{E}\equiv\frac{4\eta_s\mathcal{H}}{a\rho_m}=\frac{4}{3}\frac{1}{\Omega_m}\frac{\eta_s}{H}
\end{equation}
where \(\mathcal{H}(a)=a\,H(a)\) is the conformal Hubble function, $\Omega_m(a)=\frac{\rho_m a^2}{3\mathcal{H}^2}$ is the density fraction of viscous dark matter, and \(a\) is the scale factor. We work in units where $8\pi G_{\rm N}=c=1$. At the background level, the Einstein equations yield the modified Friedmann equation, while the conservation of the energy-momentum tensor, $\nabla_{\!\mu}T^{\mu}_{~\nu}=0$, gives the continuity equation:
\begin{align}
    &2\frac{\mathcal{H}'}{\mathcal{H}}=-1+3\Omega_m \mathcal{Z} - 3\sum_i w_i \Omega_i \label{eq:Friedmann2} \\
    &\rho_m'+3(1-\mathcal{Z})\rho_m=0 \label{eq:continuity} \, .
\end{align}
The sum in Eq. \eqref{eq:Friedmann2} accounts for additional matter components with equation of state parameters $w_i$ and fractional densities $\Omega_i$. Throughout the text, we use a prime to denote a derivative with respect to $N=\mathrm{log}\,a$. 

To derive the linear perturbation equations, we work in the Newtonian gauge, in which the perturbed metric reads
\begin{equation}\label{eq:pert_metric}
    ds^2 = a^2\bigl[-(1+2\Psi)d\tau^2+(1-2\Phi)\delta_{ij}dx^idx^j\bigr] \, ,
\end{equation}
where the two scalar degrees of freedom are encoded in the gravitational potentials $\Phi$ and $\Psi$. To first order in perturbations, and setting the dark matter sound speed to zero, we have:
\begin{align}
    u^{\mu}&=\left\{\frac{1}{a}\left(1-\Psi,~ V^i\right)\right\} \\
    T^0_{~0}&=-\rho_m(1+\delta_m) \\
    T^0_{~i}&=-T^i_{~0}=\rho_m(1-\mathcal{Z})V_i \\
    T^i_{~j}
    &=-\rho_m\mathcal{Z}\Bigl[1+\frac{1}{3}\partial_kv^k-\Phi'-\Psi\Bigr]\delta^i_j-\frac{1}{4}\rho_m\mathcal{E}\Bigl[\partial_j v^i+\partial^i v_j-\frac{2}{3}\partial_k v^k\delta^i_j\Bigr]  ~ .
\end{align}
Here, $V^i=\frac{dx^i}{d\tau}=\mathcal{H}v^i$ are the components of the peculiar velocity of dark matter. The quantity $\delta_m = \delta\rho_m/\rho_m$ denotes the matter density contrast, and $\rho_m$ is the background energy density of dark matter, evolving according to Eq. \eqref{eq:continuity}. Here and henceforth, we assumed that the viscosities $\zeta$ and $\eta_s$ remain constant. 
In Appendix \ref{app:time_dep}, the effects of a generic time dependence are discussed and shown to be negligible.
The perturbed conservation equations at first order yield the following perturbed continuity and Euler equations in Fourier space,\footnote{Throughout this work, we adopt the Fourier transform convention
\begin{equation}
\tilde{f}(\mathbf{k},t) = \int \mathrm{d}^3\mathbf{x} \, f(\mathbf{x},t)\, e^{i \mathbf{k} \cdot \mathbf{x}}\,, \qquad
f(\mathbf{x},t) = \int \frac{\mathrm{d}^3\mathbf{k}}{(2\pi)^3} \, \tilde{f}(\mathbf{k},t)\, e^{-i \mathbf{k} \cdot \mathbf{x}} \nonumber \,,
\end{equation}}
\begin{align}
    \delta_m'&=-(1-2\mathcal{Z})(\theta_m-3\Phi')-3\mathcal{Z}(\delta_m+\Psi) \label{eq:pert_continuity} \\
    \theta_m'&=-\left(\frac{1}{1-\mathcal{Z}}+\frac{1-2\mathcal{Z}}{1-\mathcal{Z}}\frac{\mathcal{H}'}{\mathcal{H}}+\frac{1}{3}\frac{\mathcal{Z}+\mathcal{E}}{1-\mathcal{Z}}\lambda^{-2}\right)\theta_m
    +\lambda^{-2}\biggl(\frac{\Psi}{1-\mathcal{Z}}+\frac{\mathcal{Z}}{1-\mathcal{Z}}\Phi'\biggr)\label{eq:pert_Euler} \, ,
\end{align}
 in which $\theta_m = -i\, \mathbf{k} \cdot \mathbf{V} / \mathcal{H}$ is the rescaled dark matter velocity divergence in Fourier space, and $\lambda = \mathcal{H}/k$ characterizes the scale dependence. The perturbed Euler equation agrees with the one presented in \cite{Barbosa:2017ojt}, for constant viscosities and if $\Phi'$ is neglected. In this work, we keep $\Phi'$-terms for generality, even if later we find them to be negligible. 
 In addition, from the perturbed Einstein equations, the gravitational potentials are given by 
\begin{align}
    &\Phi=-\frac{3}{2}\lambda^2\Omega_m\bigl[\delta_m+3\lambda^2(1-\mathcal{Z})\theta_m\bigr] - \frac{3}{2}\lambda^2\sum_i\Omega_i\bigl(\delta_i+3\lambda^2(1+w_i)\theta_i\bigr) \label{eq:Phi} \\
    &\Psi=\Phi-\frac{3}{2}\lambda^2\Omega_m\mathcal{E}\theta_m \label{eq:Psi} \\
    &\Phi'=-\Phi+\frac{3}{2}\lambda^2\Omega_m(1-\mathcal{Z}+\mathcal{E})\theta_m + \frac{3}{2}\lambda^2\sum_i\Omega_i(1+w_i)\theta_i \label{eq:PhiPrime} \, .
\end{align}
Note that the presence of shear viscosity induces an anisotropic stress ($\Psi \neq \Phi$), which was also noted in \cite{Barbosa:2017ojt}. This constitutes a degeneracy between shear viscosity and some modified gravity models, further investigated in \cite{Barbosa_2018}. Degeneracies between modifications of gravity and imperfect dark matter were also studied in \cite{AparicioResco:2020shd}. In this work, however, we will follow a much more model-independent approach.

From now on, we neglect all matter components other than the viscous dark matter and work in the sub-horizon limit $\lambda\ll 1$. 
The baryonic component is subdominant and might be neglected in a first approximation.  
In Appendix \ref{app:baryons}, however, we explicitly differentiate between dark matter, which is viscous and feels EP-violating modified gravity, and baryons, assumed to be well described as a perfect fluid governed by standard gravity. In that appendix, we show how the results of this paper can be immediately extended to this more general scenario.
 
To the lowest order in $\lambda$, the gravitational potentials in eqs.~\eqref{eq:Phi}-\eqref{eq:PhiPrime} reduce to
\begin{align}
    &\Phi= -\frac{3}{2}\lambda^2\Omega_m\delta_m\label{eq:Phi_smallscale} \\
    &\Psi=-\frac{3}{2}\lambda^2\Omega_m\left(\delta_m+\mathcal{E}\theta_m\right)\label{eq:Psi_smallscale} \\
    &\Phi'=\frac{3}{2}\lambda^2\Omega_m\bigl[\delta_m +(1-\mathcal{Z}+\mathcal{E})\theta_m \bigr] \label{eq:PhiPrime_smallscale} \, .
\end{align}
Substituting these into eqs. \eqref{eq:pert_continuity} and \eqref{eq:pert_Euler} and neglecting terms of order $\lambda^2$ or higher, because they are important only at horizon scales, yields the following perturbed continuity and Euler equations: 
\begin{align}
    \delta_m'&=-3\mathcal{Z}\delta_m - (1-2\mathcal{Z})\theta_m \equiv f \delta_m\label{eq:pert_continuity_2} \\
    \theta_m'&=-\left[\frac{1}{1-\mathcal{Z}}+\frac{1-2\mathcal{Z}}{1-\mathcal{Z}}\frac{\mathcal{H}'}{\mathcal{H}}+\frac{3}{2}\Omega_m (\mathcal{E}-\mathcal{Z})+\frac{1}{3}\frac{\mathcal{Z}+\mathcal{E}}{1-\mathcal{Z}}\lambda^{-2}\right]\theta_m-\frac{3}{2}\Omega_m\delta_m \label{eq:pert_Euler_2} \, .
\end{align}
The Euler equation for $\theta_m'$ displays the main effect of viscosity, namely the $\lambda^{-2}$ term in the ``friction'' coefficient, i.e.\ in the square brackets of Eq. \ref{eq:pert_Euler_2}. While all the other contributions from $\mathcal{Z},\mathcal {E}$ will be shown to be negligible, the last term in the square brackets plays an important role since at the observable scales $k\approx 0.1 h/$Mpc, the factor $\lambda^{-2}$ is very large.

The perturbation equations can be combined into a second order growth equation for $\delta_m$, or a first order equation for the growth rate $f=\delta_m'/\delta_m$. We discuss this equation in Appendix \ref{app:growth_eq}. In Appendix \ref{app:app_growth}, an approximate solution is derived, where the growth rate is divided into a ``non-viscous part'' and a scale-dependent viscosity contribution. Finally, we also provide an analytic parametrization of the growth rate for $\Lambda$CDM with viscous dark matter in Appendix \ref{app:growthParam}.

%%%%%%%%%%%%%%%%%%%%%%%%%%%%%%%%%%%%%%%%%%%%%%%%%%%%%%%%%%%%%%%%%%%%%%%%%%%
%%%%%%%%%%%%%%%%%%%%%%%%%%%%%%%%%%%%%%%%%%%%%%%%%%%%%%%%%%%%%%%%%%%%%%%%%%%

\section{Viscous dark matter, the equivalence principle, and modified gravity}\label{sec:vis_EP}

On the other hand, in the scenario of modified gravity and EP violation (see e.g.\ \cite{Amendola:2001rc,Castello:2022uuu, Castello:2024lhl}), the perturbed continuity equation remains the same as in Eq. \eqref{eq:pert_continuity_2}, while the Euler equation now becomes 
\begin{equation}\label{eq:pert_Euler_EP}
    \theta_m' = -\left[\frac{1}{1-\mathcal{Z}}+\frac{1-2\mathcal{Z}}{1-\mathcal{Z}}\frac{\mathcal{H}'}{\mathcal{H}} +\frac{1}{3}\frac{\mathcal{Z}+\mathcal{E}}{1-\mathcal{Z}}\lambda^{-2}+\frac{\Theta}{1-\mathcal{Z}}\right]\theta_m
    +\lambda^{-2}\biggl[(1+\Gamma)\frac{\Psi}{1-\mathcal{Z}}+\frac{\mathcal{Z}}{1-\mathcal{Z}}\Phi'\biggr] \, ,
\end{equation}
in which the quantities $\Theta(k,N)$ and $\Gamma(k,N)$ are arbitrary functions, which could in principle depend on both time and scale. They encode a friction term and a fifth force acting on dark matter, respectively. In the case where the equivalence principle is respected, both quantities vanish. In particular, the standard terms for including coupled quintessence in our model are entirely contained in $\Theta$ and $\Gamma$; see Appendix \ref{app:AppendixCQ} for details of the interplay of coupled quintessence and viscosity. Apart from $\Theta$ and $\Gamma$, we also consider general deviations from standard gravity, which are captured by two other free functions \( \mu_{\rm G}(k, N) \) and \( \eta(k,N) \) in the two Poisson equations: 
\begin{equation}\label{eq:Possion_EP}
\Psi = 
-\frac{3}{2}\lambda^2\Omega_m\bigl(\mu_{\rm G}\delta_m  + \mathcal{E}\theta_m + 3\lambda^2(1-\mathcal{Z})\theta_m\bigr) 
\,,
\end{equation}
while the equation for $\Phi$ now takes the form
\begin{equation}\label{eq:Phi_EP}
    \Phi=
    -\frac{3}{2}\eta\lambda^2\Omega_m\bigl(\mu_{\rm G}\delta_m+3\lambda^2(1-\mathcal{Z})\theta_m\bigr)
    \, ,
\end{equation}
 where $\eta(k,N)$ encodes a possible anisotropic stress due to modified gravity, which is in addition to the anisotropic stress from shear viscosity.
Taking the time derivative of Eq. \eqref{eq:Phi_EP} (including the $\lambda^4$-terms, which are important because of the $\lambda^{-2}$-terms in Eq. \eqref{eq:pert_Euler_EP}) and, in the end, neglecting $\lambda^4$-terms, yields 
\begin{equation}\label{Phi_prime_EP}
    \Phi'=\frac{3}{2}\eta\lambda^2\Omega_m\biggl[\Bigl(1-\eta'/\eta-\mu_{\rm G}'/\mu_{\rm G}\Bigr)\mu_{\rm G}\delta_m+\Bigl(\mathcal{Z}+\mathcal{E}+(1-2\mathcal{Z})\mu_{\rm G}\Bigr)\theta_m\biggr] \, .
\end{equation}
This leads to the final form of the perturbed Euler equation in a modified gravity scenario with viscous matter:
\begin{align}\label{eq:pert_Euler_EP_2}
     \theta_m'=& -\left[\frac{1}{1-\mathcal{Z}}+\frac{1-2\mathcal{Z}}{1-\mathcal{Z}}\frac{\mathcal{H}'}{\mathcal{H}}+\frac{3}{2}\frac{\Omega_m}{1-\mathcal{Z}}\biggl(\mathcal{E}(1+\Gamma)-\mathcal{Z}\eta\Bigl(\mathcal{Z}+\mathcal{E}+(1-2\mathcal{Z})\mu_{\rm G}\Bigr)\biggr)+\frac{1}{3}\frac{\mathcal{Z}+\mathcal{E}}{1-\mathcal{Z}}\lambda^{-2}+\frac{\Theta}{1-\mathcal{Z}}\right]\theta_m \nonumber \\
     &-\frac{3}{2}\frac{\Omega_m}{1-\mathcal{Z}}\biggl[1+\Gamma-\mathcal{Z}\eta\bigl(1-\eta'/\eta-\mu_{\rm G}'/\mu_{\rm G}\bigr)\biggr]\mu_{\rm G}\delta_m \, . 
\end{align}
It can be rewritten as 
\begin{equation}\label{eq:pert_Euler_EP_3}
     \theta_m'= -\left[1+\frac{\mathcal{H'}}{\mathcal{H}}+\frac{1}{3}\frac{\mathcal{Z}+\mathcal{E}}{1-\mathcal{Z}}\lambda^{-2}+\tilde{\Theta}\right]\theta_m -\frac{3}{2}\Omega_m(1+\tilde{\Gamma})\mu_{\rm G}\delta_m 
\end{equation}
with 
\begin{align}
    \tilde{\Theta}&=\frac{\Theta}{1-\mathcal{Z}}+\frac{\mathcal{Z}}{1-\mathcal{Z}}\left(1-\frac{\mathcal{H}'}{\mathcal{H}}\right)+\frac{3}{2}\frac{\Omega_m}{1-\mathcal{Z}}\biggl[\mathcal{E}(1+\Gamma)-\mathcal{Z}\eta\Bigl(\mathcal{Z}+\mathcal{E}+(1-2\mathcal{Z})\mu_{\rm G}\Bigr)\biggr] \label{eq:Theta_tilde} \\
    \tilde{\Gamma}&=\frac{\Gamma}{1-\mathcal{Z}}+\frac{\mathcal{Z}}{1-\mathcal{Z}}\biggl[1-\eta\bigl(1-\eta'/\eta-\mu_{\rm G}'/\mu_{\rm G}\bigr)\biggr] \, . \label{eq:Gamma_tilde}
\end{align}
These effective parameters encapsulate the combined effects of EP violation, modified gravity, and dark matter viscosity. They correspond to the observable quantities that can, in principle, be constrained through the Euler equations. In the next section, we connect them to observations through relativistic galaxy number counts.

%%%%%%%%%%%%%%%%%%%%%%%%%%%%%%%%%%%%%%%%%%%%%%%%%%%%%%
%%%%%%%%%%%%%%%%%%%%%%%%%%%%%%%%%%%%%%%%%%%%%%%%%%%%%%

\section{Viscous dark matter and galaxy clustering} \label{sec:vis_GC}

In this work, we employ galaxy clustering as our observational probe. In particular, galaxy redshift surveys quantify fluctuations in the observed number density of galaxies, with the key observable defined as
\begin{equation}\label{eq:Delta}
    \Delta\equiv\frac{N(\hat{\textbf{n}},z)-\bar{N}(z)}{\bar{N}(z)} \,,
\end{equation}
where \( N(\hat{\mathbf{n}}, z) \) is the number of galaxies detected in a given pixel, in the direction \( \hat{\mathbf{n}} \) and at redshift \( z \), and \( \bar{N}(z) \) denotes the average number of galaxies per pixel at that redshift. In the linear regime, keeping terms up to order $\lambda$, the observed galaxy number count fluctuation $\Delta$, including relativistic corrections, is given by (see e.g.\ \cite{Bonvin_2011, Challinor_2011, Yoo_2009}),
\begin{equation}\label{eq:Delta_galaxies}
\Delta(\mathbf{\hat{n}}, z)=b_g\,\delta_m-\frac{1}{\mathcal{H}}\partial_r(\textbf{V}\cdot\mathbf{\hat{n}})+\frac{1}{\mathcal H}\partial_r\Psi+{\textbf{V}'}\cdot \mathbf{\hat{n}} +\alpha \mathbf V\cdot \mathbf{\hat{n}}\,,
\end{equation}
where $r$ is the comoving distance, $b_g$ is the galaxy bias, and $\alpha$ is given by
\begin{equation}\label{eq:alpha}
    \alpha \equiv 1-5s+\frac{5s-2}{\mathcal H r}-\frac{{{\mathcal{H}}'}}{\mathcal H}+f^{\rm evol}\,,
\end{equation}
with $s$ denoting the magnification bias arising from the flux-limited nature of surveys and $f^{\rm evol}$ representing the evolution bias, which accounts for changes in the comoving number density due to galaxy formation, evolution, and selection effects. For a more detailed discussion on these biases, see e.g.\ \cite{Maartens:2021dqy}.

Working in Fourier space, $\Delta$ takes the form
\begin{equation} \label{eq:Delta_Fourier}
    \Delta(\textbf{k},\hat{\textbf{n}},z) = b_g\delta_m - \mu^2\theta_m - i\frac{\mu}{\lambda}\Psi + i\mu\lambda\theta'_m + i\mu\lambda\alpha\theta_m + i\mu\lambda\frac{\mathcal{H}'}{\mathcal{H}}\theta_m \,    ,
\end{equation}
where \( \mu \equiv \hat{\mathbf{k}} \cdot \hat{\mathbf{n}} \) is the cosine of the angle between the wavevector and the line of sight.

Substituting \( \theta \), \( \theta' \) and \( \Psi \) using eqs.~\eqref{eq:pert_continuity_2}, \eqref{eq:pert_Euler_EP_3}, and \eqref{eq:Possion_EP}, respectively, and introducing the growth rate $f=\delta_m'/\delta_m$, we obtain 
\begin{align}
\Delta(\mathbf{k}, \hat{\mathbf{n}}, z) &= \delta_m b_g\Bigg\{ 
1 + \mu^2 \frac{f+ 3\mathcal{Z}}{(1-2\mathcal{Z})b_g} \nonumber \\
&+ i \mu \lambda \Bigg[
\left( 1 + \tilde{\Theta} - \alpha - \frac{3}{2}\Omega_m\mathcal{E}\right)  \frac{f + 3\mathcal{Z}}{(1-2\mathcal{Z})b_g}
- \frac{3}{2} \frac{\Omega_m\mu_{\rm G} \tilde{\Gamma}}{b_g}
\Bigg] \nonumber \\ 
&+ i \mu \lambda^{-1} 
  \,\frac{1}{3} \frac{\mathcal{Z} + \mathcal{E}}{1-\mathcal{Z}}\,  \frac{f+3\mathcal{Z}}{(1-2\mathcal{Z})b_g}\Bigg\}\, .
\end{align}
Equivalently, $\Delta$ can be written in a more compact form as
\begin{equation}\label{eq:compact_Delta}
\Delta(\mathbf{k}, \hat{\mathbf{n}}, z) = \delta_m b_{g}\Bigg\{ 
1 + \mu^2  \tilde{\beta}
- i \mu \lambda 
\left(\alpha - \tilde{E}_P - C_{\rm vis}\lambda^{-2} \right) \tilde{\beta}
\Bigg\} \, ,
\end{equation}
where
\begin{align}
\tilde{\beta} &= \frac{f + 3\mathcal{Z}}{b_g(1 - 2\mathcal{Z})} \\
\tilde{E}_P & = 1 + \tilde{\Theta}  - \frac{3}{2} \Omega_m \mathcal{E} - \frac{3}{2} \frac{\Omega_m \mu_{\rm G} \tilde{\Gamma}}{f+3\mathcal{Z}}(1-2\mathcal{Z}) \label{eq:tilde_Ep}\\ 
C_{\rm vis} &= \frac{1}{3}\frac{\mathcal{Z} + \mathcal{E}}{1-\mathcal{Z}}   \,    .
\end{align}
The parameter $E_P$ has been previously introduced in \cite{Castello:2024lhl}
as a model-independent way to test the equivalence principle. Here, we see that it is generalized to $\tilde E_P$ when including viscosity.
We will frequently use the parameter $C_{\mathrm{vis}}$ as a measure of viscosity in the following and denote its present value at redshift zero with $C_{\mathrm{vis},0}$, a quantity that, as we will see, plays a central role throughout this paper. In Appendix \ref{app:ObsConstraints}, we provide upper limits on the magnitude of $C_{\mathrm{vis},0}$, derived from existing observational constraints on dark matter viscosity. We remark, however, that these upper limits have been obtained assuming $\Lambda$CDM or other specific cosmologies.

Once  $\Delta$ is obtained, one can  construct the observed auto- and cross-power spectra. As in \cite{Castello:2024lhl}, we consider two galaxy populations, namely a bright (B) and a faint (F) galaxy population. The expressions for $\Delta_B, \Delta_F$ (keeping terms up to the order of $\lambda$) are given by
\begin{align}
\Delta_B(\mathbf{k}, \hat{\mathbf{n}}, z) &= \delta_m b_{g,B}\Bigg\{ 
1 + \mu^2  \tilde{\beta}_{B}
- i \mu \lambda 
\left( \alpha_B - \tilde{E}_P - C_{\rm vis}\lambda^{-2} \right) \tilde{\beta}_{B}
\Bigg\} \, ,
\\
\Delta_F(\mathbf{k}, \hat{\mathbf{n}}, z) &= \delta_m b_{g,F}\Bigg\{ 
1 + \mu^2  \tilde{\beta}_{F}
- i \mu \lambda 
\left( \alpha_F - \tilde{E}_P - C_{\rm vis}\lambda^{-2} \right) \tilde{\beta}_{F}
\Bigg\} \, ,
\end{align}
and the power spectra (again keeping terms up to the order of $\lambda$) read as follows:
\begin{align}
P_{\Delta_B\Delta_B}& \equiv \langle \Delta_B \Delta_B^* \rangle =  \Big[ \left( 1 + \mu^2 \tilde{\beta}_{B} \right)^2 - 2\mu^2 C_{\rm vis} \left(\alpha_B-\tilde{E}_P\right) \tilde{\beta}_{B}^2  + \mu^2 \lambda^{-2}  C_{\rm vis}^2 \tilde{\beta}_{B}^2  \Big]\frac{\tilde{\beta}_{F}}{\tilde{\beta}_{B}}S_g^2BP \,    ,  \label{eq:PBB}\\
P_{\Delta_F\Delta_F}& \equiv \langle \Delta_F \Delta_F^* \rangle =
\Big[ \left( 1 + \mu^2 \tilde{\beta}_{F} \right)^2 - 2\mu^2 C_{\rm vis} \left(\alpha_F-\tilde{E}_P\right) \tilde{\beta}_{F}^2 + \mu^2 \lambda^{-2}  C_{\rm vis}^2 \tilde{\beta}_{F}^2  \Big]\frac{\tilde{\beta}_{B}}{\tilde{\beta}_{F}}S_g^2BP \,    ,  \\
P_{\Delta_B \Delta_F} &\equiv \langle \Delta_B \Delta_F^* \rangle = \Bigg\{
(1 + \mu^2  \tilde{\beta}_B)(1 + \mu^2\tilde{\beta}_F ) - \mu^2 C_{\rm vis}(\alpha_B + \alpha_F-2\tilde{E}_P) \tilde{\beta}_B \tilde{\beta}_F + \mu^2 \lambda^{-2} C_{\rm vis}^2 \tilde{\beta}_B \tilde{\beta}_F 
\nonumber \\
&\quad + i \mu \lambda \Big[
\alpha_F\tilde{\beta}_{F} - \alpha_B\tilde{\beta}_{B} - \tilde{E}_P(\tilde{\beta}_{F} - \tilde{\beta}_{B}) 
+ \mu^2 \tilde{\beta}_B \tilde{\beta}_F ( \alpha_F - \alpha_B )\Big]
+ i\mu\lambda^{-1} C_{\rm vis} ( \tilde{\beta}_B - \tilde{\beta}_F )
\Bigg\}S_g^2BP \,    .
\end{align}
Here, $P$ denotes the matter power spectrum evaluated at redshift $z=0$, and $B\equiv b_{g,B}b_{g,F}G^2(k,z)= f^2 G^2/\beta_B\beta_F$, where $G(k, z)\equiv\delta_m(k, z)/\delta_m(k, 0)$ is the linear growth factor and $\beta_T(k,z)=f(k,z)/b_{g,T}$ with $T=B,F$. The function $S_g(k,\mu,z)$ encodes a damping factor $\sigma_g$ that accounts for non-linear redshift-space distortions, and here we employ the form suggested  in \cite{Koda:2013eya, Howlett:2017asw},
\begin{equation} \label{eq:Sg_sigmag}
    S_g^2 = e^{-\frac{1}{2}k^2\mu^2\sigma_g^2} \, .
\end{equation}
As we will see in Sec. \ref{sec:Fisher}, $\sigma_g$ will be left as a free parameter to vary at each redshift bin. 

In Appendix \ref{app:growth} we show that viscosity will introduce scale-dependence in the growth of structure, and in the limit of small viscosity, one can expand $\tilde{\beta}_B, \tilde{\beta}_F,\tilde{E}_P, B$ as  
\begin{align}
    \tilde{\beta}_B &= \tilde{\beta}_{B,z} + \epsilon\lambda^{-2}\tilde{\beta}_{B,v} \label{eq:beta_bz}\\
    \tilde{\beta}_F &= \tilde{\beta}_{F,z} + \epsilon\lambda^{-2}\tilde{\beta}_{F,v} \label{eq:beta_fz}\\
    \tilde{E}_P &= \tilde{E}_{P,z} + \epsilon\lambda^{-2}\tilde{E}_{P,v}       \label{eq:Ep_exp}       \\
     B &= B_z + \epsilon\lambda^{-2}B_v   \label{eq:exp_B}
\end{align}
Here, $\epsilon$ is an order parameter that will be set to unity at the end. The terms proportional to $\lambda^{-2}$ capture the leading-order $k$-dependence introduced by the viscous effect. The explicit approximations we used in our analysis for $B_v, \tilde{\beta}_{B,v}, \tilde{\beta}_{F,v}$ are provided in Appendix \ref{app:app_growth}. As detailed there, to avoid a singular Fisher matrix, we must set $\tilde{E}_{P,v}=0$ in our analysis. However, we keep this quantity in the expressions below to explicitly show the dependence.
Substituting these expressions into the auto- and cross-power spectra given above and neglecting terms of order \( \mathcal{O}(\epsilon^2) \) and higher, \footnote{we note that we also consider $C_{\rm vis}$ to be of order $\epsilon$.} the power spectra reduce to: 
\begin{align} \label{eq:P_BB}
P_{\Delta_B\Delta_B} &= \frac{\tilde{\beta}_{F,z}}{\tilde{\beta}_{B,z}}\, S_g^2\, P\, B \Bigg[
(1 + \mu^2\tilde{\beta}_{B,z} )^2 
- 2\mu^2 C_{\mathrm{vis}}(\alpha_B - \tilde{E}_{P,z})\tilde{\beta}_{B,z}^2 
 + 2\lambda^{-2}  
\mu^2 \tilde{\beta}_{B,v}(1+\mu^2\tilde{\beta}_{B,z})
\Bigg]  
\\
\label{eq:P_FF}
P_{\Delta_F\Delta_F} &= \frac{\tilde{\beta}_{B,z}}{\tilde{\beta}_{F,z}} S_g^2 P B\Bigg[ 
(1 + \mu^2\tilde{\beta}_{F,z} )^2 
- 2\mu^2 C_{\mathrm{vis}}(\alpha_F - \tilde{E}_{P,z})\tilde{\beta}_{F,z}^2  
+2\lambda^{-2}\mu^2 \tilde{\beta}_{F,v}\left(1+\mu^2\tilde{\beta}_{F,z}\right)
\Bigg]\, 
\\
\label{eq:P_BF}
P_{\Delta_B\Delta_F}
&= S_g^2\,B\,P \Bigg[
1 + \mu^2(\tilde{\beta}_{B,z}+\tilde{\beta}_{F,z})
+ \mu^4 \tilde{\beta}_{B,z}\tilde{\beta}_{F,z} \nonumber \\[0.25em]
&\quad - \mu^2 C_{\mathrm{vis}}(\alpha_B+\alpha_F-2\tilde{E}_{P,z})\,\tilde{\beta}_{B,z}\tilde{\beta}_{F,z}
+ i\mu\lambda\Big(\alpha_F\tilde{\beta}_{F,z}-\alpha_B\tilde{\beta}_{B,z}
-\tilde{E}_{P,z}(\tilde{\beta}_{F,z}-\tilde{\beta}_{B,z}) \nonumber \\[0.25em]
&\qquad\qquad\quad
+\, \mu^2\tilde{\beta}_{B,z}\tilde{\beta}_{F,z}(\alpha_F-\alpha_B)\Big)
+ i\mu\lambda^{-1} C_{\mathrm{vis}}(\tilde{\beta}_{B,z}-\tilde{\beta}_{F,z}) \nonumber \\[0.5em]
&\quad + \lambda^{-2}\Big\{
\mu^2(\tilde{\beta}_{B,v}+\tilde{\beta}_{F,v})
+ \mu^4(\tilde{\beta}_{B,z}\tilde{\beta}_{F,v}+\tilde{\beta}_{F,z}\tilde{\beta}_{B,v}) \nonumber \\[0.25em]
&\qquad\qquad
+\, i\mu\lambda\Big[
\alpha_F\tilde{\beta}_{F,v}-\alpha_B\tilde{\beta}_{B,v}
-\tilde{E}_{P,z}(\tilde{\beta}_{F,v}-\tilde{\beta}_{B,v})
-\tilde{E}_{P,v}(\tilde{\beta}_{F,z}-\tilde{\beta}_{B,z}) \nonumber \\[0.25em]
&\qquad\qquad\quad
+\, \mu^2(\tilde{\beta}_{B,z}\tilde{\beta}_{F,v}+\tilde{\beta}_{F,z}\tilde{\beta}_{B,v})(\alpha_F-\alpha_B)
\Big]\Big\}
\Bigg].
\end{align}
The above power spectra can be grouped into polynomials of $\mu^a\lambda^b$ as follows:
\begin{align}
P_{\Delta_B \Delta_B} &= \frac{\tilde{\beta}_{F,z}}{\tilde{\beta}_{B,z}} S_g^2 BP  \sum_{a,b} \mu^a \lambda^{b} \mathcal{C}^{(BB)}_{a,b} \,  , \label{eq:co_1} \\
P_{\Delta_F \Delta_F} &= \frac{\tilde{\beta}_{B,z}}{\tilde{\beta}_{F,z}} S_g^2 BP  \sum_{a,b} \mu^a \lambda^{b} \mathcal{C}^{(FF)}_{a,b} \,  , \label{eq:co_2} \\
P_{\Delta_B \Delta_F} &= S_g^2 BP  \sum_{a,b} \mu^a \lambda^{b} \mathcal{C}^{(BF)}_{a,b} \,  ,  \label{eq:co_3}
\end{align}
with the coefficients of $\mathcal{C}^{(BB)}_{ab}, \mathcal{C}^{(FF)}_{ab}, \mathcal{C}^{(BF)}_{ab}$ listed in Appendix \ref{sec:coefficients}\footnote{Here we leave \(B\) as an overall factor to show its degeneracy with \(P\) more explicitly. Nevertheless, in our later Fisher analysis, it is expanded according to Eq.~\eqref{eq:exp_B}. This expansion gives rise to additional terms at order \(\mathcal{O}(\epsilon^2)\), which we have verified numerically to be subdominant.}.

\begin{figure*}
\centering
\includegraphics[width=0.40\textwidth]{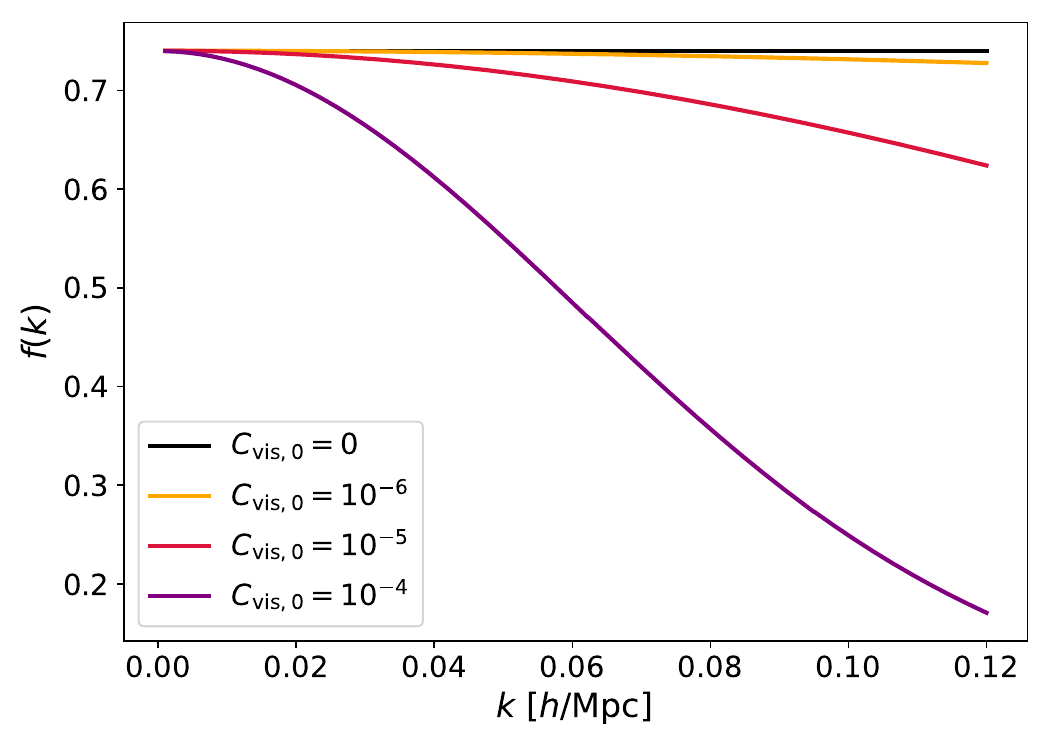}
\hfil
\includegraphics[width=0.40\textwidth]{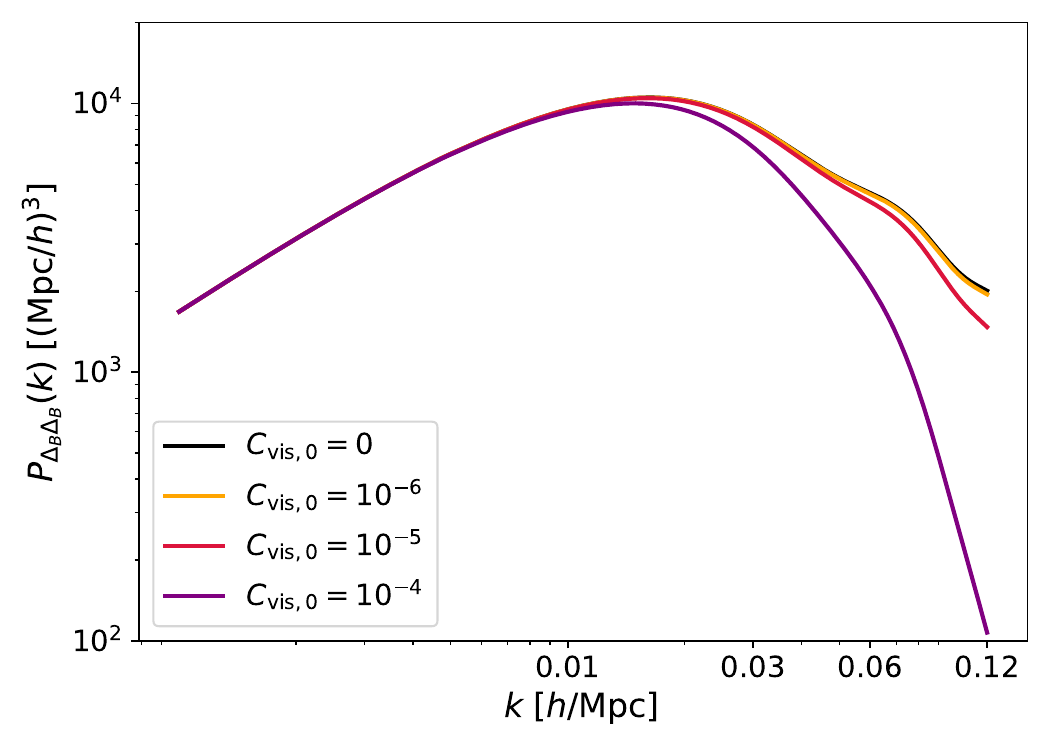}
\medskip \\
\caption{ 
Scale-dependent impact of the viscosity parameter \(C_{\rm vis,0}\) on large-scale structure observables. 
Here we consider only bulk viscosity (with \(C_{\rm vis,0} \simeq \bar{\zeta} / 9\Omega_{m,0}\)). Results are shown at redshift \(z = 0.45\) for a line-of-sight angle \(\mu = 1\), with $\Theta = \Gamma = 0, \mu_{\rm G} = \eta = 1$, assuming DESI survey specifications as listed in Table \ref{tab:DESI_specs}.
\textit{Left}: scale-dependent growth rate \(f(k)\). 
\textit{Right}: linear power spectrum \(P_{\Delta_B\Delta_B}(k)\), computed using Eq.~\eqref{eq:PBB}. 
}
\label{fig:cases} 
\end{figure*}

\begin{table} 
\centering
\renewcommand\arraystretch{1.5} % more vertical space overall
\begin{tabular}{|c|c|c|}
\hline
scales& single tracer  & two tracers $B,F$ \\ 
\hline\hline
$\lambda$ 
& \rule{0pt}{2.5ex} -- 
& $\tilde{\beta}_{B,z},\; \tilde{\beta}_{F,z},\; \tilde{E}_{P,z},\;  C_{\rm vis,0}$ \\
\hline
$\lambda^{0}$ 
& \rule{0pt}{2.5ex} $\tilde{\beta}_{z}$ 
& $\tilde{\beta}_{B,z},\; \tilde{\beta}_{F,z}$ \\
\hline
$\lambda^{-1}$ 
& \rule{0pt}{2.5ex} $\tilde{\beta}_{z}$ 
& $\tilde{\beta}_{B,z},\; \tilde{\beta}_{F,z}$ \\
\hline
$\lambda^{-2}$ 
& \rule{0pt}{2.5ex} $\tilde{\beta}_z$ 
& $\tilde{\beta}_{B,z},\; \tilde{\beta}_{F,z}$ \\
\hline
\end{tabular}
\caption{
Measurable parameters for different tracer sets at various orders in $\lambda$, including auto-correlations for the two-tracer case.}
\label{tab:measureable_quan}
\end{table}

The power spectra in Eqs. \eqref{eq:co_1}-\eqref{eq:co_3} are based on the flat-sky approximation. That is, they are derived under the assumption that the separation between the two galaxies in a pair is much smaller than their distance from the observer. A full consideration requires the so-called wide-angle (WA) correction, which is calculated by performing a multipole expansion of the galaxy two-point correlation function in real space and depends on the geometrical configuration. As calculated in \cite{Castello:2024lhl}, at the order of $\lambda$, the WA term arises only from the cross-power spectra, and it reads
\begin{equation}
P_{\Delta_\B\Delta_\F}^{\mathrm{wa}}=\frac{2}{5}\frac{1}{{\mathcal H}r}(\tilde{\beta}_{\F}-\tilde{\beta}_{\B})i\lambda S_g^2BP\times\left(5\mu^{3}-\frac{5}{2}\mu(\mu^{2}-1)\frac{d\log P}{d\log k}\right) \, ,\label{eq:wa}
\end{equation}
where $\mu$ is defined as the cosine of the angle between the galaxy pair and the direction from the observer to the midpoint of the galaxies. Therefore, the full cross-correlation is obtained by summing Eqs. \eqref{eq:wa} and \eqref{eq:co_3}, and likewise for $P_{\Delta_\F\Delta_\B}^{\mathrm{wa}}$. 

By fitting the theoretical spectra to galaxy clustering data in each redshift bin, one can measure each coefficient $\mathcal{C}_{a,b}$ in Eqs. \eqref{eq:co_1}-\eqref{eq:co_3} without any further assumptions. As will be detailed later on, in our analysis, the spectral shape is left free to vary in several $k$-bands, and the final results are marginalized over the $k$-bin values. In this way, the coefficients $\mathcal{C}_{a,b}$ are constrained in a model-independent way.   
Note that \(b_g\) is almost degenerate with \(\sigma_g\); this degeneracy is discussed in more detail in Appendix~\ref{app:degeneracy}. 

In the power spectra, the model-independent observables that can be identified at the various orders of $\lambda$ are summarized in Table \ref{tab:measureable_quan}. In Appendix \ref{app:app_growth}, we show that when the viscosity parameters are constant, $C_{\rm vis}$ depends on its present value \( C_{\rm vis,0} \), defined as
\begin{equation}
    C_{\rm vis,0} =\frac{\zeta+\frac{4}{3}\eta_s}{3\Omega_{m0}H_0}
=\frac{\mathcal{Z}_0+\mathcal{E}_0}{3}\,,\label{eq:cvis0a}
    \end{equation}
    where the subscript $0$ refers to the present value.
This is therefore the parameter that we are going to constrain. In our model-independent approach, we find that \( C_{\rm vis,0} \)
can be measured only by cross-correlating two galaxy populations, along with the inclusion of relativistic corrections.
Additionally, $C_{\rm vis}$ arises from the Doppler correction \( \mathbf{V}' \cdot \hat{\mathbf{n}} \), a term that is induced by the relativistic corrections. 
The individual contributions from bulk and shear viscosity, however, remain degenerate and cannot be disentangled.
Since $C_{\rm vis,0}$ is multiplied by $\lambda^{-2}$, which at $z=0$ amounts to $\approx 10^5$ for $k=0.1 h/$Mpc, we expect that even a small viscosity coefficient ${\mathcal O}(10^{-5})$ induces a large effect on the power spectra (see Fig. \ref{fig:cases}). Consequently, we expect to find quite strong constraints on $C_{\rm vis,0}$.

Next, we examine whether $E_P$ remains a smoking gun for EP violation in the context of viscous dark matter. Notice that $\tilde{\beta}_{B}$ essentially replaces the standard RSD coefficient $\beta_{B}$, and likewise for $\tilde{\beta}_{F}$. In \cite{Castello:2024lhl},
the coefficient of the dipole term $i\mu\lambda$ in $P_{\Delta_B\Delta_F}$ was denoted by $\tau_1$, while the coefficient of the octupole term $i\mu^3 \lambda$ was denoted by $\tau_2$.
Here, we find that   \( \tau_2 = \tilde{\beta}_{F}\tilde{\beta}_{B} \left(\alpha_F - \alpha_B\right) \) maintains the same structure and, similarly, $\tau_1$ reads
\begin{equation}
\tau_1=\alpha_F\tilde{\beta}_{F} - \alpha_B\tilde{\beta}_{B} - \tilde{E}_P(\tilde{\beta}_{F} - \tilde{\beta}_{B}) \, ,
\end{equation}
in which \( \tilde E_P \) is the parameter that can be measured  in the context of viscous dark matter and is given in explicit form  by
\begin{align} \label{eq:new_EP}
\tilde{E}_P &= 1 + \tilde{\Theta} - \frac{3}{2}\Omega_m\mathcal{E} - \frac{3}{2}\frac{\Omega_m\mu_{\rm G}\tilde{\Gamma}(1-2\mathcal{Z})}{f+3\mathcal{Z}} 
= 1+\frac{\mathcal{Z}}{1-\mathcal{Z}}\biggl(1-\frac{\mathcal{H}'}{\mathcal{H}}\biggr)+\frac{\Theta}{1-\mathcal{Z}} \nonumber\\
&\quad-\frac{3}{2}\frac{\Omega_m}{1-\mathcal{Z}}\biggl[\frac{1-2\mathcal{Z}}{f+3\mathcal{Z}}\mu_{\rm G}\biggl(\Gamma+\mathcal{Z}\Bigl(1-\eta(1-\eta'/\eta-\mu_{\rm G}'/\mu_{\rm G})\Bigr)\biggr)+\mathcal{Z}\eta\Bigl((1-2\mathcal{Z})\mu_{\rm G}+\mathcal{Z}\Bigr)-\mathcal{E}\Bigl(\Gamma+\mathcal{Z}(1-\eta)\Bigr)\biggr]
\,.
\end{align}
While it is more common in the literature to neglect \( \mathcal{E} \) rather than \( \mathcal{Z} \), i.e.\ to consider only bulk viscosity, in the case \( \mathcal{Z} = 0 \), the parameter \( \tilde E_P \) remains a probe of weak equivalence principle violation even for \( \mathcal{E} \ne 0 \), since $\tilde E_P$ differs from unity only whenever either $\Theta$ or $\Gamma$ do not vanish. 
However, in the general case,  we find directly from Eq. \eqref{eq:new_EP} that \( \mathcal{Z} \) and $\mathcal{E}$ are fully degenerate with both \( \Theta \) and \( \Gamma \), and consequently, \( \tilde E_P \) no longer serves as a direct signature of EP violation, as expected. 
\footnote{In particular, in the absence of EP violation or modified gravity (i.e.\ $\Theta=\Gamma=0$ and $\mu_{\rm G}=\eta=1$), the parameter $\tilde{E}_P$ simplifies to $1+\frac{\mathcal{Z}}{1-\mathcal{Z}}\bigl(1-\frac{\mathcal{H}'}{\mathcal{H}}\bigr)-\frac{3}{2}\Omega_m\mathcal{Z}$. Note that it is not affected by shear viscosity. For $\Lambda$CDM with viscous dark matter, employing the Friedmann equation \eqref{eq:Friedmann2} reduces the expression to $\tilde{E}_P=1+\frac{3}{2}\frac{\mathcal{Z}}{1-\mathcal{Z}}(1-\Omega_m+w_i\Omega_i)$. For small bulk viscosity, the value is still close to 1.}

On the other hand, as we show in the forecast analysis below, the viscosity parameter $C_{\mathrm{vis},0}$ is expected to be constrained to very small values, implying that both $\mathcal{Z}$ and $\mathcal{E}$ are very small at all epochs. In this case, and also assuming for simplicity that $\eta,\mu_G$ are constant, we can rewrite $\tilde E_P$ as \footnote{Here we neglect terms quadratic in viscosities and where $\mathcal{Z}$ or $\mathcal{E}$ is multiplied by $\Theta$, $\Gamma$ or $1-\eta$.}
\begin{equation}
    \tilde{E}_{P}\simeq E_P+\mathcal{Z}\biggl(1-\frac{\mathcal{H}'}{\mathcal{H}}\biggr)-\frac{3}{2}\Omega_{m}\mu_{{\rm G}}\mathcal{Z}\eta
\end{equation}
where $E_P=1+\Theta-\frac{3}{2}\frac{\Omega_{m}\mu_{\rm G}\Gamma}{f}$ is the original $E_P$-parameter. Note that viscosity still enters $E_P$ and induces a scale-dependence through the growth rate $f$. However, the measurable, purely time-dependent quantity $\tilde{E}_{P,z}\simeq 1+\Theta-\frac{3}{2}\frac{\Omega_{m}\mu_{\rm G}\Gamma}{f_z}$ reduces to $E_P$ when $\mathcal{Z,E}$ become negligible (see Appendix \ref{app:app_growth} for details). 
It is therefore clear that, unless the EP-violating terms $\Theta,\Gamma$ are themselves very small (i.e.\ comparable at $\mathcal{O}(10^{-6})$), the viscosity correction to $E_P$ remains negligible, and $\tilde{E}_{P,z}$ continues to serve as a valid null test of the equivalence principle. 

For convenience, we summarize the viscous generalizations of $\Theta, \Gamma, \mu_{\rm G}, \eta$ and $E_P$ in Appendix \ref{app:summary}.

%%%%%%%%%%%%%%%%%%%%%%%%%%%%%%%%%%%%%%%%%%%%%%%%%%%%%%%%
%%%%%%%%%%%%%%%%%%%%%%%%%%%%%%%%%%%%%%%%%%%%%%%%%%%%%%%%
%%%%%%%%%%%%%%%%%%%%%%%%%%%%%%%%%%%%%%%%%%%%%%%%%%%%%%%%
%%%%%%%%%%%%%%%%%%%%%%%%%%%%%%%%%%%%%%%%%%%%%%%%%%%%%%%%

\section{Fisher Analysis} \label{sec:Fisher}

Following the methodology described in Refs.~\cite{Quartin:2021dmr, Castello:2024lhl}, we construct the data covariance matrix in bins of wavenumber $k=|\mathbf{k}|$ and cosine angle $\mu$. The covariance matrix then takes the form of a $2\times2$ matrix that includes the auto- and cross-power spectra of the bright and faint galaxy populations, together with their shot noise contributions,
\begin{equation} \label{eq:Cov_Mat}
C=\left(\begin{array}{cc}
P_{\Delta_{\mathrm{F}}\Delta_{\mathrm{F}}} + P_{{\rm sn},\Delta_{\mathrm{F}}} & P_{\Delta_{\mathrm{F}}\Delta_{\mathrm{B}}} \\
P_{\Delta_{\mathrm{B}} \Delta_{\mathrm{F}}} & P_{\Delta_{\mathrm{B}}\Delta_{\mathrm{B}}} + P_{{\rm sn},\Delta_{\mathrm{B}}}
\end{array}\right)  \, ,
\end{equation}
where $P_{\rm sn, \Delta_{B,F}}=n_n/n_{B,F}$, and the shot noise amplitude $n_n$ is left free to vary in each redshift bin in our Fisher analysis.  By construction, the covariance matrix is Hermitian since the cross-power spectra are conjugate symmetric. The Fisher matrix (FM) element for the $n-$th $k$ bin and the $m-$th $\mu$ bin is given by the trace of products of Hermitian matrices, and thus yields a real result:
\begin{equation}
F^{nm}_{\alpha\beta}=\frac{1}{2}\frac{\partial C_{ij}}{\partial\theta_{\alpha}}C_{jp}^{-1}\frac{\partial C_{pq}}{\partial\theta_{\beta}}C_{qi}^{-1}\, ,
\end{equation}
where $\theta_\alpha$ denotes the set of model parameters. Summing over all $k$ and $\mu$ bins, the total Fisher matrix at a given $z$ bin takes the form
\begin{equation}
    F_{\alpha\beta,\rm tot}=\frac{V_s}{8\pi^2}\,\sum_{n,m} k_{n}^{2}\Delta k_n\Delta\mu_m F^{nm}_{\alpha\beta}\, ,
\end{equation}
where $V_s$ denotes the survey volume at each $z$. In the analysis, we adopt a constant bin size, with $\Delta\mu_m=0.1$ and $\Delta k_n=0.01 \, h/$Mpc.

In principle, our  $\mu$-binning  or a sufficiently high multipole expansion (often adopted in previous analyses)  contain the same information.
However, the $\mu$-binning leaves us more flexibility in experimenting with different number of bins without introducing very high multipoles that likely will not be measured effectively in future surveys. Secondly, the multipole expansion is usually adopted to justify discarding the higher multipoles on the ground that they contain less and less information. This is of course  true in most cases, but in the spirit of our model-independent approach we chose not to enforce this additional assumption. 

Mapping redshifts and angular positions into three-dimensional scales requires assuming a specific cosmological model as a reference. If the true cosmology deviates from this choice, the inferred wavenumbers and their orientations undergo geometric distortions known as the Alcock-Paczynski (AP) effect. In the true cosmology, the transformations are given by
\begin{equation}
    k = \alpha_{\rm AP}k_r, \qquad \mu=\frac{\mu_rh}{\alpha_{\rm AP}} \,  ,
\end{equation}
where $k_r$ and $\mu_r$ denote quantities in the reference cosmology, and $\alpha_{\rm AP}$ is given by
\begin{equation}
    \alpha_{\rm AP}\,=\,\frac{1}{d}\sqrt{\mu_{r}^{2}(h^{2}d^2-1)+1}\,.\label{eq:alphaAP}
\end{equation}
with the two parameters $h$ and $d$ characterizing the deviation from the fiducial, defined as
\begin{equation} \label{eq:AP_h_d}
    h \equiv \frac{E}{E_r}; \qquad d \equiv \frac{L_A(z)}{L_{A,r}(z)} \, .
\end{equation}
Here, $E(z)\equiv H(z)/H_0$ is the dimensionless Hubble function, $L_A(z)=H_0D_A(z)$, and $D_A(z)$ is the comoving angular diameter distance in the true cosmology. We use the subscript $r$ to refer to the fiducial model. As in \cite{Castello:2024lhl}, we use the parameter combination $h_d\equiv hd$ as a parameter due to the degeneracy between $d$ and $P(k)$ at the linear level. Moreover, we remark that $\mathcal{H}r$ can be expressed using AP parameters, namely $\mathcal{H}r=E_rL_{A,r}h_d$. 

Therefore, apart from varying the constant parameter $C_{\rm vis, 0}$, our analysis includes ten free parameters for each redshift bin: 
\begin{equation}
    \theta = \{\tilde{\beta}_{\mathrm{B},z}, \tilde{\beta}_{\mathrm{F},z}, \alpha_{\mathrm{B}}, \alpha_{\mathrm{F}}, h_d, d, B_z, \sigma_g,\tilde{E}_{P,z}, n_n\} \, .
\end{equation}
For convenience, we summarize in Table \ref{tab:varied_sum} the primary definitions of these parameters as introduced in the text. To preserve independence from the initial conditions of the Universe, we model the shape of the power spectrum using a set of wavebands defined in the first redshift bin. Specifically, we sample \( P(k) \) over the range \( k \in (0.01, 0.12)\, h\,\mathrm{Mpc}^{-1} \) with steps of \( \Delta k = 0.01\, h\,\mathrm{Mpc}^{-1} \), resulting in a total of twelve waveband parameters. 
The parameters $d,s_F,s_B$ can be measured independently of galaxy clustering. We can in fact measure $d$ directly with supernovae Ia, and $s_{B,F}$ via number densities as a function of flux (see Appendix \ref{sec:mag} for details). We can use the  constraints on these parameters as prior in the Fisher matrix analysis.
For the distance parameter \( d \), we impose a 3\% prior uncertainty, consistent with current constraints from the Hubble diagram and the Pantheon+ supernova sample~\cite{Brout:2022vxf}. 
The priors  on \( \alpha_B \) and \( \alpha_F \) 
are derived  by propagating the expected uncertainties on \( s_B \) and \( s_F \), as detailed in Sec.~\ref{sec:specifications}. Note that in this propagation, we fix $H,r$ to their fiducial values. Moreover, to eliminate the full degeneracy between $P(k)$ and $B$ in the first redshift bin, we fix \( B_z(z_1) = 1 \) and use \( B_z(z) \) to quantify the ratio of the combination \( b_{g,B}\, b_{g,F}\, G_z(z)^2 \) relative to that bin. Finally, to ensure the validity of the perturbative expansion, we adopt the same scale cut procedure as in~\cite{Castello:2024lhl}, namely excluding bins where the $\lambda$-hierarchy breaks down. Specifically, we require that the leading-order terms ($\lambda^0$) dominate over the first-order corrections ($\lambda^1$), thereby avoiding contamination from higher-order terms. In practice, this condition is enforced by keeping only bins where the Fisher matrix is positive definite, which results in the exclusion of a few low-$k$ bins in a few low-redshift intervals. 

\begin{table} 
\centering
\renewcommand\arraystretch{1.5} % more vertical space overall
\begin{tabular}{|c|c|c|}
\hline
parameters & description  & definitions in the text  \\ 
\hline\hline
$\tilde{\beta}_{\mathrm{B},z}, \tilde{\beta}_{\mathrm{F},z}$
& \rule{0pt}{2.5ex} purely time-dependent component of $\tilde{\beta}_{B}$ and $\tilde{\beta}_{F}$
&  Eqs. \eqref{eq:beta_bz} and \eqref{eq:beta_fz} \\
\hline
$\alpha_{B}, \alpha_{F}$ 
& \rule{0pt}{2.5ex} encoding the magnification and evolution bias
& Eq. \eqref{eq:alpha} \\
\hline
$h_d,d$ 
& \rule{0pt}{2.5ex} AP effect parameters
& Eq. \eqref{eq:AP_h_d} and the paragraph below \\
\hline
$B_z$ 
& \rule{0pt}{2.5ex} purely time-dependent component of $B$
& Eq. \eqref{eq:exp_B} \\
\hline
$\sigma_g$ 
& \rule{0pt}{2.5ex} non-linear RSD damping factor
& Eq. \eqref{eq:Sg_sigmag} \\
\hline
$\tilde{E}_{P,z}$ 
& \rule{0pt}{2.5ex} parameter indicating EP violation 
& Eqs. \eqref{eq:tilde_Ep} and \eqref{eq:Ep_exp} \\
\hline
$n_n$ 
& \rule{0pt}{2.5ex} shot noise amplitude
& Eq. \eqref{eq:Cov_Mat} \\

\hline
\end{tabular}
\caption{List of free parameters varied in each redshift bin, along with their definitions in the text.}
\label{tab:varied_sum}
\end{table}

%%%%%%%%%%%%%%%%%%%%%%%%%%%%%%%%%%%%%%%%%%%%%%%%%%%%%%%%
%%%%%%%%%%%%%%%%%%%%%%%%%%%%%%%%%%%%%%%%%%%%%%%%%%%%%%%%
%%%%%%%%%%%%%%%%%%%%%%%%%%%%%%%%%%%%%%%%%%%%%%%%%%%%%%%%
%%%%%%%%%%%%%%%%%%%%%%%%%%%%%%%%%%%%%%%%%%%%%%%%%%%%%%%%

\section{Survey Specifications and Fiducials} \label{sec:specifications}

We present forecasts for three upcoming galaxy surveys: the DESI Bright Galaxy Sample (BGS) \cite{DESI:2016fyo}, a stage IV H$\alpha$ spectroscopic survey similar to Euclid \cite{Euclid:2024yrr}, and the SKA2 survey \cite{Bull:2015lja}. In each case, we adopt a flat $\Lambda$CDM cosmology consistent with \textit{the Planck} 2018 results \cite{Planck:2018vyg} to set the  background evolution, linear growth rate, and the shape of the matter power spectrum at fiducial values.

In our forecast, each survey is analyzed by splitting the galaxy sample into two galaxy populations, namely bright and faint, determined by a redshift-dependent flux threshold. This threshold is chosen so that each population contains the same number of galaxies in each redshift bin. In our analysis, we assume a fiducial bias ratio of $b_{g,F}/b_{g,B}=0.5$ and choose the fiducial biases estimated by $b_{g,B}=1.4b_g(z)$ and $b_{g,F}=0.7b_g(z)$. This setup is closely aligned with the measurements reported in \cite{Bonvin:2023jjq}. Moreover, a significant advantage of using the bias ratio is that it is observable, which can be derived by evaluating $\tilde{\beta}_B/\tilde{\beta}_F$, allowing us to directly obtain best-fit values and  constraints once data become available.

For galaxy redshift uncertainties, they are modeled using a Gaussian velocity dispersion of $\sigma_g = 4.24h^{-1}$ Mpc, following \cite{Howlett:2017asw}. Magnification biases $s_{B,F}$ are computed based on the formalism in \cite{Bonvin:2023jjq} and \cite{Maartens:2021dqy}, while the evolution bias is set to zero at the fiducial, as it has been shown to produce subdominant effects from the observables in this work \cite{Bonvin:2023jjq, Castello:2023zjr}. Throughout our analysis, we assume a fiducial cosmology with no violation of the EP and where $C_{\rm vis, 0}=0$, i.e.\ $\tilde{E}_P=1$ at the fiducial. Moreover, the viscosity parameters are set to zero at the fiducial, namely $\mathcal{Z}=\mathcal{E}=0$.

For DESI BGS, the fiducial galaxy bias is set by the requirement $b_g^{\rm BGS}\,G(z) = 1.34$, where $G(z)$ is the growth factor normalized to unity at the present epoch. The magnification biases for this sample are computed following~\cite{Bonvin:2023jjq,Maartens:2021dqy}. The adopted DESI BGS specifications are listed in Table~\ref{tab:DESI_specs}. 

\begin{table*}
\caption{Survey Specifications for DESI BGS. Here and in the table below, the galaxy number density, $n_g$, is given in units of $10^{-3}\,(h/\mathrm{Mpc})^3$, assuming $h = 0.67$. Survey volumes $V$ are expressed in $(\mathrm{Gpc}/h)^3$. We assume a fiducial bias ratio of $b_{g,B}/b_{g,F} = 0.5$. The galaxy number density $n_g$ and the magnification bias $s_B, s_F$, are obtained from \cite{Maartens:2021dqy}.  The observed absolute magnitude limit is chosen to be \( m_c = 19.5 \). The evolution bias, $f_{evol}$, is set to zero.} \label{tab:DESI_specs} 
\centering
\renewcommand{\arraystretch}{1.5} 
\begin{tabular}{ccccccccccc}
$z$ & $V$ & $n_g$ & $b_g$ & $s_B$ & $s_F$ & $\tilde{\beta}_B$ & $\tilde{\beta}_F$ & $\alpha_B$ & $\alpha_F$  & $B$ \\
\hline
0.05 & 0.04 & 56.7 & 1.38 & 0.243 & 0.11 & 0.289 & 0.578 & -17 & -30.1 & 1.00 \\
0.15 & 0.23 & 16.9 & 1.45 & 0.473 & 0.219 & 0.302 & 0.604 & 0.938 & -7.16 & 0.992 \\
0.25 & 0.58 & 5.54& 1.53 & 0.779 & 0.441 & 0.309 & 0.618 & 5.66 & -0.544 & 0.994 \\
0.35 & 1.04 & 1.42 & 1.61 & 1.21 & 0.812 & 0.313 & 0.625 & 8.9 & 3.93 & 0.99 \\
0.45 & 1.55 & 0.25& 1.7 & 1.81 & 1.37 & 0.312 & 0.624 & 11.8 & 7.74 & 0.996 \\
\end{tabular}
\end{table*}

For Euclid, survey volume, fiducial galaxy biases, and the galaxy number density are taken from the Euclid paper~\cite{Euclid:2019clj}. Magnification biases are computed from the relations in~\cite{Maartens:2021dqy}, assuming a detector cut of $F_c = 3 \times 10^{-16} \,\mathrm{erg}\,\mathrm{cm}^{-2}\,\mathrm{s}^{-1}
$. The Euclid specifications used in our forecasts are given in Table~\ref{tab:Euclid_specs}.

For SKA Phase~2, we adopt the aggressive survey configuration described in~\cite{Quartin:2021dmr}. Fiducial galaxy biases are derived using fitting functions for HI-selected galaxies~\cite{Yahya:2014yva}. Magnification biases are computed from the relations in~\cite{Maartens:2021dqy}, assuming a flux sensitivity limit of $5\,\mu\mathrm{Jy}$. The SKA Phase~2 specifications used in our forecasts are given in Table~\ref{tab:SKA2_specs}.

\begin{table*}
\caption{\label{tab:Euclid_specs} Specifications for the Euclid spectroscopic survey, obtained from \cite{Euclid:2019clj} and \cite{Maartens:2021dqy}.} 
\centering
\renewcommand{\arraystretch}{1.5} 
\begin{tabular}{cccccccccccc}
$z$ & $V$ & $n_g$ & $b_g$ & $s_B$ & $s_F$ & $\tilde{\beta}_B$ & $\tilde{\beta}_F$ & $\alpha_B$ & $\alpha_F$  & $B$ \\
\hline
1.0 & 7.94 & 0.686 & 1.46 & 1.04 & 0.904 & 0.428 & 0.856 & 0.664 & 0.344   & 1 \\
1.2& 9.15 & 0.558 & 1.61 & 1.08 & 0.974 & 0.4 & 0.8 & 0.122 & -0.01  & 1.03 \\
1.4 & 10.1 & 0.421 & 1.75  & 1.1 & 1.03 & 0.376 & 0.753 & -0.32 & -0.36   & 1.03 \\
1.65 & 16.2 & 0.261 & 1.9 & 1.12 & 1.07 & 0.354 & 0.707 &  -0.762 & -0.755 & 1.01 \\
\end{tabular}
\end{table*}

\begin{table*}
\caption{\label{tab:SKA2_specs}Specifications for SKA Phase~2, obtained from \cite{Bull:2015lja} and \cite{Maartens:2021dqy}.}
\centering
\renewcommand{\arraystretch}{1.5} 
\begin{tabular}{cccccccccccc}
$z$ & $V$ & $n_g$ & $b_g$ & $s_B$ & $s_F$ & $\tilde{\beta}_B$ & $\tilde{\beta}_F$ & $\alpha_B$ & $\alpha_F$  & $B$ \\
\hline
0.25 & 1.2 & 121. & 0.674 & 0.364 & -0.026 & 0.701 & 1.4 & -1.96 & -9.12  & 1 \\
0.35 & 2.1 & 71.8 & 0.73 & 0.469 & 0.0235 & 0.69 & 1.38 & -0.35 & -5.91  & 1.06 \\
0.45 & 3.09 & 43.6 & 0.79 & 0.541 & 0.139 & 0.671 & 1.34 & 0.157 & -3.52  & 1.12 \\
0.55 & 4.11 & 26.8 & 0.854 & 0.608 & 0.276 & 0.649 & 1.3 & 0.389 & -1.92  & 1.18 \\
0.65 & 5.11 & 17. & 0.922 & 0.68 & 0.412 & 0.623 & 1.25 & 0.523 & -0.93  & 1.24 \\
0.75 & 6.06 & 10.9 & 0.996 & 0.755 & 0.539 & 0.594 & 1.19 & 0.576 & -0.346  & 1.32 \\
\end{tabular}
\end{table*}

These survey specifications define the framework for the Fisher matrix forecasts presented in the following sections.

The magnification bias $s_{B,F}$ can be measured directly from the galaxy luminosity function (GLF). We follow the prescriptions in \cite{Maartens:2021dqy}  for all three surveys, and estimate the magnification bias for the B,F populations. We also estimate the uncertainty on $s_{B,F}$ by numerically propagating the errors of the GLF as provided in \cite{Pozzetti_2016} for Euclid and \cite{Loveday_2011} for DESI. We obtain the results of Table \ref{tab:desi-s}-\ref{tab:SKA2-s}. For SKA2 we could not find the uncertainties of the GLF parameters, so we adopted the largest uncertainty of the DESI bins. Since in $\alpha_{F,B}$ we fixed $H$ to the fiducial,  the corresponding errors are entirely due to $s_{F,B}$, and  are estimated by standard error propagation. These uncertainties on $\alpha_{F,B}$, are then employed as priors in the Fisher analysis, since we treat $\alpha_{F,B}$ as free parameters here (instead of fixing them at the fiducials, as in \cite{Castello:2024lhl}), which are fully degenerate with $\tilde{E}_P$.

\begin{table}
\caption{Values of $s_{F,B}$ and $\alpha_{F,B}$, and their 1$\sigma$ relative errors for DESI BGS.\label{tab:desi-s}}
\begin{tabular}{cccccc}
& $z=0.05$ & 0.15 & 0.25 & 0.35 & 0.45\tabularnewline
\hline 
$s_B$ & $0.243\pm0.003$ & $0.473\pm0.007$ & $0.779\pm0.015$ & $1.21\pm0.028$ &$1.81\pm0.05$\tabularnewline
$s_F$ & $0.11\pm0.004$ & $0.219\pm0.005$ & $0.441\pm0.012$ & $0.812\pm0.025$ & $1.37\pm0.05$\tabularnewline
$\alpha_B$ & $-17\pm0.3$ & $0.938\pm0.22$ & $5.66\pm0.28$ & $8.9\pm0.35$ & $11.8\pm0.46$ \tabularnewline
$\alpha_F$ & $-30.1\pm0.4$ & $-7.16\pm0.16$ & $-0.544\pm0.22$ & $3.93\pm0.31$ & $7.74\pm0.46$ \tabularnewline
\end{tabular}
\end{table}

\begin{table}
\caption{Values of $s_{F,B}$ and $\alpha_{F,B}$, together with their 1$\sigma$ relative errors for Euclid.\label{tab:euclid-s}}
\begin{tabular}{ccccc}
 & $z=1.0$ & 1.2 & 1.4 & 1.65\tabularnewline
\hline 
$s_B$ & $1.04\pm0.1$ & $1.08\pm0.09$ & $1.1\pm0.09$ & $1.12\pm0.08$\tabularnewline
$s_F$ & $0.90\pm0.15$ & $0.97\pm0.14$ & $1.03\pm0.12$ & $1.07\pm0.11$\tabularnewline
$\alpha_B$ & $0.664\pm0.23$ & $0.122\pm0.12$ & $-0.32\pm0.05$ & $-0.762\pm0.011$\tabularnewline
$\alpha_F$ & $0.344\pm0.35$ & $-0.01\pm0.18$ & $-0.36\pm0.067$ & $-0.755\pm0.016$ \tabularnewline
\end{tabular}
\end{table}

\begin{table}
\caption{Values of $s_{F,B}$ and $\alpha_{F,B}$, and their 1$\sigma$ relative errors for SKA2.\label{tab:SKA2-s}}
\begin{tabular}{ccccccc}
& $z=0.25$ & 0.35 & 0.45 & 0.55 & 0.65 & 0.75 \tabularnewline
\hline 
$s_B$ & $0.364\pm0.036$ & $0.469\pm0.047$ & $0.541\pm0.054$ & $0.608\pm0.061$ &$0.68\pm0.068$ & $0.755\pm0.076$ \tabularnewline
$s_F$ & $-0.026\pm0.0026$ & $0.0235\pm0.0024$ & $0.139\pm0.014$ & $0.276\pm0.028$ & $0.412\pm0.041$ & $0.539\pm0.054$ \tabularnewline
$\alpha_B$ & $-1.96\pm0.67$ & $-0.35\pm0.59$ & $0.157\pm0.49$ & $0.389\pm0.42$ & $0.523\pm0.37$ & $0.576\pm0.32$ \tabularnewline
$\alpha_F$ & $-9.12\pm0.048$ & $-5.91\pm0.029$ & $-3.52\pm0.13$ & $-1.92\pm0.19$ & $-0.93\pm0.22$ & $-0.346\pm0.23$ \tabularnewline
\end{tabular}
\end{table}
%%%%%%%%%%%%%%%%%%%%%%%%%%%%%%%%%%%%%%%%%%%%%%%%%%%%%%%%
%%%%%%%%%%%%%%%%%%%%%%%%%%%%%%%%%%%%%%%%%%%%%%%%%%%%%%%%
%%%%%%%%%%%%%%%%%%%%%%%%%%%%%%%%%%%%%%%%%%%%%%%%%%%%%%%%
%%%%%%%%%%%%%%%%%%%%%%%%%%%%%%%%%%%%%%%%%%%%%%%%%%%%%%%%

\section{Results and Discussion} \label{sec:res}

\begin{table}
\caption{\label{tab:DESI32} DESI \(1\sigma\) absolute errors for the parameters, assuming the fiducial bias ratio \(b_{g,F}/b_{g,B} = 0.5\). Throughout this and the following tables, we report only one value for constraints on \(C_{\rm vis,0}\), since it is constant along all the redshift bins.}
\centering
\begin{tabular}{ccccccccc}
$z$ & $\log\tilde{\beta}_{\mathrm{F},z}$ & $\log\tilde{\beta}_{\mathrm{B},z}$ & $\log h_d$ & $\log \tilde{E}_{P,z}$ & $C_{\rm vis,0}$ & $\log B_z$ & $\log\sigma_g$ & $\log n_n$ \\ \hline
 0.05 & 0.071 & 0.067 & 0.081 & 6.2 &\multirow{5}{5em}{\num{1.44e-6}\rule{0pt}{2.5ex}} & - & 0.92 & 0.054  \\
 0.15 & 0.054 & 0.051 & 0.052 & 3.7 & & 0.096 & 0.43 & 0.022  \\
 0.25 & 0.056 & 0.052 & 0.043 & 4.4 & & 0.093 & 0.32 & 0.014 \\
 0.35 & 0.072 & 0.066 & 0.041 & 6.7 & & 0.092 & 0.3 & 0.011 \\
 0.45 & 0.12 & 0.11 & 0.048 & 14 &  & 0.096 & 0.35 & 0.009 \\
\end{tabular}
\end{table}

\begin{table}
\caption{\label{tab:Euclid32} Euclid $1\sigma$ absolute errors for the parameters, assuming the fiducial bias ratio  \(b_{g,F}/b_{g,B} = 0.5\).}
\centering
\begin{tabular}{ccccccccc}
$z$ & $\log \tilde{\beta}_{\mathrm{F},z}$ & $\log\tilde{\beta}_{\mathrm{B},z}$ & $\log h_d$ & $\log \tilde{E}_{P,z}$ & $C_{\rm vis,0}$ & $\log B_z$ & $\log\sigma_g$ & $\log n_n$ \\ \hline
 1.0 & 0.029 & 0.025 & 0.016 & 2.9 &\multirow{5}{5em}{\num{3.32e-7}\rule{0pt}{2.5ex}} & - & 0.12 & 0.0038 \\
 1.2 & 0.029 & 0.026 & 0.015 & 2.9 & & 0.024 & 0.11  & 0.0036\\
 1.4 & 0.032 & 0.028 & 0.016 & 3.3 & & 0.025 & 0.11 &  0.0034\\
 1.65 & 0.033 & 0.029 & 0.015 & 3.4 & & 0.026 & 0.13  & 0.0027\\
\end{tabular}
\end{table}

\begin{table}
\caption{\label{tab:SKA32}  SKA $1\sigma$ absolute errors for the parameters, assuming the fiducial bias ratio  \(b_{g,F}/b_{g,B} = 0.5\)..}
\centering
\begin{tabular}{ccccccccc}
$z$ & $\log\tilde{\beta}_{\mathrm{F},z}$ & $\log\tilde{\beta}_{\mathrm{B},z}$ & $\log h_d$ & $\log \tilde{E}_{P,z}$ & $C_{\rm vis,0}$ & $\log B_z$ & $\log\sigma_g$ & $\log n_n$ \\ \hline
 0.25 & 0.0052 & 0.0047 & 0.0058 & 1.2 &\multirow{5}{5em}{\num{1.08e-7}\rule{0pt}{2.5ex}} & - & 0.16 & 0.01\\
 0.35 & 0.0051 & 0.0046 & 0.0057 & 1.1 &  & 0.037 & 0.12 & 0.0075\\
 0.45 & 0.0051 & 0.0046 & 0.0058 & 0.89 &  & 0.035 & 0.1 & 0.0061\\
 0.55 & 0.0055 & 0.005 & 0.0062 & 0.9 &  & 0.033 & 0.09 & 0.0053\\
 0.65 & 0.0061 & 0.0054 & 0.0068 & 0.94 &  & 0.033 & 0.085 & 0.0048\\
 0.75 & 0.0068 & 0.006 & 0.0075 & 0.95 &  & 0.033 & 0.082 & 0.0044\\
\end{tabular}
\end{table}

\begin{table}
\caption{\label{tab:SKA32_2}  SKA $1\sigma$ absolute errors for the parameters, asssuming the fiducial bias ratio $b_F/b_B=0.2$, with $b_F = 0.7b_g$ and $b_B = 3.5b_g$.}
\centering
\begin{tabular}{ccccccccc}
$z$ & $\log\tilde{\beta}_{\mathrm{F},z}$ & $\log\tilde{\beta}_{\mathrm{B},z}$ & $\log h_d$ & $\log \tilde{E}_{P,z}$ & $C_{\rm vis,0}$ & $\log B_z$ & $\log\sigma_g$ & $\log n_n$ \\ \hline
 0.25 & 0.004 & 0.0032 & 0.0044 & 0.43 &\multirow{5}{5em}{\num{7.46e-8}\rule{0pt}{2.5ex}} & - & 0.15 & 0.0098 \\
 0.35 & 0.0038 & 0.003 & 0.0042 & 0.31 &  & 0.036 & 0.12 & 0.0074\\
 0.45 & 0.004 & 0.0032 & 0.0045 & 0.37 &  & 0.034 & 0.098 & 0.0061\\
 0.55 & 0.0043 & 0.0034 & 0.0049 & 0.44 &  & 0.033 & 0.086 & 0.0053\\
 0.65 & 0.0048 & 0.0038 & 0.0055 & 0.49 &  & 0.032 & 0.08 & 0.0048\\
 0.75 & 0.0054 & 0.0043 & 0.0062 & 0.53 &  & 0.032 & 0.076 & 0.0044\\
\end{tabular}
\end{table}

\begin{table}
\caption{\label{tab:SKA32_3}  SKA \(1\sigma\) errors for the parameters, assuming the fiducial bias ratio \(b_F/b_B = 0.7\), with \(b_F = 0.7\,b_g\) and \(b_B = b_g\).}
\centering
\begin{tabular}{ccccccccc}
$z$ & $\log\tilde{\beta}_{\mathrm{F},z}$ & $\log\tilde{\beta}_{\mathrm{B},z}$ & $\log h_d$ & $\log \tilde{E}_{P,z}$ & $C_{\rm vis,0}$ & $\log B_z$ & $\log\sigma_g$ & $\log n_n$ \\ \hline
 0.25 & 0.0079 & 0.0073 & 0.0087 & 2. &\multirow{5}{5em}{\num{1.73e-7}\rule{0pt}{2.5ex}} & - & 0.16  & 0.0099\\
 0.35 & 0.0081 & 0.0076 & 0.0086 & 2.1 &  & 0.037 & 0.13 & 0.0075\\
 0.45 & 0.008 & 0.0075 & 0.0085 & 1.5 &  & 0.034 & 0.11 & 0.0061\\
 0.55 & 0.0087 & 0.0082 & 0.0089 & 1.7 &  & 0.033 & 0.1 & 0.0053\\
 0.65 & 0.0095 & 0.0088 & 0.0093 & 1.8 &  & 0.033 & 0.097 & 0.0048\\
 0.75 & 0.01 & 0.0096 & 0.0097 & 1.8 &  & 0.033 & 0.096 & 0.0044\\
\end{tabular}
\end{table}

We present the forecasting results for the baseline bias configuration with \( b_{g,F}/b_{g,B} = 0.5 \) in Tables~\ref{tab:DESI32}, \ref{tab:Euclid32}, and~\ref{tab:SKA32}, corresponding to the DESI, Euclid, and SKA2 cases, respectively. 
All results displayed below are obtained by marginalizing over all the remaining parameters.
We begin by commenting on the measurability of the viscosity parameter \( C_{\rm vis,0} \). From these tables, we find that upcoming galaxy surveys will be capable of placing robust and remarkably precise constraints on the viscosity of dark matter. Namely, the present-day value of the dimensionless viscosity parameter \( C_{\rm vis,0} \) can be constrained at the level of \( \mathcal{O}(10^{-7}) \), with the exact sensitivity depending on the specific survey configurations. We remark that our constraints are of a comparable order of magnitude to previous constraints (for a more detailed discussion on these constraints, see Appendix \ref{app:ObsConstraints}), but with many fewer model assumptions. 

Among the surveys considered, SKA2 offers the most stringent constraints, owing to its large sky coverage and high galaxy number density. For this survey, the $1\sigma$ uncertainty on \( C_{\rm vis,0} \) can reach below \( 1.1 \times 10^{-7} \). In addition, Euclid also achieves tight constraints, reaching the few-\(10^{-7}\) level. DESI BGS, although limited to lower redshifts and affected by larger shot noise in its higher bins, still yields competitive bounds, with the \(1\sigma\) uncertainty on \( C_{\rm vis,0} \) reaching approximately \( 1.4 \times 10^{-6} \).

An additional parameter included in our Fisher analysis, which is not present in the previous work \cite{Castello:2024lhl}, is the shot noise amplitude \( n_n \). To allow for a more flexible and data-driven modeling of noise contributions that evolve with cosmic time, we treat \( n_n \) as a free parameter in each redshift bin. This allows the noise contribution to deviate from the standard assumption that galaxies follow a Poisson sampling of the underlying matter distribution. This effective noise term can capture additional stochastic effects, such as observational systematics, selection effects (e.g., the Malmquist bias), or complex masking, which may vary across different redshifts and cannot be accurately inferred from galaxy number densities alone. As shown in Tables~\ref{tab:DESI32}--\ref{tab:SKA32}, we find that \( n_n \) can be constrained to better than 1\% across all redshift bins for both Euclid and SKA2. For DESI BGS, the constraints are slightly weaker, ranging from approximately 5\% at low redshift to sub-percent precision at higher redshifts.

To assess the impact of the bias ratio on the constraining power, we repeat the SKA2 forecast for three different bias configurations: the baseline choice \( b_{g,F}/b_{g,B} = 0.5 \) (Table~\ref{tab:SKA32}), a more asymmetric split with \( b_{g,F}/b_{g,B} = 0.2 \) (Table~\ref{tab:SKA32_2}), and a more symmetric case with \( b_{g,F}/b_{g,B} = 0.7 \) (Table~\ref{tab:SKA32_3}). We find that, consistent with expectations from our earlier analysis in \cite{Castello:2024lhl}, larger differences in galaxy bias between the two tracers lead to tighter constraints on both \(\tilde{E}_{P,z}\) and \( C_{\rm vis,0} \). Specifically, the most asymmetric configuration (\( b_{g,F}/b_{g,B} = 0.2 \)) yields the tightest constraint on the viscosity parameter, with a \(1\sigma\) uncertainty of \( C_{\rm vis,0} < 7.5 \times 10^{-8} \), while the more symmetric case (\( b_{g,F}/b_{g,B} = 0.7 \)) leads to a weaker constraint of \( \sim 1.7 \times 10^{-7} \). These results highlight the importance of optimizing tracer selection to maximize sensitivity to relativistic effects and to break parameter degeneracies. We also find that the precision of \(\tilde{E}_{P,z}\) is more sensitive to the bias ratio than that for \( C_{\rm vis,0} \), reflecting its direct dependence on the dipole amplitude. Nevertheless, the viscosity parameter can be robustly constrained across a wide range of bias configurations.

Lastly, we find that the constraints on \(\tilde{E}_{P,z}\) are significantly weaker than those obtained in \cite{Castello:2024lhl} for the viscous-free counterpart $E_P$. 
The key reason is that, in the present analysis, the parameters \(\alpha_B\) and \(\alpha_F\) are treated as free in each redshift bin. 
Since these parameters enter the dipole with the same structure as \(\tilde{E}_{P,z}\), they are fully degenerate with it; thus, only external information can break the degeneracy. 
To remain conservative, we adopt broad Gaussian priors on \(\alpha_B\) and \(\alpha_F\), estimated from the uncertainties of the galaxy magnification biases (see, e.g., Table \ref{tab:SKA2-s}). 
Compared to fixed $\alpha$ values, this choice ensures robustness but inevitably weakens the constraining power on \(\tilde{E}_{P,z}\).
Future improvements in the determination of magnification biases will, therefore, be crucial for tightening the bounds on \(\tilde{E}_{P,z}\).

%%%%%%%%%%%%%%%%%%%%%%%%%%%%%%%%%%%%%%%%%%%%%%%%%%%%%%%%
%%%%%%%%%%%%%%%%%%%%%%%%%%%%%%%%%%%%%%%%%%%%%%%%%%%%%%%%
%%%%%%%%%%%%%%%%%%%%%%%%%%%%%%%%%%%%%%%%%%%%%%%%%%%%%%%%
%%%%%%%%%%%%%%%%%%%%%%%%%%%%%%%%%%%%%%%%%%%%%%%%%%%%%%%%

\section{Conclusion} \label{sec:con}

In this work, we presented a thorough study of the cosmological Euler equation, including the effects of modified gravity and viscous dark matter, characterized by both bulk and shear viscosities, on cosmological observables relevant for testing gravity. We derived the modified background and linear perturbation equations. Although the viscous effects are negligible at the background level, the viscous terms enhance the effective friction in the growth of matter perturbations, leading to a scale-dependent suppression of the observed power spectrum.  
Furthermore, viscosity creates terms in the perturbed Euler and Einstein equations that can mimic the effects induced by modifications to gravity.

Motivated by the question of whether viscosity would also interfere with the EP test proposed in \cite{Castello:2024lhl}, we extended the previous model-independent analysis to incorporate viscosity. In addition to standard Newtonian terms, we took into account relativistic corrections in galaxy number counts, preserving all terms up to the order of $\lambda=\mathcal{H}/k$ in the galaxy power spectra. In this context, we identified the parameter $\tilde{E}_P$, which generalizes the original EP estimator $E_P$ in the presence of viscosity. Moreover, we showed that in the small-viscosity limit, the scale-independent component $\tilde{E}_{P,z}$ reduced to $E_P$ and can still serve as a smoking gun for EP violation.

Notably, we identified an observable $C_{\rm vis,0}$ that encodes the present-day viscosity and can be measured using a model-independent approach. Performing a Fisher forecast for Stage-IV surveys, we found that this parameter can be tightly constrained for all three surveys considered here—DESI, Euclid, and SKA2—with a precision of order $\mathcal{O}(10^{-6})$ or better under the baseline bias configuration, thanks to the scale-dependence of the perturbation growth induced by viscosity (boosted by $\lambda^{-2}$ in the power spectra) and the Doppler effect encoded in the relativistic corrections (at $\mathcal{O}(\lambda^{-1})$ and $\mathcal{O}(\lambda^{0})$ in the power spectra). This constraint can be further tightened if the two tracer populations show a larger bias difference. We note, however, that the bulk and shear components of viscosity cannot be disentangled within our model-independent framework. 

Looking ahead, the main conclusion of this paper points in two directions. First, the weak equivalence principle can still be tested using the null test on $\tilde{E}_{P,z}$ in the presence of dark matter viscosity, unless the \textit{true} EP violation occurs only at the $10^{-6}$ level. In that latter case, one cannot disentangle EP violation from the effects of viscosity. Secondly, viscosity itself can be regarded as a promising quantity that can be robustly measured with the forthcoming Stage-IV surveys. Once real data become available, one can place constraints on it through an MCMC analysis. It should be noted that the Fisher formalism is exact only if the likelihood is Gaussian and the observables depend linearly on
all parameters. In practice, nonlinearities in the model mean the Fisher matrix provides only a local estimate of the covariance around the fiducial model. The true posteriors may have non-Gaussian tails or more complex features; thus, our Fisher forecast should be interpreted as a lower bound on the expected uncertainties from real data, representing the best-case statistical constraining power. We leave these investigations for future work.

%%%%%%%%%%%%%%%%%%%%%%%%%%%%%%%%%%%%%%%%%%%%%
%%%%%%%%%%%%%%%%%%%%%%%%%%%%%%%%%%%%%%%%%%%%%

\vspace{0.5cm}
\noindent {\bf Acknowledgments }
\newline
\newline
LA acknowledges support by DFG  under Germany's Excellence 
Strategy EXC 2181/1 - 390900948 (the Heidelberg STRUCTURES Excellence 
Cluster) and under Project  554679582 "GeoGrav: Cosmological Geometry
and Gravity with nonlinear physics". The Mathematica code we used to produce the Fisher matrix results is publicly available at \cite{code}.

%%%%%%%%%%%%%%%%%%%%%%%%%%%%%%%%%%%%%%%%%%%%%
%%%%%%%%%%%%%%%%%%%%%%%%%%%%%%%%%%%%%%%%%%%%%

\appendix 

\section{Growth equation and growth rate}\label{app:growth}

\subsection{Growth equation of viscous dark matter\label{app:growth_eq}}
The perturbed continuity and Euler equations \eqref{eq:pert_continuity_2} and \eqref{eq:pert_Euler_EP_3} can be combined into a second order equation for the density contrast, which has the form
\begin{equation}
    \delta_m''+\bigl(\mathcal{A}+\mathcal{B}\lambda^{-2}\bigr)\delta_m'-\bigl(\mathcal{C}+\mathcal{D}\lambda^{-2}\bigr)\delta_m=0 ~,
\end{equation}
with 
\begin{align}
    \mathcal{A}&=1+\frac{\mathcal{H'}}{\mathcal{H}}+3\mathcal{Z}+\frac{2\mathcal{Z'}}{1-2\mathcal{Z}}+\tilde{\Theta}\\
    \mathcal{B}&=\frac{1}{3}\frac{\mathcal{Z}+\mathcal{E}}{1-\mathcal{Z}}\equiv C_{\rm vis}\\
    \mathcal{C}&=\frac{3}{2}\Omega_m(1-2\mathcal{Z})(1+\tilde{\Gamma})\mu_{\rm G}-3\mathcal{Z}\left(1+\frac{\mathcal{H}'}{\mathcal{H}}+\tilde{\Theta}\right)-\frac{3\mathcal{Z}'}{1-2\mathcal{Z}}\\
    \mathcal{D}&=-\frac{\mathcal{Z}(\mathcal{Z}+\mathcal{E})}{1-\mathcal{Z}} ~.
\end{align}
Equivalently, the equation for the growth factor $f\equiv\delta_m'/\delta_m$ is
\begin{equation}
    f'+f^2+(\mathcal{A}+\mathcal{B}\lambda^{-2})f-\mathcal{C}-\mathcal{D}\lambda^{-2}=0\, .\label{eq:evolution_f_g}
\end{equation}
In these general expressions, $\mathcal{Z}'$ has not yet been specified. 
Using the background continuity equation \eqref{eq:continuity}, it follows that\footnote{For the scenario of coupled quintessence, see the note at the end of Appendix \ref{app:AppendixCQ}.} $\mathcal{Z}'=\mathcal{Z}\left(2+\frac{\mathcal{H}'}{\mathcal{H}}-3\mathcal{Z}\right)$. Inserting this and also $\tilde{\Theta}$ and $\tilde{\Gamma}$ yields
\begin{align}
    \mathcal{A}&=\frac{1}{1-2\mathcal{Z}}\left(1+\frac{\mathcal{H'}}{\mathcal{H}}+5\mathcal{Z}-12\mathcal{Z}^2\right)+\frac{\mathcal{Z}}{1-\mathcal{Z}}\left(1-\frac{\mathcal{H}'}{\mathcal{H}}\right)\nonumber\\
    &\qquad+\frac{3}{2}\frac{\Omega_m}{1-\mathcal{Z}}\biggl(\mathcal{E}(1+\Gamma)-\mathcal{Z}\eta\Bigl(\mathcal{Z}+\mathcal{E}+(1-2\mathcal{Z})\mu_{\rm G}\Bigr)\biggr)+\frac{\Theta}{1-\mathcal{Z}}\\
    \mathcal{B}&=\frac{1}{3}\frac{\mathcal{Z}+\mathcal{E}}{1-\mathcal{Z}}\equiv C_{\rm vis}\\
    \mathcal{C}&=\frac{3}{2}\Omega_m\biggl\{(1-2\mathcal{Z})\left(1+\frac{\Gamma}{1-\mathcal{Z}}+\frac{\mathcal{Z}}{1-\mathcal{Z}}\Bigl(1-\eta\bigl(1-\eta'/\eta-\mu_{\rm G}'/\mu_{\rm G}\bigr)\Bigr)\right)\mu_{\rm G} \nonumber\\
    &\qquad\qquad\qquad\qquad\qquad\qquad\qquad\qquad\qquad\qquad\qquad -\frac{3\mathcal{Z}}{1-\mathcal{Z}}\biggl(\mathcal{E}(1+\Gamma)-\mathcal{Z}\eta\Bigl(\mathcal{Z}+\mathcal{E}+(1-2\mathcal{Z})\mu_{\rm G}\Bigr)\biggr)\biggr\}\nonumber\\
    &\qquad -\frac{3\mathcal{Z}}{1-2\mathcal{Z}}\left(3+2(1-\mathcal{Z})\frac{\mathcal{H}'}{\mathcal{H}}-5\mathcal{Z}\right)-\frac{3\mathcal{Z}^2}{1-\mathcal{Z}}\left(1-\frac{\mathcal{H}'}{\mathcal{H}}\right)-\frac{3\mathcal{Z}}{1-\mathcal{Z}}\Theta\\
    \mathcal{D}&=-\frac{\mathcal{Z}(\mathcal{Z}+\mathcal{E})}{1-\mathcal{Z}} ~.
\end{align}
It is important to notice that the term $\mathcal{B}$ is always positive for $0<{\mathcal Z}<1$ and for positive ${\mathcal E}$. Since $\mathcal{B}$ is multiplied by the   factor $\lambda^{-2}$, which can be very large at small scales,  the ``friction'' term $\mathcal{A}+\mathcal{B}\lambda^{-2}$ can increase a lot and, as a consequence, the growth factor $f$ at small scales is strongly reduced even for small values of the viscosity coefficients.
Since $\mathcal{Z,E}$ are of order $10^{-5}$ or smaller (see Appendix \ref{app:ObsConstraints}), we can neglect many terms in the coefficients, so that we obtain:
\begin{align}
    \mathcal{A} & \simeq1+\frac{\mathcal{H'}}{\mathcal{H}}+\Theta\\
    \mathcal{B} & \simeq\frac{1}{3}(\mathcal{Z}+\mathcal{E})\simeq C_{{\rm vis}}\\
    \mathcal{C} & \simeq\frac{3}{2}\Omega_{m}\bigl(1+\Gamma\bigr)\mu_{G}\\
    \mathcal{D} & \simeq0~.
\end{align}
We see that viscosity affects the growth rate $f$ only through $C_{{\rm vis}}\simeq\frac{1}{3\Omega_{m}H}\bigl(\zeta+\frac{4}{3}\eta_{s}\bigr)$ and that its dominant effect is suppression of structure growth on small scales.

\subsection{Approximate solution} \label{app:app_growth}
In this limit, and further assuming $\mathcal{B}\lambda^{-2}$ to be a small correction at the linear scales we are interested in, we can derive an approximate solution for the growth rate by decomposing it into a $z$-dependent part and a $(k,z)$-dependent correction: $f \approx f_{z}(z)+\epsilon\lambda_{0}^{-2}\tilde{f}_{v}(z)$, where $\lambda_0 = H_0/k$ and recall that $\epsilon$ is an order parameter that will be put to unity at the end. We assume $C_{\rm vis}$ to be of order $\epsilon$. Then for $f_{z}$ we have at zero-th order in $\epsilon$,
\begin{equation}
    f_{z}'+f_{z}^{2}+\mathcal{A}f_{z}-\mathcal{C}=0\,.\label{eq:evolution_f_z}
\end{equation}
Since for $\Theta=\Gamma=0$ and $\mu_{G}=1$ this is just the standard growth equation, in $\Lambda$CDM we have $f_{z}\approx\Omega_{m}^{\gamma}$ with $\gamma=0.545$. At first order in $\epsilon$ we obtain
\begin{equation}
    \tilde{f}'_{v}+\tilde{f}_{v}(\mathcal{A}+2f_{z})=-\frac{1}{a^{2}E^{2}}C_{{\rm vis}}f_{z}\,.\label{eq:evolution_f_k}
\end{equation}
where $E\equiv H/H_{0}$. After imposing the boundary condition that scale-dependence is negligible at high redshifts, $\tilde{f}_{v}(N_{0})=0$ as $N_{0}\to-\infty$, the solution to Eq.~\eqref{eq:evolution_f_k} in terms of $N=\log a$ is 
\begin{equation}
    \tilde{f}_{v}(N)=-e^{-I_{1}(N)}\int_{N_{0}}^{N}e^{I_{1}(x)}\frac{C_{{\rm vis}}}{e^{2x}E^{2}}f_{z}\mathrm{d}x\,,\label{eq:fk_solution}
\end{equation}
where $I_{1}(N)=\int_{N_{0}}^{N}(2f_{z}+\mathcal{A})\mathrm{d}x$. The initial time $N_0$ should be set far enough in the past that the viscosity correction is initially negligible. 
If $\zeta,\eta_{s}$ are constant, we can write $C_{\rm vis}=C_{\rm vis,0}\,e^{3N}E\,$ in terms of its present value 
\begin{equation}
    C_{\rm vis,0} =\frac{\zeta+\frac{4}{3}\eta_s}{3\Omega_{m0}H_0}
=\frac{\mathcal{Z}_0+\mathcal{E}_0}{3}\,,\label{eq:cvis0}
    \end{equation} 
    and Eq. \eqref{eq:fk_solution} becomes
\begin{equation}
    \tilde{f}_{v}(N)=-C_{\rm vis,0}\,\mathcal{I}(N) \,,\label{eq:fk_solution-2}
\end{equation}
with $\mathcal{I}(N)=e^{-I_{1}(N)}\int_{N_{0}}^{N}e^{I_{1}(x)}\frac{e^{x}}{E}f_{z}\mathrm{d}x$.
The full linear growth rate at a given scale $k$ is then given by 
\begin{equation}
    f=f_{z}-\frac{k^{2}}{H_{0}^{2}}C_{\mathrm{vis},0}\,\mathcal{I}=f_z-\lambda^{-2}C_{\mathrm{vis},0}\,E^2\,\mathcal{I} \,.
\end{equation}
In the range $C_{\mathrm{vis},0}\leq 10^{-6}$ and $0.05\,h/\mathrm{Mpc}\leq k \leq 0.12\,h/\mathrm{Mpc}$, this approximation agrees with the numerical solution of the growth equation within 1\% (neglecting the modified gravity parameters and assuming $\Lambda$CDM background evolution). For $C_{\mathrm{vis},0}=10^{-5}$ and $k=0.12\,h/\mathrm{Mpc}$, the relative error can become as large as 8\% for small redshifts.

Under this decomposition of the growth rate, the parameter $\tilde\beta$ becomes 
\begin{equation}
   \tilde{\beta}=\frac{f+3\mathcal{Z}}{b_{g}(1-2\mathcal{Z})}=\tilde{\beta}_{z}+\lambda^{-2}\tilde{\beta}_{v} \,  ,
\end{equation}
with $\tilde{\beta}_z=\frac{f_z+3\mathcal{Z}}{b_g(1-2\mathcal{Z})}\simeq f_z/b_g$ and $\tilde{\beta}_v=\frac{E^2\tilde{f}_v}{b_g(1-2\mathcal{Z)}}\simeq E^2\tilde{f}_v/b_g=-C_{\rm vis,0}\,E^2\,\mathcal{I}/b_g$. 
Analogously, we can define $\tilde{\beta}_{B/F,z}$ and $\tilde{\beta}_{B/F,v}$ by using $b_{g,B/F}$. 
The growth function normalized to unity today can be written as $G(N)=\exp\int_{N}^{0}f\,dx = G_z+\lambda^{-2}G_v$
with
\begin{align}
    G_z&=\exp\int_{N}^{0}f_z\,\mathrm{d}x \label{eq:G_z}\\
    G_v&=-C_{\mathrm{vis},0}G_zE^2\int_{N}^{0}\mathcal{I}\,\mathrm{d}x \label{eq:G_v}\,,
\end{align}
keeping terms up to \(\mathcal{O}(\epsilon)\). Furthermore, $B=b_{g,B}b_{g,F}G^2=B_z+\lambda^{-2}B_v$ with
\begin{align}
    B_z&=b_{g,B}b_{g,F}G_z^2\equiv\frac{G_z^2f_z^2}{\tilde{\beta}_{B,z}\tilde{\beta}_{F,z}}\\
    B_v&=-2C_{\mathrm{vis},0}B_zE^2\int_{N}^{0}\mathcal{I}\,\mathrm{d}x \,.
\end{align}
The $\tilde{E}_P$ parameter \eqref{eq:new_EP} becomes $\tilde{E}_P=\tilde{E}_{P,z}+\lambda^{-2}\tilde{E}_{P,v}$ with
\begin{align}
    \tilde{E}_{P,z}&=1+\tilde{\Theta}-\frac{3}{2}\Omega_m\mathcal{E}-\frac{3}{2}\frac{\Omega_m\mu_{\rm G}\tilde{\Gamma}(1-2\mathcal{Z})}{f_z+3\mathcal{Z}}\simeq 1+\Theta-\frac{3}{2}\frac{\Omega_m\mu_{\rm G}\Gamma}{f_z}\\
    \tilde{E}_{P,v}&=-\frac{3}{2}\Omega_m\mu_{\rm G}\tilde{\Gamma}(1-2\mathcal{Z})\frac{C_{\mathrm{vis},0}E^2\mathcal{I}}{(f_z+3\mathcal{Z})^2}\simeq  -\frac{3}{2}\Omega_m\mu_{\rm G}\Gamma C_{\mathrm{vis},0}E^2\mathcal{I}/f_z^2 \,.
\end{align}
The dominant contribution of viscosity comes indirectly from the suppressed growth rate on small scales. In our analysis, we set $\tilde{E}_{P,v}=0$ as the derivative $\frac{\partial \tilde{E}_{P,v}}{\partial C_{\rm vis, 0}}$ vanishes at the fiducial value $\Gamma =0$.

\subsection{Parametrization of the growth rate}
\label{app:growthParam}
In addition to the approximate solution for the growth rate given above, we found an analytic parametrization that extends the well-known relation $f=\Omega_m^\gamma$ (with $\gamma=0.545$) to the viscous case. We restrict ourselves to the case $\Theta=\Gamma=0$, $\eta=\mu_{\rm G}=1$. 
Because the actual viscosity is so small (according to the constraints in Appendix \ref{app:ObsConstraints} and the results of our analysis, the upper bound on $\mathcal{Z}$ is of order $10^{-5}$ to $10^{-6}$), the contribution of viscosity to the background evolution was neglected. 

To fit a parametrization, we sampled values of $f$ in the range $-1.5\leq N \leq 0$ while varying the parameters between $0.01\,h/\mathrm{Mpc}\leq k \leq 0.12\,h/\mathrm{Mpc}$, $10^{-7}\leq C_{\mathrm{vis},0}\leq 10^{-5}$, and $0.2\leq \Omega_{m0}\leq 0.4$. The parametrization we considered is
\begin{equation}
    f=\Omega_m^{\gamma\Bigl[1+(b_0 +b_1 \Omega_{m0})C_{\mathrm{vis},0}\,\bigl(\frac{k}{H_0}\bigr)^2\Bigr]} ~.
\end{equation}
 As best-fit values for the fit parameters $b_0$ and $b_1$, we obtained
\begin{equation}
    b_0=0.186 ~,\qquad b_1=0.905 ~.
\end{equation}
This fit is precise to better than 2\% in the entire range of $z,k,C_{\mathrm{vis},0}$ and $\Omega_{m0}$ we considered.

\section{Time dependent viscosities}
\label{app:time_dep}
In this work, we assumed constant viscosities $\zeta$ and $\eta_s$. In this appendix, we investigate how a generic time dependence $\zeta(N)$, $\eta_s(N)$ modifies the relevant equations. The viscosities are still assumed to be unperturbed. In that case, the perturbed continuity equation \eqref{eq:pert_continuity_2} remains unchanged, while in the perturbed Euler equation \eqref{eq:pert_Euler_2}, there is an additional term $-\frac{\mathcal{Z}}{1-\mathcal{Z}}\frac{\zeta'}{\zeta}$ in the ``friction part''. In the notation of section \ref{sec:vis_EP}, this means that $\tilde{\Theta}\rightarrow\tilde{\Theta}-\frac{\mathcal{Z}}{1-\mathcal{Z}}\frac{\zeta'}{\zeta}$. An additional term that arises in Eq. \eqref{eq:PhiPrime} is of order $\lambda^4$. The $\tilde{E}_p$-parameter is also modified as $\tilde{E}_p\rightarrow \tilde{E}_P-\frac{\mathcal{Z}}{1-\mathcal{Z}}\frac{\zeta'}{\zeta}$. Note that a possible time derivative of $\eta_s$ does not enter the equations. Finally, in the growth equation (see Appendix \ref{app:growth}), one has to substitute $\mathcal{A}\rightarrow\mathcal{A}-\frac{\mathcal{Z}}{1-\mathcal{Z}}\frac{\zeta'}{\zeta}$ and $\mathcal{C}\rightarrow\mathcal{C}+3\mathcal{Z}\bigl(\frac{\mathcal{Z}}{1-\mathcal{Z}}-\frac{1}{1-2\mathcal{Z}}\bigr)\frac{\zeta'}{\zeta}$. Since these corrections are not multiplied by the large factor $\lambda^{-2}$, and assuming $\zeta$ does not vary  too strongly with time, they are small and can be neglected. The results of Appendix \ref{app:growth} therefore remain valid.

\section{Coefficients of the power spectra}\label{sec:coefficients}

\begin{table} \label{tab:co_BB}
\centering
\renewcommand{\arraystretch}{1.4}
\begin{tabular}{c|c}
\hline
Power \( \mu^a \lambda^b \) & Coefficient \( \mathcal{C}^{(BB)}_{a,b} \) \\
\hline
\( \mu^0 \lambda^0 \)     & \( 1 \) \\
\( \mu^2 \lambda^0 \)     & \( 2\tilde{\beta}_{B,z} - 2 C_{\mathrm{vis}}(\alpha_B - \tilde{E}_{P,z})\tilde{\beta}_{B,z}^2 \) \\
\( \mu^4 \lambda^0 \)     & \( \tilde{\beta}_{B,z}^2 \) \\
\( \mu^2 \lambda^{-2} \)  & \( 2 \tilde{\beta}_{B,v} \) \\
\( \mu^4 \lambda^{-2} \)  & \(2 \tilde{\beta}_{B,z} \tilde{\beta}_{B,v} \) \\
\hline
\end{tabular}
\caption{Coefficients \( \mathcal{C}^{(BB)}_{a,b} \) in the expansion of \( P_{\Delta_B\Delta_B} \) in powers of \( \mu^a \lambda^b \).}
\label{tab:C_P_BB}
\end{table}

In this appendix, we present the coefficients \( \mathcal{C}^{(XY)}_{a,b} \) that appear in the expansion of the power spectra \( P_{\Delta_X \Delta_Y} \) in Eqs.~\eqref{eq:co_1}--\eqref{eq:co_3} in powers of \( \mu^a \lambda^b \), where \( X, Y \in \{B, F\} \).

\paragraph{\(P_{\Delta_B\Delta_B}\):}
The coefficients \( \mathcal{C}^{(BB)}_{a,b} \) appearing in the expansion of the power spectrum \( P_{\Delta_B \Delta_B} \) are listed in Table~\ref{tab:C_P_BB}.

\paragraph{\(P_{\Delta_F\Delta_F}\):}
The coefficients \( \mathcal{C}^{(FF)}_{a,b} \) for the power spectrum \( P_{\Delta_F \Delta_F} \) can be directly obtained from Table~\ref{tab:C_P_BB} by replacing all subscripts and labels \( B \rightarrow F \).

\paragraph{\(P_{\Delta_B\Delta_F}\):}
The coefficients \( \mathcal{C}^{(BF)}_{a,b} \) corresponding to the cross-spectrum \( P_{\Delta_B \Delta_F} \) are given in Table~\ref{tab:C_P_BF}.

\begin{table} \label{tab:tab:C_P_BF}
\centering
\renewcommand{\arraystretch}{1.4}
\begin{tabular}{c|c}
\hline
Power \( \mu^a \lambda^b \) & Coefficient \( \mathcal{C}^{(BF)}_{a,b} \) \\
\hline
\( \mu^0 \lambda^0 \) & \( 1 \) \\
\( \mu^2 \lambda^0 \) & 
\( (\tilde{\beta}_{B,z} + \tilde{\beta}_{F,z}) - C_{\mathrm{vis}}(\alpha_B + \alpha_F - 2\tilde{E}_{P,z}) \tilde{\beta}_{B,z} \tilde{\beta}_{F,z} \) \\
\( \mu^4 \lambda^0 \) & \(  \tilde{\beta}_{B,z} \tilde{\beta}_{F,z} \) \\
\( \mu^1 \lambda^{1} \) & 
\( i \left[ \alpha_F \tilde{\beta}_{F,z} - \alpha_B \tilde{\beta}_{B,z} - \tilde{E}_{P,z} (\tilde{\beta}_{F,z} - \tilde{\beta}_{B,z}) \right] \) \\
\( \mu^3 \lambda^{1} \) & 
\( i  \tilde{\beta}_{B,z} \tilde{\beta}_{F,z} (\alpha_F - \alpha_B) \) \\
\( \mu^1 \lambda^{-1} \) & 
\begin{tabular}[t]{@{}c@{}}
\( i C_{\mathrm{vis}} (\tilde{\beta}_{B,z} - \tilde{\beta}_{F,z}) \) \\
\(+ i \left[ \alpha_F \tilde{\beta}_{F,v} - \alpha_B \tilde{\beta}_{B,v} - \tilde{E}_{P,z} (\tilde{\beta}_{F,v} - \tilde{\beta}_{B,v}) - \tilde{E}_{P,v} (\tilde{\beta}_{F,z} - \tilde{\beta}_{B,z}) \right] \)
\end{tabular} \\
\( \mu^2 \lambda^{-2} \) & 
\( \tilde{\beta}_{B,v} + \tilde{\beta}_{F,v} \) \\
\( \mu^4 \lambda^{-2} \) & 
\( \tilde{\beta}_{B,z} \tilde{\beta}_{F,v} + \tilde{\beta}_{F,z} \tilde{\beta}_{B,v} \) \\
\( \mu^3 \lambda^{-1} \) & 
\( i (\tilde{\beta}_{B,z} \tilde{\beta}_{F,v} + \tilde{\beta}_{F,z} \tilde{\beta}_{B,v}) (\alpha_F - \alpha_B) \) \\
\hline
\end{tabular}
\caption{Coefficients \( \mathcal{C}^{(BF)}_{a,b} \) in the expansion of \( P_{\Delta_B\Delta_F} \) in powers of \( \mu^a \lambda^b \).}
\label{tab:C_P_BF}
\end{table}

\section{Full expansion of the power spectra and parameters degeneracies}\label{app:degeneracy}

The expansions of the power spectra presented in Eqs.~\eqref{eq:co_1}--\eqref{eq:co_3} do not yet reflect the full structure of the expressions, as there are additional scale-dependent contributions encoded within the prefactors \( S_g^2 \) and \( B \). These terms must be expanded consistently to extract all contributions in powers of \( \mu^a \lambda^b \). The expansions of $S_g^2$ and $B$ are given by:
\begin{align}
    S_g^2 &= e^{-\frac{1}{2}k^2\mu^2\sigma_g^2}\approx 1 -\frac{1}{2}k^2\mu^2\sigma_g^2 = 1 -\frac{1}{2}\lambda^{-2}\mu^2\mathcal{H}^2\sigma_g^2 \,   , \label{eq:expSg}  \\
    B &= B_z + \epsilon \lambda^{-2}B_v \,    \label{eq:expB} .
\end{align}
Plugging Eqs.~\eqref{eq:expSg} and \eqref{eq:expB} into the auto- and cross-power spectra given in Eqs.~\eqref{eq:P_BB}--\eqref{eq:P_BF}, the resulting expressions can be grouped according to powers of \( \mu^a \lambda^b \) as
\begin{align}
P_{\Delta_B \Delta_B} &= \frac{\tilde{\beta}_{F,z}}{\tilde{\beta}_{B,z}} P \sum_{a,b} \mu^a \lambda^{b} \mathcal{K}^{(BB)}_{a,b} \,  , \label{eq:fullco_1} \\
P_{\Delta_F \Delta_F} &= \frac{\tilde{\beta}_{B,z}}{\tilde{\beta}_{F,z}}P \sum_{a,b} \mu^a \lambda^{b} \mathcal{K}^{(FF)}_{a,b} \,  , \label{eq:fullco_2} \\
P_{\Delta_B \Delta_F} &= P  \sum_{a,b} \mu^a \lambda^{b} \mathcal{K}^{(BF)}_{a,b} \,  .  \label{eq:fullco_3}
\end{align}

The coefficients \( \mathcal{K}^{(BB)}_{a,b} \) and \( \mathcal{K}^{(BF)}_{a,b} \), corresponding to the auto- and cross-power spectra \( P_{\Delta_B \Delta_B} \) and \( P_{\Delta_B \Delta_F} \), are listed in Tables~\ref{tab:K_P_BB} and \ref{tab:K_P_BF}, respectively. From these tables, we observe that, at first order in the expansion of \( S_g^2 \), the parameter \( \sigma_g \) is fully degenerate with \( \tilde{\beta}_{v} \) and $B_v$, and thus the \( \sigma_g \) encoded in the exponential of \( S_g \) is strongly degenerate with \( b_{g} \). Therefore, in our analysis, we fix the \( b_{g} \) encoded in $\tilde{\beta}_{v}$ to its fiducials. For a similar reason, we also fix $\mathcal{I}$ (as encoded in $B_v, \tilde{\beta}_{B,v}, \tilde{\beta}_{F,v}$, see Appendix \ref{app:app_growth}) to its fiducial values.

\begin{table}
\centering
\renewcommand{\arraystretch}{1.4}
\begin{tabular}{c|c}
\hline
Power \( \mu^a \lambda^b \) & Coefficient \( \mathcal{K}^{(BB)}_{a,b} \) \\
\hline
\( \mu^0 \lambda^0 \) & \( B_z \) \\
\( \mu^2 \lambda^0 \) & 
\( B_z \left( 2 \tilde{\beta}_{B,z} - 2 C_{\rm vis}(\alpha_B - \tilde{E}_{P,z}) \tilde{\beta}_{B,z}^2 \right) \) \\
\( \mu^4 \lambda^0 \) & 
\( B_z \tilde{\beta}_{B,z}^2 \) \\
\( \mu^2 \lambda^{-2} \) & 
\(  B_v + 2 \tilde{\beta}_{B,v} B_z - \tfrac{1}{2} \mathcal{H}^2 \sigma_g^2 B_z 
- 2  C_{\rm vis} (\alpha_B - \tilde{E}_{P,z}) \tilde{\beta}_{B,z}^2 B_v \) \\
\( \mu^4 \lambda^{-2} \) & 
\( 2 \tilde{\beta}_{B,z} \tilde{\beta}_{B,v} B_z 
+  \tilde{\beta}_{B,z}^2 B_v 
- \tilde{\beta}_{B,z} \mathcal{H}^2 \sigma_g^2 B_z \) \\
\hline
\end{tabular}
\caption{Coefficients \( \mathcal{K}^{(BB)}_{a,b} \) in the full expansion of 
\( P_{\Delta_B\Delta_B} \) in powers of \( \mu^a \lambda^b \), evaluated at the same orders as listed in Table.~\ref{tab:C_P_BB}.
}
\label{tab:K_P_BB}
\end{table}

\begin{table}
\centering
\renewcommand{\arraystretch}{1.4}
\begin{tabular}{c|c}
\hline
Power \( \mu^a \lambda^b \) & Coefficient \( \mathcal{K}^{(BF)}_{a,b} \) \\
\hline
\( \mu^0 \lambda^0 \) & \( B_z \) \\
\( \mu^2 \lambda^0 \) & 
\begin{tabular}[t]{@{}c@{}}
\( B_z(\tilde{\beta}_{B,z} + \tilde{\beta}_{F,z}) - B_z C_{\mathrm{vis}}(\alpha_B + \alpha_F - 2\tilde{E}_{P,z}) \tilde{\beta}_{B,z} \tilde{\beta}_{F,z} \)
\end{tabular} \\
\( \mu^4 \lambda^0 \) & 
\( B_z \tilde{\beta}_{B,z} \tilde{\beta}_{F,z} \) \\
\( \mu^1 \lambda^{1} \) & 
\( i B_z \left[ \alpha_F \tilde{\beta}_{F,z} - \alpha_B \tilde{\beta}_{B,z} - \tilde{E}_{P,z} (\tilde{\beta}_{F,z} - \tilde{\beta}_{B,z}) \right] \) \\
\( \mu^3 \lambda^{1} \) & 
\( i B_z \tilde{\beta}_{B,z} \tilde{\beta}_{F,z} (\alpha_F - \alpha_B) \) \\
\( \mu^1 \lambda^{-1} \) & 
\begin{tabular}[t]{@{}c@{}}
\( i B_z \left[C_{\mathrm{vis}}(\tilde{\beta}_{B,z} - \tilde{\beta}_{F,z}) + \alpha_F \tilde{\beta}_{F,v} - \alpha_B \tilde{\beta}_{B,v} \right. \) 
\( \left. - \tilde{E}_{P,z}(\tilde{\beta}_{F,v} - \tilde{\beta}_{B,v}) - \tilde{E}_{P,v}(\tilde{\beta}_{F,z} - \tilde{\beta}_{B,z}) \right] \)\\
\(+ i B_z B_v\left[ \alpha_F \tilde{\beta}_{F,z} - \alpha_B \tilde{\beta}_{B,z} - \tilde{E}_{P,z} (\tilde{\beta}_{F,z} - \tilde{\beta}_{B,z}) \right] \)
\end{tabular} \\
\( \mu^2 \lambda^{-2} \) & 
\begin{tabular}[t]{@{}c@{}}
\( B_z (\tilde{\beta}_{B,v} + \tilde{\beta}_{F,v}) - \frac{1}{2} \mathcal{H}^2 \sigma_g^2 B_z \) 
\(+  B_v \left[\tilde{\beta}_{B,z} + \tilde{\beta}_{F,z} - C_{\mathrm{vis}}(\alpha_B + \alpha_F - 2\tilde{E}_{P,z}) \tilde{\beta}_{B,z} \tilde{\beta}_{F,z}\right] \)
\end{tabular} \\
\( \mu^4 \lambda^{-2} \) & 
\begin{tabular}[t]{@{}c@{}}
\( B_z (\tilde{\beta}_{B,z} \tilde{\beta}_{F,v} + \tilde{\beta}_{F,z} \tilde{\beta}_{B,v}) - \frac{1}{2} \mathcal{H}^2 \sigma_g^2 B_z (\tilde{\beta}_{B,z} + \tilde{\beta}_{F,z}) \) 
\(+ B_v \tilde{\beta}_{B,z} \tilde{\beta}_{F,z} \)
\end{tabular} \\
\( \mu^3 \lambda^{-1} \) & 
\begin{tabular}[t]{@{}c@{}}
\( i (\alpha_F - \alpha_B)(\tilde{\beta}_{B,z} \tilde{\beta}_{F,v} + \tilde{\beta}_{F,z} \tilde{\beta}_{B,v})B_z \) \\
\(- \frac{1}{2} i \mathcal{H}^2 \sigma_g^2 B_z \left[ \alpha_F \tilde{\beta}_{F,v} - \alpha_B \tilde{\beta}_{B,v} \right. \) 
\( \left. - \tilde{E}_{P,z}(\tilde{\beta}_{F,v} - \tilde{\beta}_{B,v}) - \tilde{E}_{P,v}(\tilde{\beta}_{F,z} - \tilde{\beta}_{B,z}) \right] \)
\end{tabular} \\
\hline
\end{tabular}
\caption{Coefficients \( \mathcal{K}^{(BF)}_{a,b} \) in the full expansion of 
\( P_{\Delta_B\Delta_F} \) in powers of \( \mu^a \lambda^b \), evaluated at the same orders as listed in Table~\ref{tab:C_P_BF}.}
\label{tab:K_P_BF}
\end{table}

\section{Viscous dark matter and baryons}
\label{app:baryons}
Up to now, the only matter species considered was viscous dark matter. Here, we give the relevant equations for the case in which baryons are included as well. For the baryonic component, a pressureless perfect fluid (i.e.\ no viscosity) is assumed and that the baryons are subject to standard gravity, i.e.\ $\mu_{G(\mathrm{b})}=\eta_{(b)}=1$ and $\Theta_{(b)}=\Gamma_{\!(b)}=0$. The subscripts $b$ and $dm$ are used to denote baryonic and viscous dark matter quantities, respectively. The gravitational potentials are now sourced by both components and read to lowest order in $\lambda$:
\begin{align}
    \Psi&=-\frac{3}{2}\lambda^2\Bigl[\Omega_{\rm dm}(\mu_{\rm G}\delta_{\rm dm}+\mathcal{E}\theta_{\rm dm})+\Omega_{\rm b}\delta_{\rm b}\Bigr] \label{eq:Psi_vb} \\
    \Phi&=-\frac{3}{2}\lambda^2\bigl(\eta\mu_{\rm G}\Omega_{\rm dm}\delta_{\rm dm}+\Omega_{\rm b}\delta_{\rm b}\bigr) \\
    \Phi'&=\frac{3}{2}\eta\lambda^2\Omega_{\rm dm}\biggl[\Bigl(1-\eta'/\eta-\mu_{\rm G}'/\mu_{\rm G}\Bigr)\mu_{\rm G}\delta_{\rm dm}+\Bigl(\mathcal{Z}+\mathcal{E}+(1-2\mathcal{Z})\mu_{\rm G}\Bigr)\theta_{\rm dm}\biggr]+\frac{3}{2}\lambda^2\Omega_{\rm b}\Bigl[\delta_{\rm b}+\theta_{\rm b}\Bigr] ~.
\end{align}
Note that only the dark matter terms contain the modified gravity parameters. The dark matter perturbed continuity equation remains the same as in \eqref{eq:pert_continuity_2}, $\delta_{\rm dm}'=-3\mathcal{Z}\delta_{\rm dm}-(1-2\mathcal{Z})\theta_{\rm dm}\equiv f_{\rm dm}\delta_{\rm dm}$, while the one for baryons is simply $\delta_{\rm b}'=-\theta_{\rm b}\equiv f\delta_{\rm b}$. The perturbed Euler equation of dark matter \eqref{eq:pert_Euler_EP_3} acquires an additional term:
\begin{equation}
    \theta_{\rm dm}'= -\left[1+\frac{\mathcal{H'}}{\mathcal{H}}+\frac{1}{3}\frac{\mathcal{Z}+\mathcal{E}}{1-\mathcal{Z}}\lambda^{-2}+\tilde{\Theta}\right]\theta_{\rm dm} -\frac{3}{2}\Omega_{\rm dm}(1+\tilde{\Gamma})\mu_{\rm G}\delta_{\rm dm} -\frac{3}{2}\frac{\Omega_{\rm b}}{1-\mathcal{Z}}\Bigl(1+\Gamma-\mathcal{Z}(1-f_{\rm b})\Bigr)\delta_{\rm b} ~.
\end{equation}
The baryonic counterpart reads
\begin{align}
    \theta_{\rm b}'&=-\left[1+\frac{\mathcal{H'}}{\mathcal{H}}\right]\theta_{\rm b}+\lambda^{-2}\Psi \nonumber\\
    &=-\left[1+\frac{\mathcal{H'}}{\mathcal{H}}\right]\theta_{\rm b} -\frac{3}{2}\Omega_{\rm b}\delta_{\rm b} -\frac{3}{2}\Omega_{\rm dm}\biggl(1-\frac{\mathcal{E}(f_{\rm dm}+3\mathcal{Z})}{(1-2\mathcal{Z})\mu_{\rm G}}\biggr)\mu_{\rm G}\delta_{\rm dm} ~.
\end{align}
The growth equations (cf. Appendix \ref{app:growth_eq}) become
\begin{align}
    &\delta_{\rm dm}''+\bigl(\mathcal{A}+\mathcal{B}\lambda^{-2}\bigr)\delta_{\rm dm}'-\bigl(\mathcal{C}+\mathcal{D}\lambda^{-2}\bigr)\delta_{\rm dm} - \frac{3}{2}\Omega_{\rm b}\frac{1-2\mathcal{Z}}{1-\mathcal{Z}}\Bigl(1+\Gamma-\mathcal{Z}(1-f_{\rm b})\Bigr)\delta_{\rm b}=0 \\
    &\delta_{\rm b}''+\left(1+\frac{\mathcal{H}'}{\mathcal{H}}\right)\delta_{\rm b}'-\frac{3}{2}\Omega_{\rm b}\delta_{\rm b}-\frac{3}{2}\Omega_{\rm dm}\left(1-\frac{\mathcal{E}(f_{\rm dm}+3\mathcal{Z)}}{(1-2\mathcal{Z})\mu_{\rm G}}\right)\mu_{\rm G}\delta_{\rm dm}=0 ~.
\end{align}
In the limit of very small viscosities, these simplify to
\begin{align}
    &\delta_{\rm dm}''+\left(1+\frac{\mathcal{H}'}{\mathcal{H}}+\Theta+C_{\mathrm{vis}}\lambda^{-2}\right)\delta_{\rm dm}'-\frac{3}{2}\Omega_{\rm dm}(1+\Gamma)\mu_{\rm G}\delta_{\rm dm} - \frac{3}{2}\Omega_{\rm b}(1+\Gamma)\delta_{\rm b}=0 \label{eq:delta_vis_dm}\\
    &\delta_{\rm b}''+\left(1+\frac{\mathcal{H}'}{\mathcal{H}}\right)\delta_{\rm b}'-\frac{3}{2}\Omega_{\rm b}\delta_{\rm b}-\frac{3}{2}\Omega_{\rm dm}\mu_{\rm G}\delta_{\rm dm}=0 ~, \label{eq:delta_baryon}
\end{align}
where the source terms that couple dark matter and baryon growth are the standard ones, including the modified gravity parameters. 

\begin{figure*}
\centering
\includegraphics[width=0.40\textwidth]{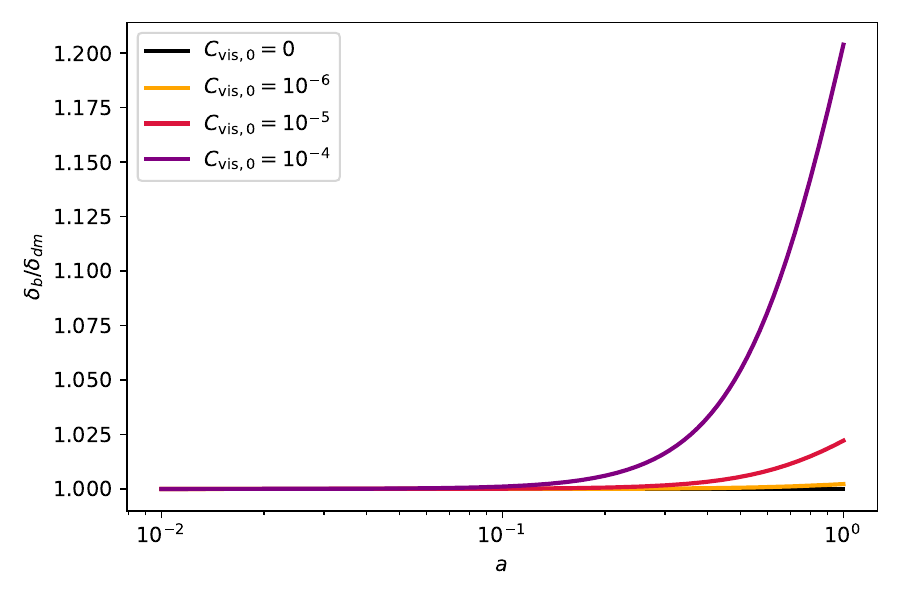}
\hfil
\includegraphics[width=0.40\textwidth]{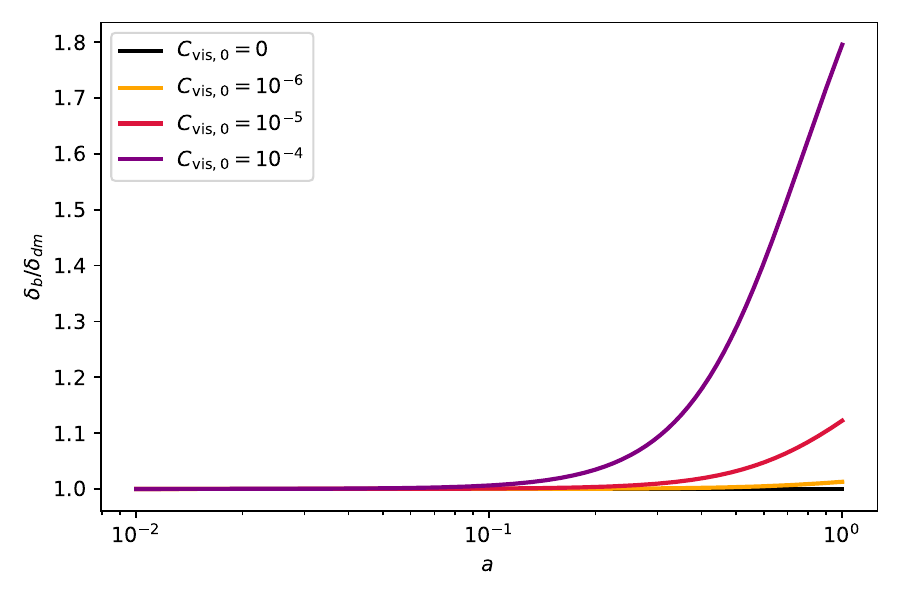}
\medskip \\
\caption{
Impact of the viscosity parameter \( C_{\rm vis,0}\) on the growth of the density contrast of viscous dark matter and baryons, numerically solved from Eqs.~\eqref{eq:delta_vis_dm} and \eqref{eq:delta_baryon}, evaluated at $k=0.05$ $h$/Mpc (left) and $k=0.12$ $h$/Mpc (right). We show the ratio of the density contrast \(\delta_{\rm b} / \delta_{\rm dm}\). As expected, the growth of viscous dark matter is suppressed relative to that of baryons. For illustrative purpose, we consider only bulk viscosity and set $\Theta=\Gamma=0, \mu_{\rm G}=\eta =1$.}
\label{fig:baryons} 
\end{figure*}

The galaxy density contrast \eqref{eq:Delta_Fourier},
\begin{equation}
    \Delta(\textbf{k},\hat{\textbf{n}},z) = b_g\delta_m - \mu^2\theta_m - i\frac{\mu}{\lambda}\Psi + i\mu\lambda\theta'_m + i\mu\lambda\alpha\theta_m + i\mu\lambda\frac{\mathcal{H}'}{\mathcal{H}}\theta_m ~,
\end{equation}
contains the total quantities 
\begin{align}
    \delta_m&=\frac{1}{\Omega_{\rm dm}+\Omega_{\rm b}}(\Omega_{\rm dm}\delta_{\rm dm}+\Omega_{\rm b}\delta_{\rm b}) \\
    %\theta_m&=\frac{1}{1+w_{\mathrm{eff}}}(\Omega_{\rm dm}\theta_{\rm dm}+\Omega_{\rm b}\theta_{\rm b})=\frac{1}{\Omega_{\rm dm}+\Omega_{\rm b}}
    \theta_m&=\frac{1}{\Omega_{\rm dm}+\Omega_{\rm b}}(\Omega_{\rm dm}\theta_{\rm dm}+\Omega_{\rm b}\theta_{\rm b}) \\
    \theta_m'&=\frac{1}{\Omega_{\rm dm}+\Omega_{\rm b}}(\Omega_{\rm dm}\theta_{\rm dm}'+\Omega_{\rm b}\theta_{\rm b}')
\end{align}
where we used $\Omega_{{\rm dm}/\rm b}'=-\Omega_{{\rm dm}/\rm b}(1+2\frac{\mathcal{H}'}{\mathcal{H}})$, assuming the continuity equation \eqref{eq:continuity} without a coupling term (cf. Appendix \ref{app:AppendixCQ}) and neglecting the viscous contribution on the background level. 
This yields
\begin{align}
\Delta(\mathbf{k}, \hat{\mathbf{n}}, z) &= \delta_{\rm dm} \frac{\Omega_{\rm dm}}{\Omega_{\rm dm}+\Omega_{\rm b}} b_g\bigg\{ 
1 + \mu^2  \tilde{\beta}_{\rm dm}
- i \mu \lambda 
\left(\alpha - \tilde{E}_{P,{\rm dm}} - C_{\rm vis}\lambda^{-2} \right) \tilde{\beta}_{\rm dm}
\bigg\} \nonumber\\
&+\delta_{\rm b} \frac{\Omega_{\rm b}}{\Omega_{\rm dm}+\Omega_{\rm b}} b_g\bigg\{ 
1 + \mu^2  \tilde{\beta}_{\rm b}
- i \mu \lambda 
\left(\alpha - \tilde{E}_{P,\rm b}\right) \tilde{\beta}_{\rm b}
\bigg\}
\end{align}
where 
\begin{align}
    \tilde{\beta}_{\rm dm} &= \frac{f_{\rm dm} + 3\mathcal{Z}}{b_g(1 - 2\mathcal{Z})} \, ,\qquad \tilde{\beta}_{\rm b} = \frac{f_{\rm b} }{b_g} \\
    \tilde{E}_{P,{\rm dm}} &= 1 + \tilde{\Theta}  - \frac{3}{2} \Omega_{\rm dm} \mathcal{E} - \frac{3}{2} \frac{\Omega_{\rm dm} \mu_{\rm G} \tilde{\Gamma}}{f_{\rm dm}+3\mathcal{Z}}(1-2\mathcal{Z})\, , \qquad \tilde{E}_{P,\rm b}=1-\frac{3}{2}\Omega_{\rm dm}\mathcal{Z}-\frac{3}{2}\frac{\Omega_{\rm dm}\Gamma}{f_{\rm b}(1-\mathcal{Z})}
\end{align} 
In Fig. \ref{fig:baryons} it is illustrated that for small enough viscosities ($\tilde{\zeta}\lesssim10^{-5}$), we can set $\delta_{\rm b}\simeq\delta_{\rm dm}$. Under this approximation, and in addition $\tilde\beta_{\rm b}\simeq\tilde\beta_{\rm dm}\simeq\beta_m=\frac{f_m}{b_g}$, we obtain
\begin{equation}
\Delta(\mathbf{k},\hat{\mathbf{n}},z)=\delta_{m}b_{g}\bigg\{1+\mu^{2}\tilde{\beta}_{m}-i\mu\lambda\left(\alpha-\tilde{E}_{P,\mathrm{eff}}-C_{{\rm vis, eff}}\lambda^{-2}\right)\tilde{\beta}_{m}\bigg\}
\end{equation}
where
\begin{align}
    C_{{\rm vis, eff}}&=\frac{\Omega_{\rm dm}}{\Omega_{\rm dm}+\Omega_{\rm b}} C_{{\rm vis}}  \label{eq:C_vis_eff}\\
    \tilde{E}_{P,\mathrm{eff}}&=\frac{1}{\Omega_{\rm dm}+\Omega_{\rm b}}\left(\Omega_{\rm dm}\tilde{E}_{P,{\rm dm}}+\Omega_{\rm b}\tilde{E}_{P,{\rm b}}\right) \nonumber\\
    &=1+\frac{\Omega_{\rm dm}}{\Omega_{\rm dm}+\Omega_{\rm b}}\tilde{\Theta}-\frac{\Omega_{\rm dm}}{\Omega_{\rm dm}+\Omega_{\rm b}}\left(\frac{3}{2}\Omega_{\rm dm}\mathcal{E}+\frac{3}{2}\frac{\Omega_{\rm dm}\mu_{\rm G}\tilde{\Gamma}}{f_m+3\mathcal{Z}}(1-2\mathcal{Z})\right)-\frac{\Omega_{\rm b}}{\Omega_{\rm dm}+\Omega_{\rm b}}\left(\frac{3}{2}\Omega_{\rm dm}\mathcal{Z}+\frac{3}{2}\frac{\Omega_{\rm dm}\Gamma}{f_m(1-\mathcal{Z})}\right) ~.
\end{align}
When neglecting the terms containing $\mathcal{Z}$ and $\mathcal{E}$ directly, the last equation becomes
\begin{equation}
    \tilde{E}_{P,\mathrm{eff}}\simeq 1+\frac{\Omega_{\rm dm}}{\Omega_{\rm dm}+\Omega_{\rm b}}\Theta -\frac{3}{2}\frac{\Omega_{\rm dm}\Gamma}{f_m}~\frac{\Omega_{\rm dm}\mu_{\rm G}+\Omega_{\rm b}}{\Omega_{\rm dm}+\Omega_{\rm b}} ~.
\end{equation}
The remaining (and dominant) effect of viscosity is its contribution to the growth rate $f_m$, cf. Appendix \ref{app:growth_eq} and \ref{app:app_growth}.

We conclude that, to take baryons into account, $\tilde E_P$ and $C_{\rm vis}$ have to be redefined accordingly. The quantities that are actually measured are then $\tilde{E}_{P,\mathrm{eff}}$ and $C_{{\rm vis, eff}}$. Nevertheless, baryons are anticipated to have a negligible effect on the constraints on $C_{\rm vis, 0}$. To verify this, we fixed the ratio $\frac{\Omega_{\rm dm}}{\Omega_{\rm dm}+\Omega_{\rm b}}$ to its $\Lambda$CDM value and reran the analysis. We find that the $C_{\rm vis, 0}$ constraints, as reported in Tables \ref{tab:DESI32}-\ref{tab:SKA32_3}, remain unchanged.

It is then explicit that, in the limit of very small viscosities, the parameter $\tilde{E}_{P,\mathrm{eff}}$ can still be used as an EP test, even in the presence of baryons.

\section{Viscous dark matter and coupled quintessence}\label{app:AppendixCQ}
Here we consider the coupled quintessence model where the conservation equations of dark matter and dark energy are of the form 
\begin{equation}
    \nabla_{\!\!\mu}T^{\mu}_{~~\nu}=Q(\phi) \ T \ \nabla_{\!\!\nu}\phi ~, \qquad \nabla_{\!\!\mu}T^{\mu}_{\!(\phi)\nu}=-Q(\phi) \ T \ \nabla_{\!\!\nu}\phi ~ ,
\end{equation}
where $\phi$ is a canonical scalar field with potential $V$ and energy-momentum tensor $T^{\mu}_{\!(\phi)\nu}$. The quantity $T=T^{\mu}_{~~\mu}$ denotes the trace of the viscous matter energy-momentum tensor defined in Eq. \eqref{eq:EMT}. Assuming $w=c_s^2=0$ for the viscous dark matter and neglecting other matter species, the background evolution equations are given by (defining $\hat{V}=\frac{a^2V}{\mathcal{H}^2}$):
\begin{align}
    1 &=\Omega_m+\frac{1}{3}\left(\frac{1}{2}\phi'^2+\hat{V}\right)\\
    2\frac{\mathcal{H'}}{\mathcal{H}}&=-1+3\Omega_m\mathcal{Z}-\frac{1}{2}\phi'^2+\hat{V} 
\\
    \rho_m'+3(1-\mathcal{Z})\rho_m &=(1+3\mathcal{Z})Q\rho_m\phi' \\
    \phi''+\left(2+\frac{\mathcal{H'}}{\mathcal{H}}\right)\phi'+\hat{V}_{\!,\phi}&=-3\Omega_m(1+3\mathcal{Z})Q \, .
\end{align}
At the perturbation level, one can derive the perturbed Einstein equations and the perturbed conservation equations (with $\varphi\equiv\delta\phi$) 
\begin{align}
    \Phi&=-\frac{\tfrac{1}{2}\lambda^2}{1-\tfrac{1}{2}\lambda^2\phi'^2}\biggl[\phi'(3\varphi+\varphi')+\hat{V}_{\!,\phi}\varphi+3\Omega_m\Bigl(\delta_m+3\lambda^2\bigl(1-\mathcal{Z}+\tfrac{1}{6}\mathcal{E}\phi'^2\bigr)\theta_m\Bigr)\biggr] \\
    \Phi'&=\frac{1}{2}\phi'\varphi+\frac{\tfrac{1}{2}\lambda^2}{1-\tfrac{1}{2}\lambda^2\phi'^2}\biggl[\phi'(3\varphi+\varphi')+\hat{V}_{\!,\phi}\varphi+3\Omega_m\Bigl(\delta_m+(1-\mathcal{Z}+\mathcal{E})\theta_m+3\lambda^2(1-\mathcal{Z})\bigl(1-\tfrac{1}{6}\phi'^2\bigr)\theta_m\Bigr)\biggr] \\
    \Psi&=\Phi-\frac{3}{2}\Omega_m\mathcal{E}\lambda^2\theta_m \qquad\qquad \Psi'=\Phi'-\frac{3}{2}\Omega_m\mathcal{E}\lambda^2\Bigl(\Bigl(1+\frac{\mathcal{H}'}{\mathcal{H}}\Bigr)\theta_m+\theta_m'\Bigr) \\
    \delta_m'&=-\bigl(1-\mathcal{Z}(2+Q\phi')\bigr)\bigl(\theta_m-3\Phi'\bigr)-3\mathcal{Z}\bigl(1+Q\phi'\bigr)\bigl(\delta_m+\Psi\bigr)+(1+3\mathcal{Z})\bigl(Q_{\!,\phi}\phi'\varphi+Q\varphi'\bigr) \\
    (1-\mathcal{Z})\theta_m'&=-\left[1+(1-2\mathcal{Z})\frac{\mathcal{H}'}{\mathcal{H}}+(1+3\mathcal{Z})Q\phi'+\frac{1}{3}(\mathcal{Z}+\mathcal{E})\lambda^{-2}\right]\theta_m
    +\lambda^{-2}\biggl[\Psi+\mathcal{Z}\Phi'+(1+3\mathcal{Z})Q\varphi\biggr] \\
    \varphi''&+\left(2+\frac{\mathcal{H'}}{\mathcal{H}}\right)\varphi'+\left(\lambda^{-2}+\hat{V}_{\!\!,\phi\phi}\right)\varphi-(3\Phi'+\Psi')\phi'+2\hat{V}_{\!\!,\phi}\Psi \nonumber\\
    &\qquad = -3\Omega_m\Bigl[Q\delta_m+(1+3\mathcal{Z})Q_{\!,\phi}\varphi+(2+3\mathcal{Z})Q\Psi+\mathcal{Z}Q(\theta_m-3\Phi')\Bigr]
\end{align}
where $\hat{V}_{\!\!,\phi}\equiv d\hat{V}/d\phi, ~ Q_{\!,\phi}\equiv dQ/d\phi$. Replacing $\Phi,\Psi,\Phi',\Psi'$ and taking the sub-horizon limit $\lambda\ll 1$ yields 
\begin{align}
    \varphi&=-3\Omega_m\lambda^2\Bigl[Q(\delta_m+\mathcal{Z}\theta_m)-\frac{1}{6}\mathcal{E}\frac{\mathcal{Z}+\mathcal{E}}{1-\mathcal{Z}}\phi'\theta_m\Bigr] \\
    \delta_m'&=-\bigl(1-\mathcal{Z}(2+Q\phi')\bigr)\theta_m-3\mathcal{Z}(1+Q\phi')\delta_m \label{eq:continuity_CQ}\\
    \theta_m'&=-\Biggl[\frac{1}{1-\mathcal{Z}}+\frac{1-2\mathcal{Z}}{1-\mathcal{Z}}\frac{\mathcal{H'}}{\mathcal{H}}+\frac{1+3\mathcal{Z}}{1-\mathcal{Z}}Q\phi'+\frac{1}{3}\frac{\mathcal{Z}+\mathcal{E}}{1-\mathcal{Z}}\lambda^{-2} \nonumber\\
    &\qquad +\frac{3}{2}\Omega_m\Biggl(\mathcal{E}-\mathcal{Z}+\frac{\mathcal{Z}}{1-\mathcal{Z}}\Bigl((1+3\mathcal{Z})2Q^2+\mathcal{Z}Q\phi'\Bigr)-\frac{1}{3}\frac{(\mathcal{Z}+\mathcal{E})\mathcal{E}}{(1-\mathcal{Z})^2}\Bigl((1+3\mathcal{Z})Q\phi'+\frac{1}{2}\mathcal{Z}\phi'^2\Bigr)\Biggr)\Biggr]\theta_m \nonumber \\
    &\qquad -\frac{3}{2} \Omega_m\biggl[1+\frac{1+3\mathcal{Z}}{1-\mathcal{Z}}2Q^2+\frac{\mathcal{Z}}{1-\mathcal{Z}}Q\phi'\biggr]\delta_m
\end{align}
In terms of the notation introduced in Eqs. \eqref{eq:pert_Euler_EP_3}-\eqref{eq:Gamma_tilde}, we find
\begin{align}
    \eta&=\mu_{\rm G}=1\\
    \tilde{\Gamma}&=\frac{1+3\mathcal{Z}}{1-\mathcal{Z}}2Q^2+\frac{\mathcal{Z}}{1-\mathcal{Z}}Q\phi' ~~ \simeq ~(1+4\mathcal{Z})2Q^2+\mathcal{Z}Q\phi'\\
    \tilde{\Theta}&=\frac{\mathcal{Z}}{1-\mathcal{Z}}\left(1-\frac{\mathcal{H'}}{\mathcal{H}}\right)+\frac{1+3\mathcal{Z}}{1-\mathcal{Z}}Q\phi'\nonumber\\
    &\quad+\frac{3}{2}\Omega_m\Biggl(\mathcal{E}-\mathcal{Z}+\frac{\mathcal{Z}}{1-\mathcal{Z}}\Bigl((1+3\mathcal{Z})2Q^2+\mathcal{Z}Q\phi'\Bigr)-\frac{\mathcal{E}}{3}\frac{\mathcal{Z}+\mathcal{E}}{(1-\mathcal{Z})^2}\Bigl((1+3\mathcal{Z})Q\phi'+\frac{1}{2}\mathcal{Z}\phi'^2\Bigr)\Biggr) \\
    &\simeq \mathcal{Z}\left(1-\frac{\mathcal{H'}}{\mathcal{H}}\right)+(1+4\mathcal{Z})Q\phi'+\frac{3}{2}\Omega_m\bigl(\mathcal{E}-\mathcal{Z}(1-2Q^2)\bigr)
\end{align}
or writing in terms of $\Gamma,\Theta$ and $\tilde{\eta},\tilde{\mu}_{\rm G}$ (as defined in Appendix \ref{app:summary}),
\begin{align}
    \tilde{\eta}&=\frac{1}{1-\mathcal{E}\frac{f+3\mathcal{Z}}{1-2\mathcal{Z}}}~\simeq~1+\mathcal{E}f\\
    \tilde{\mu}_{\rm G}&=1-\mathcal{E}\frac{f+3\mathcal{Z}}{1-2\mathcal{Z}}~\simeq~1-\mathcal{E}f\\
    \Gamma &=  (1+3\mathcal{Z})2Q^2 + \mathcal{Z}Q\phi'\\
    \Theta &= (1+3\mathcal{Z})Q\phi'+\frac{3}{2}\Omega_m\left[(\mathcal{Z}-\mathcal{E})\Bigl((1+3\mathcal{Z})2Q^2+\mathcal{Z}Q\phi'\Bigr)-\frac{\mathcal{E}}{3}\frac{\mathcal{Z}+\mathcal{E}}{1-\mathcal{Z}}\Bigl((1+3\mathcal{Z})Q\phi'+\frac{1}{2}\mathcal{Z}\phi'^2\Bigr)\right] \\
    &\simeq (1+3\mathcal{Z})Q\phi'+3\Omega_m(\mathcal{Z}-\mathcal{E})Q^2 ~.
\end{align}
We also provided approximate expressions valid in the case $\mathcal{Z},\mathcal{E}\ll 1$, where terms of order $\mathcal{O}(\mathcal{Z}^2,\mathcal{E}^2,\mathcal{Z}\mathcal{E})$ were neglected. It is interesting to note that even if there is no explicit coupling ($Q=0$), there is a contribution to $\tilde{\Theta}$ respectively to $\Theta$, proportional to $\phi'^2$ and of cubic order in viscosities. 

In the case where it is possible to neglect the ``mixed terms'' between viscosity and quintessence, i.e.\ $\mathcal{Z}Q^2$, $\mathcal{Z}Q\phi'$ and $\mathcal{Z}\phi'^2$ (and analogously with $\mathcal{E}$ as prefactor), one is left only with the standard modifications of coupled quintessence:
\begin{equation}
    \Gamma=2Q^2 ~,\qquad \Theta=Q\phi' ~.
\end{equation}
The expressions for $\Theta$ and $\Gamma$ could be inserted in our analysis above if the modification of gravity is solely due to coupled quintessence. To derive a growth equation, one has to start from the continuity equation \eqref{eq:continuity_CQ} and use $\mathcal{Z}'=\mathcal{Z}\left(2+\frac{\mathcal{H}'}{\mathcal{H}}-3\mathcal{Z}-(1+3\mathcal{Z})Q\phi'\right)$, which can however be approximated as $\mathcal{Z}'\simeq \mathcal{Z}\left(2+\frac{\mathcal{H}'}{\mathcal{H}}-3\mathcal{Z}\right)$.

\section{Observational constraints for viscosity}
\label{app:ObsConstraints}
In table \ref{tab:constraints} we present a selection of observational constraints on dark matter viscosity from the literature. To obtain the constraints, all of these papers assume a $\Lambda$CDM cosmology, where the dark matter fluid has bulk and/or shear viscosity, described in the Eckart framework ($\Lambda$vCDM model). The cosmological parameters have been fixed, except in \cite{Ashoorioon:2023jwf}, \cite{AparicioResco:2020shd}, and \cite{Anand:2017wsj}, where some parameters were left to vary in the statistical analysis. The constraints cited here have been obtained assuming constant viscosity, although some authors also investigate different parameterizations of time dependent viscosity, typically power laws in the dark matter density. In many cases, only upper bounds are given, because the result was compatible with zero viscosity. The numbers are presented in the form of dimensionless values, defined as $\tilde{\zeta}\coloneqq\frac{3\zeta}{H_0}\equiv\frac{24\pi G\zeta}{c^2 H_0}$ and analogously for $\tilde{\eta}_s$. The relation to our (time dependent) parameters $\mathcal{Z}$ and $\mathcal{E}$ is: $\mathcal{Z}=\frac{\tilde{\zeta}}{3\Omega_m H/H_0}$, $\mathcal{E}=\frac{4\tilde{\eta}_s}{9\Omega_m H/H_0}$. The present-day values $\mathcal{Z}_0=\frac{\tilde{\zeta}}{3\Omega_{m0}}$, $\mathcal{E}_0=\frac{4\tilde{\eta}_s}{9\Omega_{m0}}$ are of the same order of magnitude as $\tilde{\zeta}$ and $\tilde{\eta}_s$. Moreover, $C_{\mathrm{vis}}=\frac{1}{3}\frac{\tilde{\zeta}+\frac{4}{3}\tilde{\eta}_s}{3\Omega_m H/H_0 -\tilde{\zeta}} \simeq\frac{\tilde{\zeta}+\frac{4}{3}\tilde{\eta}_s}{9\Omega_m H/H_0}$ and $C_{\mathrm{vis},0}=\frac{1}{3}\frac{\tilde{\zeta}+\frac{4}{3}\tilde{\eta}_s}{3\Omega_{m0} -\tilde{\zeta}}\simeq\frac{\tilde{\zeta}+\frac{4}{3}\tilde{\eta}_s}{9\Omega_{m0}}$. Using the latter formula, the numbers in the second column are translated into constraints on $C_{\mathrm{vis},0}$ in the third column, using $\Omega_{m,0}=0.315\pm0.007$ \cite{Planck:2018vyg}. In cases where only one type of viscosity is present, the other is assumed to be zero. Whenever asymmetric errors are given, the root mean square of positive and negative errors is used and then propagated in the standard way. 
An exception is case \cite{AparicioResco:2020shd}, where only the positive error was taken into account. 

\begin{table}
    \centering
    \renewcommand\arraystretch{1.3} % more vertical space
    \begin{tabular}{
        >{\centering\arraybackslash}p{1.0cm}
        >{\centering\arraybackslash}p{4.0cm}
        >{\centering\arraybackslash}p{4.0cm}
        >{\centering\arraybackslash}p{4.0cm}
        >{\centering\arraybackslash}p{3.0cm}
    }
    \hline
    Reference & Constraints on $\tilde{\zeta}$, $\tilde{\eta}_s$ & Constraints on $C_{\mathrm{vis},0}$ & Probes/Criterion  & Model \\ \hline \hline

    \cite{Normann:2016jns} 
    & $\tilde{\zeta}\lesssim2.6\times 10^{-2}$ 
    & $C_{\mathrm{vis},0}\lesssim0.9\times10^{-2}$
    & independent $H(z)$ data from Ref. \cite{Chen:2013vea}
    & $\zeta$ constant; $\zeta(z)$ 
    \\ \hline
    \multirow{4}{*}{\cite{Velten:2012uv}} 
    & $\tilde{\zeta}\leq 0.31$ ($2\,\sigma$ c.l.) 
    & $C_{\mathrm{vis},0}\leq0.11$
    & SN, BAO, CMB   
    & 
    \\ 
    & $\tilde{\zeta}\leq10^{-6}$ 
    & $C_{\mathrm{vis},0}\leq3.5\times10^{-7}$
    & $\sim$ 10 Mpc clusters 
    & $\zeta$ constant 
    \\ 
    & $\tilde{\zeta}\leq10^{-9}$ 
    & $C_{\mathrm{vis},0}\leq3.5\times10^{-10}$
    & $\sim$ 10 kpc galaxies 
    & 
    \\ 
    & $\tilde{\zeta}\leq10^{-11}$ 
    & $C_{\mathrm{vis},0}\leq3.5\times10^{-12}$
    & $\sim$ 1 kpc dwarf galaxies 
    & 
    \\ \hline

    \cite{Velten:2013pra} 
    & $\tilde{\zeta}\leq5\times10^{-11}$ 
    & $C_{\mathrm{vis},0}\leq1.8\times10^{-11}$
    & $\sim$ 1 kpc dwarf galaxies 
    & $\zeta$ constant; $\zeta(z)$; (neo-)Newtonian and relativistic approaches
    \\ \hline
    
    \cite{Velten:2014xca} 
    & $\tilde{\zeta}\approx10^{-8}$ 
    & $C_{\mathrm{vis},0}\approx10^{-9}$
    & linear power spectrum and cumulative mass function
    & $\zeta$ constant
    \\ \hline
    
    \cite{Barbosa:2017ojt} 
    & \makecell[c]{
    \rule{0pt}{2.5ex} 
    $\tilde{\zeta}\leq 1.427\times10^{-6}$ \\
    $\tilde{\eta}_s\leq 2.593\times10^{-6}$ }
    & $C_{\mathrm{vis},0}\leq1.7\times10^{-6}$
    & $f\sigma_8(z)$ data from RSD measurements
    & $\zeta$ and $\eta_s$ constant 
    \\ \hline

    \cite{Ashoorioon:2023jwf} 
    & $\tilde{\zeta}=0.26^{+0.05}_{-0.04}\times10^{-6}$
    & $C_{\mathrm{vis},0}=(0.92\pm0.16)\times10^{-7}$
    & CMB, BAO, SN, CC, LSS
    & $\zeta$ constant; $\zeta(z)$ 
    \\ \hline

    \cite{AparicioResco:2020shd} 
    & $\tilde{\eta}_s=0.261^{+6.875}_{-0.261}\times10^{-6}$ \newline
    ($2\,\sigma$ c.l.)
    & $C_{\mathrm{vis},0}=1.23^{+32.33}_{-1.23}\times10^{-7}$ \newline
    ($2\,\sigma$ c.l.)
    & Power spectrum of SDSS LRG 
    & $\eta_s$ constant; $\eta(z)$;  specific parametrization of growth function
    \\ \hline

    \cite{Anand:2017wsj} 
    & \makecell[c]{
        \rule{0pt}{2.5ex} 
        $\tilde{\zeta}=1.32^{+0.50}_{-1.00}\times10^{-6}$ \\
        $\tilde{\eta}_s=1.20^{+0.40}_{-1.00}\times10^{-6}$
    }
    & $C_{\mathrm{vis},0}=(1.03\pm0.45)\times10^{-6}$
    & CMB, SZ, lensing, BAO
    & $\zeta$ and $\eta_s$ constant
    \\ \hline

    \cite{Goswami:2016tsu} 
    & $\tilde{\eta}_s\leq59.2$
    & $C_{\mathrm{vis},0}\leq27.8$
    & dissipation of gravitational waves from GW150914
    & GW propagation through fluid with $\eta_s$ constant
    \\ \hline

    \end{tabular}
    \caption{Observational constraints on viscosity in terms of the dimensionless parameters $\tilde{\zeta}\coloneqq\frac{3\zeta}{H_0}\equiv\frac{24\pi G\zeta}{c^2 H_0}$ and $\tilde{\eta}_s$ analogously. For references \cite{Normann:2016jns} and \cite{Goswami:2016tsu}, we transformed their constraints into dimensionless form assuming a Hubble constant with $h=0.67$. The specified errors are $1\,\sigma$, unless stated otherwise.} 
    \label{tab:constraints}
\end{table}

It can be seen that constraints originating from the background dynamics, i.e.\ \cite{Normann:2016jns} and the first one of \cite{Velten:2012uv}, or from gravitational wave propagation (\cite{Goswami:2016tsu}), are rather weak. However, there are much stronger limits on viscosity obtained from studying the growth of matter perturbations. In \cite{Velten:2012uv} it can be seen that demanding the existence of smaller and smaller nonlinear structures leads to increasingly tighter constraints, because viscosity tends to suppress structure formation at very small scales. The strongest limits of $\tilde{\zeta}\sim10^{-8}-10^{-11}$ in \cite{Velten:2012uv}, \cite{Velten:2013pra}, \cite{Velten:2014xca} are obtained by referring to scales of (dwarf) galaxies, not cosmological scales. However, astrophysical aspects of galaxies were not taken into account, making it difficult to extract reliable constraints from these scales. Moreover, these values do not come from a statistical analysis using observational data and should only be considered rough estimates. Another aspect is, that a suppression of growth at galaxy scales can help solving the small scale issues of standard CDM, such as the missing satellites problem or the core-cusp problem, which is also pointed out in these three papers. We therefore do not take the very strong constraints too seriously, but rely on the remaining results \cite{Barbosa:2017ojt}, \cite{Ashoorioon:2023jwf}, \cite{AparicioResco:2020shd}, \cite{Anand:2017wsj}, based on large scale structure and statistical data analysis. The suggested values for the dimensionless viscosities, which can be interpreted as upper limits, are all of the order $10^{-6}$ to $10^{-7}$. It is interesting to compare this with the (shear) viscosity of water at room temperature, which is only about $\sim10^{-3}\,\rm Pa\,s$ or $\sim10^{-11}$ in dimensionless form.

\section{Magnification bias and galaxy luminosity function  of surveys} \label{sec:mag}

In this appendix, we summarize the computation of the magnification bias for the DESI BGS, Euclid, and SKA2 galaxy surveys, based on the galaxy luminosity functions (GLFs) provided in Ref.~\cite{Maartens:2021dqy}.

The magnification bias \( s \) quantifies the change in the observed galaxy number density induced by lensing magnification, which alters the luminosity threshold at a given redshift. As such, \( s \) depends on the specific survey characteristics. Following Ref.~\cite{Bonvin:2023jjq}, we evaluate the magnification bias separately for bright and faint galaxy populations, and outline the procedure below.

For galaxies detected above a flux threshold \( F_{\rm lim} \), the total magnification bias \( s_{\rm tot} \) is defined as
\begin{equation}
s_{\rm tot}(z, F_{\rm lim}) \equiv -\frac{2}{5} \frac{\partial \ln \bar{N}_{F+B}(z, F \geq F_{\rm lim})}{\partial \ln F_{\rm lim}} \,,
\label{eq:s_single}
\end{equation}
where \( \bar{N}_{F+B}(z, F \geq F_{\rm lim}) \) is the mean number of galaxies per pixel at redshift \( z \), including both bright and faint populations. Equivalently, $s_{\rm tot}$ can be expressed as
\begin{equation} \label{eq:alter_s}
    s_{\rm tot} = -\frac{2}{5}\frac{\partial\ln n_{\rm g}}{\partial\ln L_{\rm lim}}
    =\frac{\partial\log_{10} n_{\rm g}}{\partial M_{\rm lim}}=\frac{\partial\log_{10} N_{\rm g}}{\partial m_{\rm lim}}   \,     ,
\end{equation}
where $n_{\rm g}$ is the comoving number density at the source, $L$ the luminosity, $M$ the absolute magnitude, and $m$ the apparent magnitude.

When the galaxy sample is split into two populations by a flux cut \( F_{\rm cut} \), the magnification bias for the bright population is given by
\begin{equation}
s_B(z) = -\frac{2}{5} \frac{\partial \ln \bar{N}_B(z, F \geq F_{\rm cut})}{\partial \ln F_{\rm cut}} \,,
\label{eq:s_B}
\end{equation}
where \( \bar{N}_B \) denotes the mean number of bright galaxies per pixel.

The magnification bias for the faint population, \( s_F \), is then determined by subtracting the contribution of the bright galaxies from the total:
\begin{equation}
s_F(z) = s_{\rm tot}(z) \frac{\bar{N}_{F+B}(z)}{\bar{N}_F(z)} - s_B(z) \frac{\bar{N}_B(z)}{\bar{N}_F(z)} \,,
\label{eq:s_F}
\end{equation}
where \( \bar{N}_F (z, F_{\rm cut} > F \geq F_{\rm lim}) \) is the mean number of faint galaxies per pixel. Assuming an equal split between the two populations ( $\bar{N}_B$ =  $\bar{N}_F$), Eq.~\eqref{eq:s_F} reduces to $s_F = 2s_{\rm tot} - s_B$.

Analytical models of the GLF are adopted from Ref.~\cite{Maartens:2021dqy}. The corresponding expressions, along with the resulting galaxy number counts and magnification bias, are presented below.

\textbf{DESI BGS:} The GLF for the DESI BGS, based on Refs. \cite{Jelic-Cizmek:2020pkh, Beutler:2020evf, DESI:2016fyo}, is modeled as a redshift-dependent Schechter function denoted in terms of dimensionless magnitudes, and it is given by
\begin{equation}
\Phi(M, z) = (0.4 \ln 10) \, \phi_*(z) \, 10^{0.4 [M_*(z) - M](1 + \alpha)} \exp\left[-10^{0.4(M_*(z) - M)}\right] \,,
\end{equation}
with
\begin{align*}
\alpha &= -1.23, \\
M_*(z) &= 5 \log_{10} h - 20.64 - 0.6z, \\
\phi_*(z) &= 10^{-2.022 + 0.92z} \; h^3 \, \mathrm{Mpc}^{-3}.
\end{align*}
The threshold absolute magnitude with K-correction related to the apparent magnitude limit $m_{\rm lim}$ via
\begin{equation}
M_{\rm lim}(z) = m_{\rm lim} - 5 \log \left[ \frac{d_L(z)}{10\,\mathrm{pc}} \right] - 0.87z \,   ,
\end{equation}
and the comoving number density and the galaxy magnification bias are given by 
\begin{align}
n_{\rm g}(z, M_{\rm lim}) &= \int_{-\infty}^{M_{\rm lim}(z)} \mathrm{d}M\, \Phi(M, z) \,, \\
s_{\rm tot}(z, M_{\rm lim}) &= \frac{\partial\log_{10}n_{\rm g}(z, M_{\rm lim})}{\partial M_{\rm lim}}= \frac{1}{\ln 10} \frac{\Phi(z, M_{\rm lim})}{n_{\rm g}(z, M_{\rm lim})} \,  .
\end{align}

\textbf{Euclid}: For the Euclid H$\alpha$ spectroscopic survey, we adopt the \textit{2019 luminosity function model} (see also Refs.~(\cite{Pozzetti:2016cch,Euclid:2019clj})). This GLF covers the redshift range \( 0.9 \leq z \leq 1.8 \), and reads
\begin{align}
\Phi(z, y) &= \phi_*(z)\, g(y), \quad \text{with} \quad y \equiv \frac{L}{L_*(z)}, \\
g(y) &= \frac{y^\alpha}{1 + (\mathrm{e} - 1)\, y^\nu} \,.
\end{align}
The redshift evolution of the characteristic luminosity is given by
\begin{align}
\log L_*(z) &= \log L_{*\infty} + \left[ \frac{1.5}{1 + z} \right]^{\beta} \log \left( \frac{L_*(0.5)}{L_{*\infty}} \right), \qquad 
\phi_*(z) = \phi_{*0}, 
\end{align}
where the model parameters are:
\begin{align}
\alpha &= -1.587, \quad \nu = 2.288, \quad \beta = 1.615, \nonumber \\
L_*(0.5) &= 10^{41.733} \; \mathrm{erg\,s^{-1}}, \quad 
L_{*\infty} = 10^{42.956} \; \mathrm{erg\,s^{-1}}, \quad 
\phi_{*0} = 10^{-2.92} \; \mathrm{Mpc}^{-3}. 
\end{align}
The comoving number density and the magnification bias are then given by
\begin{align}
n_{\rm g}(z, y_{\rm lim}) &= \phi_*(z)\, G(y_{\rm lim}), \quad \text{where} \quad 
G(y_{\rm lim}) = \int_{y_{\rm lim}}^{\infty} \mathrm{d}y\, g(y), \\
s_{\rm tot}(z, y_{\rm lim}) &= \frac{2}{5}\, \frac{g(y_{\rm lim})}{G(y_{\rm lim})}y_{\rm lim} \,,
\end{align}
with \( y_{\rm lim}(z) = L_{\rm lim}(z) / L_*(z) \). For a survey with fixed flux limit \( F_{\rm lim} \), the luminosity threshold at a given redshift is 
\begin{equation}
L_{\rm lim}(z) = 4\pi F_{\rm lim} (1 + z)^2 r^2(z) \,.
\end{equation}

\textbf{SKA2:} For a futuristic SKA2-like HI galaxy survey, instead of adopting an explicit HI luminosity function, the expected galaxy number counts are modeled using the S$^3$-SAX simulation, following Ref.~\cite{Yahya:2014yva}. The resulting fitting formula is further adjusted in Ref.~\cite{Maartens:2021dqy} to incorporate the detection threshold:
\begin{equation}
N_g(z, S_{\rm lim}) = 10^{c_1(S_{\rm lim})} z^{c_2(S_{\rm lim})} \exp[-c_3(S_{\rm lim}) z] \; {\rm deg}^{-2} \,,
\end{equation}
where the coefficients \( c_i \) at a given flux sensitivity \( S_{\rm lim} \) can be obtained from interpolating Table~3 of Ref.~\cite{Yahya:2014yva}. The detection threshold is given by \( S_{\rm lim}(z) = S_{\rm rms}(z) \, N_{\rm lim}/10 \), and the rms noise reads
\begin{equation}
S_{\rm rms}(\nu) = \frac{2 k_B T_{\rm sys}(\nu)}{A_{\rm eff} N_d \sqrt{2 t_p(\nu) \delta \nu}} \,, \qquad
T_{\rm sys}(\nu) = T_{\rm rec} + 60 \left( \frac{\nu}{300\,\mathrm{MHz}} \right)^{-2.5} \, {\rm K} \,,
\end{equation}
where $k_B$ is the Boltzmann constant, \( \nu = \nu_{21}/(1 + z) \) with the rest-frame frequency \( \nu_{21} = 1420\,\mathrm{MHz} \). The time per pointing is computed as
\begin{equation}
t_p(z) = t_{\rm tot} \, \frac{\theta_{\rm b}^2(z)}{\Omega_{\rm sky}} \,, \qquad
\theta_{\rm b}^2(z) = \frac{\pi}{8} \left[ 1.3 \frac{\lambda_{21}(1 + z)}{D_d} \right]^2 \,, \qquad
A_{\rm eff} = \epsilon \frac{\pi}{4} D_d^2 \,.
\end{equation}
We adopt the SKA2 specifications: \( T_{\rm rec} = 15\,\mathrm{K} \), \( N_d = 70000 \), \( D_d = 3.1\,\mathrm{m} \), \( \epsilon = 0.81 \), \( t_{\rm tot} = 10{,}000\,\mathrm{hr} \), \( \delta \nu = 10\,\mathrm{kHz} \), \( N_{\rm lim} = 10 \), and \( \Omega_{\rm sky} = 30000\, \mathrm{deg}^2 \). The magnification bias is computed as
\begin{equation}
s_{\rm tot}(z) = -\frac{2}{5} \frac{\partial \ln N_g}{\partial \ln S_{\rm lim}} \,.
\end{equation}

For each galaxy survey, we apply a flux cut independently in each redshift bin. Specifically, we determine the values \( m_{\rm cut} \), \( F_{\rm cut} \), and \( N_{\rm cut} \) that replace the survey limits \( m_{\rm lim} \), \( F_{\rm lim} \), and \( N_{\rm lim} \), such that the number of galaxies is reduced by half in each bin. This yields the magnification bias for the bright population, \( s_B \). The magnification bias for the faint population, \( s_F \), is then obtained by subtracting \( s_B \) from the total magnification bias \( s_{\rm tot} \).

\section{Summary of degeneracies between viscosity parameters and \texorpdfstring{$\eta$, $\mu_{\rm G}$, $\Theta$, $\Gamma$, $E_P$}{eta, muG, Theta, Gamma, EP}} \label{app:summary}

In this appendix, we summarize how dark matter viscosity can mimic the effects of modified gravity parameters, leading to degeneracies in cosmological observables. As in the main text, we denote with a tilde the quantities that are actually measured, while quantities without a tilde refer to those truly associated with modified gravity. 
\begin{align}
\tilde{\Theta} 
&= \frac{\Theta}{1 - \mathcal{Z}} 
  + \frac{\mathcal{Z}}{1 - \mathcal{Z}}\left(1 - \frac{\mathcal{H}'}{\mathcal{H}}\right) 
  + \frac{3}{2} \frac{\Omega_m}{1 - \mathcal{Z}}
    \left[
      \mathcal{E}(1 + \Gamma)
      - \mathcal{Z}\,\eta\bigl(\mathcal{Z} + \mathcal{E} + (1 - 2\mathcal{Z}) \mu_{\rm G}\bigr)
    \right], \\[4pt]
\tilde{\Gamma}
&= \frac{\Gamma}{1 - \mathcal{Z}} 
  + \frac{\mathcal{Z}}{1 - \mathcal{Z}} 
    \left[
      1 - \eta\left(1 - \frac{\eta'}{\eta} - \frac{\mu_{\rm G}'}{\mu_{\rm G}}\right)
    \right], \\[4pt]
\tilde{\mu}_{\rm G} 
&= \mu_{\rm G} - \mathcal{E}\,\frac{f + 3\mathcal{Z}}{1 - 2\mathcal{Z}}, \\[4pt]
\tilde{\eta}
&= \frac{\Phi}{\Psi}
  = \frac{\eta}{1 - \mathcal{E}\,\dfrac{f + 3\mathcal{Z}}{(1 - 2\mathcal{Z})\,\mu_{\rm G}}}, \\[4pt]
\tilde{E}_P 
&= 1 + \tilde{\Theta} 
  - \frac{3}{2}\,\Omega_m\,\mathcal{E} 
  - \frac{3}{2}\,\frac{\Omega_m\,\mu_{\rm G}\,\tilde{\Gamma}(1 - 2\mathcal{Z})}{f + 3\mathcal{Z}} \nonumber \\[2pt]
&= 1 
  + \frac{\mathcal{Z}}{1 - \mathcal{Z}}\left(1 - \frac{\mathcal{H}'}{\mathcal{H}}\right)
  + \frac{\Theta}{1 - \mathcal{Z}}
  - \frac{3}{2}\,\frac{\Omega_m}{1 - \mathcal{Z}}
    \biggl[
      \frac{1 - 2\mathcal{Z}}{f + 3\mathcal{Z}}\,\mu_{\rm G}
      \Bigl(
        \Gamma + \mathcal{Z}\bigl(1 - \eta(1 - \eta'/\eta - \mu_{\rm G}'/\mu_{\rm G})\bigr)
      \Bigr) \nonumber \\[-2pt]
&\quad
      + \mathcal{Z}\,\eta\bigl((1 - 2\mathcal{Z})\,\mu_{\rm G} + \mathcal{Z}\bigr)
      - \mathcal{E}\bigl(\Gamma + \mathcal{Z}(1 - \eta)\bigr)
    \biggr].
\end{align}
Because the parameters $\mathcal{Z}$ and $\mathcal{E}$ (and accordingly $C_{\mathrm{vis},0}$) are very small, the only viscosity terms that are practically relevant come from the growth rate $f$, where the viscous contribution is ``boosted'' by the factor $\lambda^{-2}$ on small scales. Plugging in the approximate solution for the growth rate derived in Appendix \ref{app:app_growth}, and keeping only terms that are at most linear in viscosity and boosted, one arrives at the approximate expressions 
\begin{align}
    \tilde{\Theta}&\simeq\Theta\\
    \tilde{\Gamma}&\simeq\Gamma\\
    \tilde{\mu}_{\rm G}&\simeq\mu_{\rm G}\\
    \tilde{\eta}&\simeq\eta\\
    \tilde{E}_P&\simeq1+\Theta-\frac{3}{2}\Omega_m\frac{\mu_{\rm G}\Gamma}{f_z}\bigl(1+\lambda^{-2}C_{\mathrm{vis},0}E^2\mathcal{I}/f_z\bigr)\,.
\end{align}

%%%%%%%%%%%%%%%%%%%%%%%%%%%%%%%%%%%%%%%%%%%%
%%%%%%%%%%%%%%%%%%%%%%%%%%%%%%%%%%%%%%%%%%%%%

\bibliographystyle{ieeetr} 
\clearpage
\bibliography{references}

@article{Eckart:1940te,
    author = "Eckart, Carl",
    title = "{The Thermodynamics of irreversible processes. 3.. Relativistic theory of the simple fluid}",
    doi = "10.1103/PhysRev.58.919",
    journal = "Phys. Rev.",
    volume = "58",
    pages = "919--924",
    year = "1940"
}

@article{Amendola:2003wa,
    author = "Amendola, Luca",
    title = "{Linear and non-linear perturbations in dark energy models}",
    eprint = "astro-ph/0311175",
    archivePrefix = "arXiv",
    doi = "10.1103/PhysRevD.69.103524",
    journal = "Phys. Rev. D",
    volume = "69",
    pages = "103524",
    year = "2004"
}

@article{Pozzetti_2016,
   title={Modelling the number density of Hαemitters for future spectroscopic near-IR space missions},
   volume={590},
   ISSN={1432-0746},
   url={http://dx.doi.org/10.1051/0004-6361/201527081},
   DOI={10.1051/0004-6361/201527081},
   journal={Astronomy \& Astrophysics},
   publisher={EDP Sciences},
   author={Pozzetti, L. and Hirata, C. M. and Geach, J. E. and Cimatti, A. and Baugh, C. and Cucciati, O. and Merson, A. and Norberg, P. and Shi, D.},
   year={2016},
   month=apr, pages={A3} }

@article{Jelic-Cizmek:2020pkh,
    author = "Jelic-Cizmek, Goran and Lepori, Francesca and Bonvin, Camille and Durrer, Ruth",
    title = "{On the importance of lensing for galaxy clustering in photometric and spectroscopic surveys}",
    eprint = "2004.12981",
    archivePrefix = "arXiv",
    primaryClass = "astro-ph.CO",
    doi = "10.1088/1475-7516/2021/04/055",
    journal = "JCAP",
    volume = "04",
    pages = "055",
    year = "2021"
}

@article{Zheng:2023yco,
    author = "Zheng, Ziyang and Sakr, Ziad and Amendola, Luca",
    title = "{Testing the cosmological Poisson equation in a model-independent way}",
    eprint = "2312.07436",
    archivePrefix = "arXiv",
    primaryClass = "astro-ph.CO",
    doi = "10.1016/j.physletb.2024.138647",
    journal = "Phys. Lett. B",
    volume = "853",
    pages = "138647",
    year = "2024"
}

@article{Sakr:2025kgq,
    author = "Sakr, Ziad and Zheng, Ziyang and Casas, Santiago",
    title = "{Model-independent forecasts for the cosmological anisotropic stress from a combination of Euclid and DESI like surveys}",
    doi = "10.1093/mnras/staf1111",
    journal = "Mon. Not. Roy. Astron. Soc.",
    volume = "541",
    number = "4",
    pages = "2978--2991",
    year = "2025"
}

@article{Euclid:2024yrr,
    author = "Mellier, Y. and others",
    collaboration = "Euclid",
    title = "{Euclid. I. Overview of the Euclid mission}",
    eprint = "2405.13491",
    archivePrefix = "arXiv",
    primaryClass = "astro-ph.CO",
    doi = "10.1051/0004-6361/202450810",
    journal = "Astron. Astrophys.",
    volume = "697",
    pages = "A1",
    year = "2025"
}

@article{Beutler:2020evf,
    author = "Beutler, Florian and Di Dio, Enea",
    title = "{Modeling relativistic contributions to the halo power spectrum dipole}",
    eprint = "2004.08014",
    archivePrefix = "arXiv",
    primaryClass = "astro-ph.CO",
    doi = "10.1088/1475-7516/2020/07/048",
    journal = "JCAP",
    volume = "07",
    number = "07",
    pages = "048",
    year = "2020"
}

@article{Yahya:2014yva,
    author = "Yahya, S. and Bull, P. and Santos, M. G. and Silva, M. and Maartens, R. and Okouma, P. and Bassett, B.",
    title = "{Cosmological performance of SKA HI galaxy surveys}",
    eprint = "1412.4700",
    archivePrefix = "arXiv",
    primaryClass = "astro-ph.CO",
    doi = "10.1093/mnras/stv695",
    journal = "Mon. Not. Roy. Astron. Soc.",
    volume = "450",
    number = "3",
    pages = "2251--2260",
    year = "2015"
}

@article{Pozzetti:2016cch,
    author = "Pozzetti, L. and Hirata, C. M. and Geach, J. E. and Cimatti, A. and Baugh, C. and Cucciati, O. and Merson, A. and Norberg, P. and Shi, D.",
    title = "{Modelling the number density of H{\ensuremath{\alpha}} emitters for future spectroscopic near-IR space missions}",
    eprint = "1603.01453",
    archivePrefix = "arXiv",
    primaryClass = "astro-ph.GA",
    doi = "10.1051/0004-6361/201527081",
    journal = "Astron. Astrophys.",
    volume = "590",
    pages = "A3",
    year = "2016"
}

@article{DESI:2016fyo,
    author = "Aghamousa, Amir and others",
    collaboration = "DESI",
    title = "{The DESI Experiment Part I: Science,Targeting, and Survey Design}",
    eprint = "1611.00036",
    archivePrefix = "arXiv",
    primaryClass = "astro-ph.IM",
    reportNumber = "FERMILAB-PUB-16-517-AE",
    month = "10",
    year = "2016"
}

@article{Koda:2013eya,
    author = "Koda, Jun and Blake, Chris and Davis, Tamara and Magoulas, Christina and Springob, Christopher M. and Scrimgeour, Morag and Johnson, Andrew and Poole, Gregory B. and Staveley-Smith, Lister",
    title = "{Are peculiar velocity surveys competitive as a cosmological probe?}",
    eprint = "1312.1022",
    archivePrefix = "arXiv",
    primaryClass = "astro-ph.CO",
    doi = "10.1093/mnras/stu1610",
    journal = "Mon. Not. Roy. Astron. Soc.",
    volume = "445",
    number = "4",
    pages = "4267--4286",
    year = "2014"
}

@article{Loveday_2011,
   title={Galaxy and Mass Assembly (GAMA): ugriz galaxy luminosity functions: GAMA luminosity functions},
   volume={420},
   ISSN={0035-8711},
   url={http://dx.doi.org/10.1111/j.1365-2966.2011.20111.x},
   DOI={10.1111/j.1365-2966.2011.20111.x},
   number={2},
   journal={Monthly Notices of the Royal Astronomical Society},
   publisher={Oxford University Press (OUP)},
   author={Loveday, J. and Norberg, P. and Baldry, I. K. and Driver, S. P. and Hopkins, A. M. and Peacock, J. A. and Bamford, S. P. and Liske, J. and Bland-Hawthorn, J. and Brough, S. and Brown, M. J. I. and Cameron, E. and Conselice, C. J. and Croom, S. M. and Frenk, C. S. and Gunawardhana, M. and Hill, D. T. and Jones, D. H. and Kelvin, L. S. and Kuijken, K. and Nichol, R. C. and Parkinson, H. R. and Phillipps, S. and Pimbblet, K. A. and Popescu, C. C. and Prescott, M. and Robotham, A. S. G. and Sharp, R. G. and Sutherland, W. J. and Taylor, E. N. and Thomas, D. and Tuffs, R. J. and van Kampen, E. and Wijesinghe, D.},
   year={2011},
   month=dec, pages={1239–1262} }

@article{Amendola:2001rc,
    author = "Amendola, Luca and Tocchini-Valentini, Domenico",
    title = "{Baryon bias and structure formation in an accelerating universe}",
    eprint = "astro-ph/0111535",
    archivePrefix = "arXiv",
    doi = "10.1103/PhysRevD.66.043528",
    journal = "Phys. Rev. D",
    volume = "66",
    pages = "043528",
    year = "2002"
}

@article{Amendola:1999er,
    author = "Amendola, Luca",
    title = "{Coupled quintessence}",
    eprint = "astro-ph/9908023",
    archivePrefix = "arXiv",
    doi = "10.1103/PhysRevD.62.043511",
    journal = "Phys. Rev. D",
    volume = "62",
    pages = "043511",
    year = "2000"
}

@article{Weinberg:1971mx,
    author = "Weinberg, Steven",
    title = "{Entropy generation and the survival of protogalaxies in an expanding universe}",
    journal = "Astrophys. J.",
    volume = "168",
    pages = "175--194",
    year = "1971",
    doi = "10.1086/151073"
}

@article{Maartens:2021dqy,
    author = "Maartens, Roy and Fonseca, Jos{\'e} and Camera, Stefano and Jolicoeur, Sheean and Viljoen, Jan-Albert and Clarkson, Chris",
    title = "{Magnification and evolution biases in large-scale structure surveys}",
    eprint = "2107.13401",
    archivePrefix = "arXiv",
    primaryClass = "astro-ph.CO",
    doi = "10.1088/1475-7516/2021/12/009",
    journal = "JCAP",
    volume = "12",
    number = "12",
    pages = "009",
    year = "2021"
}

@article{Euclid:2019clj,
    author = "Blanchard, A. and others",
    collaboration = "Euclid",
    title = "{Euclid preparation. VII. Forecast validation for Euclid cosmological probes}",
    eprint = "1910.09273",
    archivePrefix = "arXiv",
    primaryClass = "astro-ph.CO",
    doi = "10.1051/0004-6361/202038071",
    journal = "Astron. Astrophys.",
    volume = "642",
    pages = "A191",
    year = "2020"
}

@article{Bull:2015lja,
    author = "Bull, Philip",
    title = "{Extending cosmological tests of General Relativity with the Square Kilometre Array}",
    eprint = "1509.07562",
    archivePrefix = "arXiv",
    primaryClass = "astro-ph.CO",
    doi = "10.3847/0004-637X/817/1/26",
    journal = "Astrophys. J.",
    volume = "817",
    number = "1",
    pages = "26",
    year = "2016"
}

@article{Barbosa_2018,
   title={Modified gravity versus shear viscosity: Imprints on the scalar matter perturbations},
   volume={98},
   ISSN={2470-0029},
   url={http://dx.doi.org/10.1103/PhysRevD.98.123522},
   DOI={10.1103/physrevd.98.123522},
   number={12},
   journal={Physical Review D},
   publisher={American Physical Society (APS)},
   author={Barbosa, C. M. S. and Velten, H. and Fabris, J. C. and Ramos, Rudnei O.},
   year={2018},
   month=dec }

@article{Castello:2022uuu,
    author = "Castello, Sveva and Grimm, Nastassia and Bonvin, Camille",
    title = "{Rescuing constraints on modified gravity using gravitational redshift in large-scale structure}",
    eprint = "2204.11507",
    archivePrefix = "arXiv",
    primaryClass = "astro-ph.CO",
    doi = "10.1103/PhysRevD.106.083511",
    journal = "Phys. Rev. D",
    volume = "106",
    number = "8",
    pages = "083511",
    year = "2022"
}

@article{Quartin:2021dmr,
    author = "Quartin, Miguel and Amendola, Luca and Moraes, Bruno",
    title = "{The 6~{\texttimes}~2pt method: supernova velocities meet multiple tracers}",
    eprint = "2111.05185",
    archivePrefix = "arXiv",
    primaryClass = "astro-ph.CO",
    doi = "10.1093/mnras/stac571",
    journal = "Mon. Not. Roy. Astron. Soc.",
    volume = "512",
    number = "2",
    pages = "2841--2853",
    year = "2022"
}

@article{Normann:2016jns,
    author = "Normann, Ben David and Brevik, Iver",
    title = "{General Bulk-Viscous Solutions and Estimates of Bulk Viscosity in the Cosmic Fluid}",
    eprint = "1601.04519",
    archivePrefix = "arXiv",
    primaryClass = "gr-qc",
    doi = "10.3390/e18060215",
    journal = "Entropy",
    volume = "18",
    pages = "215",
    year = "2016"
}

@article{Pogosian:2010tj,
    author = "Pogosian, Levon and Silvestri, Alessandra and Koyama, Kazuya and Zhao, Gong-Bo",
    title = "{How to optimally parametrize deviations from General Relativity in the evolution of cosmological perturbations?}",
    eprint = "1002.2382",
    archivePrefix = "arXiv",
    primaryClass = "astro-ph.CO",
    doi = "10.1103/PhysRevD.81.104023",
    journal = "Phys. Rev. D",
    volume = "81",
    pages = "104023",
    year = "2010"
}

@article{Amendola:2007rr,
    author = "Amendola, Luca and Kunz, Martin and Sapone, Domenico",
    title = "{Measuring the dark side (with weak lensing)}",
    eprint = "0704.2421",
    archivePrefix = "arXiv",
    primaryClass = "astro-ph",
    doi = "10.1088/1475-7516/2008/04/013",
    journal = "JCAP",
    volume = "04",
    pages = "013",
    year = "2008"
}

@article{Amendola:2013qna,
    author = "Amendola, Luca and Fogli, Simone and Guarnizo, Alejandro and Kunz, Martin and Vollmer, Adrian",
    title = "{Model-independent constraints on the cosmological anisotropic stress}",
    eprint = "1311.4765",
    archivePrefix = "arXiv",
    primaryClass = "astro-ph.CO",
    doi = "10.1103/PhysRevD.89.063538",
    journal = "Phys. Rev. D",
    volume = "89",
    number = "6",
    pages = "063538",
    year = "2014"
}

@article{Velten:2012uv,
    author = "Velten, Hermano and Schwarz, Dominik",
    title = "{Dissipation of dark matter}",
    eprint = "1206.0986",
    archivePrefix = "arXiv",
    primaryClass = "astro-ph.CO",
    doi = "10.1103/PhysRevD.86.083501",
    journal = "Phys. Rev. D",
    volume = "86",
    pages = "083501",
    year = "2012"
}

@article{Barbosa:2017ojt,
    author = "Barbosa, C. M. S. and Velten, H. and Fabris, J. C. and Ramos, Rudnei O.",
    title = "{Assessing the impact of bulk and shear viscosities on large scale structure formation}",
    eprint = "1702.07040",
    archivePrefix = "arXiv",
    primaryClass = "astro-ph.CO",
    doi = "10.1103/PhysRevD.96.023527",
    journal = "Phys. Rev. D",
    volume = "96",
    number = "2",
    pages = "023527",
    year = "2017"
}

@article{Challinor:2011bk,
    author = "Challinor, Anthony and Lewis, Antony",
    title = "{The linear power spectrum of observed source number counts}",
    eprint = "1105.5292",
    archivePrefix = "arXiv",
    primaryClass = "astro-ph.CO",
    doi = "10.1103/PhysRevD.84.043516",
    journal = "Phys. Rev. D",
    volume = "84",
    pages = "043516",
    year = "2011"
}

@article{Sobral-Blanco:2021cks,
    author = "Sobral-Blanco, Daniel and Bonvin, Camille",
    title = "{Measuring anisotropic stress with relativistic effects}",
    eprint = "2102.05086",
    archivePrefix = "arXiv",
    primaryClass = "astro-ph.CO",
    doi = "10.1103/PhysRevD.104.063516",
    journal = "Phys. Rev. D",
    volume = "104",
    number = "6",
    pages = "063516",
    year = "2021"
}

@article{Tutusaus:2022cab,
    author = "Tutusaus, Isaac and Sobral-Blanco, Daniel and Bonvin, Camille",
    title = "{Combining gravitational lensing and gravitational redshift to measure the anisotropic stress with future galaxy surveys}",
    eprint = "2209.08987",
    archivePrefix = "arXiv",
    primaryClass = "astro-ph.CO",
    doi = "10.1103/PhysRevD.107.083526",
    journal = "Phys. Rev. D",
    volume = "107",
    number = "8",
    pages = "083526",
    year = "2023"
}

@article{Castello:2024jmq,
    author = "Castello, Sveva and Wang, Zhuangfei and Dam, Lawrence and Bonvin, Camille and Pogosian, Levon",
    title = "{Disentangling modified gravity from a dark force with gravitational redshift}",
    eprint = "2404.09379",
    archivePrefix = "arXiv",
    primaryClass = "astro-ph.CO",
    doi = "10.1103/PhysRevD.110.103523",
    journal = "Phys. Rev. D",
    volume = "110",
    number = "10",
    pages = "103523",
    year = "2024"
}

@article{Velten:2014xca,
    author = "Velten, Hermano and Caram{\^e}s, Thiago R. P. and Fabris, J{\'u}lio C. and Casarini, Luciano and Batista, Ronaldo C.",
    title = "{Structure formation in a $\Lambda$ viscous CDM universe}",
    eprint = "1410.3066",
    archivePrefix = "arXiv",
    primaryClass = "astro-ph.CO",
    doi = "10.1103/PhysRevD.90.123526",
    journal = "Phys. Rev. D",
    volume = "90",
    number = "12",
    pages = "123526",
    year = "2014"
}

@article{Velten:2013pra,
    author = "Velten, H. and Schwarz, D. J. and Fabris, J. C. and Zimdahl, W.",
    title = "{Viscous dark matter growth in (neo-)Newtonian cosmology}",
    eprint = "1307.6536",
    archivePrefix = "arXiv",
    primaryClass = "astro-ph.CO",
    doi = "10.1103/PhysRevD.88.103522",
    journal = "Phys. Rev. D",
    volume = "88",
    number = "10",
    pages = "103522",
    year = "2013"
}

@article{Ashoorioon:2023jwf,
    author = "Ashoorioon, Amjad and Davari, Zahra",
    title = "{Dark Matter Cosmology with Varying Viscosity: A Possible Resolution to the S $_{8}$ Tension}",
    eprint = "2303.06627",
    archivePrefix = "arXiv",
    primaryClass = "astro-ph.CO",
    doi = "10.3847/1538-4357/ad0372",
    journal = "Astrophys. J.",
    volume = "959",
    number = "2",
    pages = "120",
    year = "2023"
}

@article{AparicioResco:2020shd,
    author = "Aparicio Resco, Miguel and Maroto, Antonio L.",
    title = "{Modified gravity or imperfect dark matter: a model-independent discrimination}",
    eprint = "2010.01368",
    archivePrefix = "arXiv",
    primaryClass = "astro-ph.CO",
    doi = "10.1088/1475-7516/2021/02/020",
    journal = "JCAP",
    volume = "02",
    pages = "020",
    year = "2021"
}

@article{Anand:2017wsj,
    author = "Anand, Sampurn and Chaubal, Prakrut and Mazumdar, Arindam and Mohanty, Subhendra",
    title = "{Cosmic viscosity as a remedy for tension between PLANCK and LSS data}",
    eprint = "1708.07030",
    archivePrefix = "arXiv",
    primaryClass = "astro-ph.CO",
    doi = "10.1088/1475-7516/2017/11/005",
    journal = "JCAP",
    volume = "11",
    pages = "005",
    year = "2017"
}

@article{Goswami:2016tsu,
    author = "Goswami, Gaurav and Chakravarty, Girish Kumar and Mohanty, Subhendra and Prasanna, A. R.",
    title = "{Constraints on cosmological viscosity and self interacting dark matter from gravitational wave observations}",
    eprint = "1603.02635",
    archivePrefix = "arXiv",
    primaryClass = "hep-ph",
    doi = "10.1103/PhysRevD.95.103509",
    journal = "Phys. Rev. D",
    volume = "95",
    number = "10",
    pages = "103509",
    year = "2017"
}

@article{Chen:2013vea,
    author = "Chen, Yun and Geng, Chao-Qiang and Cao, Shuo and Huang, Yu-Mei and Zhu, Zong-Hong",
    title = "{Constraints on a $\phi$CDM model from strong gravitational lensing and updated Hubble parameter measurements}",
    eprint = "1312.1443",
    archivePrefix = "arXiv",
    primaryClass = "astro-ph.CO",
    doi = "10.1088/1475-7516/2015/02/010",
    journal = "JCAP",
    volume = "02",
    pages = "010",
    year = "2015"
}

@article{Brout:2022vxf,
    author = "Brout, Dillon and others",
    title = "{The Pantheon+ Analysis: Cosmological Constraints}",
    eprint = "2202.04077",
    archivePrefix = "arXiv",
    primaryClass = "astro-ph.CO",
    doi = "10.3847/1538-4357/ac8e04",
    journal = "Astrophys. J.",
    volume = "938",
    number = "2",
    pages = "110",
    year = "2022"
}

@article{Planck:2018vyg,
    author = "Aghanim, N. and others",
    collaboration = "Planck",
    title = "{Planck 2018 results. VI. Cosmological parameters}",
    eprint = "1807.06209",
    archivePrefix = "arXiv",
    primaryClass = "astro-ph.CO",
    doi = "10.1051/0004-6361/201833910",
    journal = "Astron. Astrophys.",
    volume = "641",
    pages = "A6",
    year = "2020",
    note = "[Erratum: Astron.Astrophys. 652, C4 (2021)]"
}

@article{Bonvin:2023jjq,
    author = "Bonvin, Camille and Lepori, Francesca and Schulz, Sebastian and Tutusaus, Isaac and Adamek, Julian and Fosalba, Pablo",
    title = "{A case study for measuring the relativistic dipole of a galaxy cross-correlation with the Dark Energy Spectroscopic Instrument}",
    eprint = "2306.04213",
    archivePrefix = "arXiv",
    primaryClass = "astro-ph.CO",
    doi = "10.1093/mnras/stad2567",
    journal = "Mon. Not. Roy. Astron. Soc.",
    volume = "525",
    number = "3",
    pages = "4611--4627",
    year = "2023"
}

@article{Howlett:2017asw,
    author = "Howlett, Cullan and Robotham, Aaron S. G. and Lagos, Claudia D. P. and Kim, Alex G.",
    title = "{Measuring the growth rate of structure with Type IA Supernovae from LSST}",
    eprint = "1708.08236",
    archivePrefix = "arXiv",
    primaryClass = "astro-ph.CO",
    doi = "10.3847/1538-4357/aa88c8",
    journal = "Astrophys. J.",
    volume = "847",
    number = "2",
    pages = "128",
    year = "2017"
}

@article{Bonvin:2018ckp,
    author = "Bonvin, Camille and Fleury, Pierre",
    title = "{Testing the equivalence principle on cosmological scales}",
    eprint = "1803.02771",
    archivePrefix = "arXiv",
    primaryClass = "astro-ph.CO",
    doi = "10.1088/1475-7516/2018/05/061",
    journal = "JCAP",
    volume = "05",
    pages = "061",
    year = "2018"
}

@article{Grimm:2025fwc,
    author = "Grimm, Nastassia and Bonvin, Camille and Tutusaus, Isaac",
    title = "{Does dark matter fall in the same way as standard model particles? A direct constraint of Euler's equation with cosmological data}",
    eprint = "2502.12843",
    archivePrefix = "arXiv",
    primaryClass = "astro-ph.CO",
    month = "2",
    year = "2025"
}

@article{Castello:2023zjr,
    author = "Castello, Sveva and Mancarella, Michele and Grimm, Nastassia and Sobral-Blanco, Daniel and Tutusaus, Isaac and Bonvin, Camille",
    title = "{Gravitational redshift constraints on the effective theory of interacting dark energy}",
    eprint = "2311.14425",
    archivePrefix = "arXiv",
    primaryClass = "astro-ph.CO",
    doi = "10.1088/1475-7516/2024/05/003",
    journal = "JCAP",
    volume = "05",
    pages = "003",
    year = "2024"
}

@article{Castello:2024lhl,
    author = "Castello, Sveva and Zheng, Ziyang and Bonvin, Camille and Amendola, Luca",
    title = "{Testing the equivalence principle across the Universe: A model-independent approach with galaxy multitracing}",
    eprint = "2412.08627",
    archivePrefix = "arXiv",
    primaryClass = "astro-ph.CO",
    doi = "10.1103/1my7-zklj",
    journal = "Phys. Rev. D",
    volume = "111",
    number = "12",
    pages = "123559",
    year = "2025"
}

@article{Bonvin_2011,
   title={What galaxy surveys really measure},
   volume={84},
   ISSN={1550-2368},
   url={http://dx.doi.org/10.1103/PhysRevD.84.063505},
   DOI={10.1103/physrevd.84.063505},
   number={6},
   journal={Physical Review D},
   publisher={American Physical Society (APS)},
   author={Bonvin, Camille and Durrer, Ruth},
   year={2011},
   month=sep }

@article{Challinor_2011,
   title={Linear power spectrum of observed source number counts},
   volume={84},
   ISSN={1550-2368},
   url={http://dx.doi.org/10.1103/PhysRevD.84.043516},
   DOI={10.1103/physrevd.84.043516},
   number={4},
   journal={Physical Review D},
   publisher={American Physical Society (APS)},
   author={Challinor, Anthony and Lewis, Antony},
   year={2011},
   month=aug }

@article{Yoo_2009,
   title={New perspective on galaxy clustering as a cosmological probe: General relativistic effects},
   volume={80},
   ISSN={1550-2368},
   url={http://dx.doi.org/10.1103/PhysRevD.80.083514},
   DOI={10.1103/physrevd.80.083514},
   number={8},
   journal={Physical Review D},
   publisher={American Physical Society (APS)},
   author={Yoo, Jaiyul and Fitzpatrick, A. Liam and Zaldarriaga, Matias},
   year={2009},
   month=oct }

@misc{code,
  title        = {Fisher matrix code for probing dark matter viscosity, equivalence principle, and modified gravity
},
  note         = {Available at~\url{https://github.com/zzyheidelberg/Fisher_DMviscosity_EP}}
}
\end{document}